\documentclass[onecolumn,10pt]{IEEEtran}
\usepackage{cite}
\usepackage{amsmath,amssymb,amsfonts}
\usepackage{graphicx}
\usepackage{subcaption}
\usepackage{cleveref}
\usepackage{color, xcolor}
\usepackage{multirow}
\usepackage{pstricks}

\usepackage{amsthm}
\usepackage{nicematrix}
\usepackage{bm}
\usepackage{booktabs}
\usepackage{multirow}
\usepackage{algorithm}
\usepackage{algpseudocode}
\usepackage{caption}
\usepackage{manyfoot}
\usepackage{blkarray}
\usepackage{mathrsfs}
\usepackage{diagbox}
\newtheorem{theorem}{Theorem}
\newtheorem{lemma}{Lemma}
\newtheorem{definition}{Definition}
\newtheorem{corollary}{Corollary}

\newtheorem{example}{Example}
\newtheorem{remark}{Remark}

\newtheorem{proposition}{Proposition}

\newcommand{\tabcaption}{\def\@captype{table}\caption}

\begin{document}
% Do not fill math or special symbols in the title.
%\title{coded caching}
%: COMBINATORIAL DESIGN

\title{A Construction Framework of Coded Caching Scheme for Multi-Access MISO Systems via Knapsack Problem}
\author{Siying Luo, Youlong Wu~\IEEEmembership{Member,~IEEE,} Mingming Zhang, Minquan Cheng~\IEEEmembership{Member,~IEEE,}  Dianhua Wu
	\thanks{S. Luo, M. Zhang and M. Cheng are with the Key Lab of Education Blockchain and Intelligent Technology, Ministry of Education, and also with the Guangxi Key Lab of Multi-source Information Mining $\&$ Security, Guangxi Normal University, 541004 Guilin, China (e-mail: luosiying01@163.com, zhangmm07@gxnu.edu.cn, chengqinshi@hotmail.com). 
	}
\thanks{Y. Wu is with the School of Information Science and Technology, ShanghaiTech University, Shanghai 201210, China (e-mail: wuyl1@shanghaitech.edu.cn). The work of Y. Wu was supported in part by the Research Fund of Key Lab of Education Blockchain and Intelligent Technology, Ministry of Education under grant EBME25-F-05, and by the National Nature Science Foundation of China (NSFC) under grant 62571329.}
\thanks{D. Wu,  School of Mathematics and Statistics, Guangxi Normal University, 541004 Guilin, China (e-mail: dhwu@gxnu.edu.cn)}
}
\maketitle

\begin{abstract}
This paper investigates the coded caching problem in a multi-access multiple-input single-output (MAMISO) network with the combinatorial topology. 
The considered system consists of a server containing $N$ files, $\Lambda$ cache nodes, and $K$ cache-less users, where each user can access a unique subset of $r$ cache nodes. The server is equipped with $L$ transmit antennas. 
Our objective is to design a caching scheme that simultaneously achieves a high sum Degree of Freedom (sum-DoF) and low subpacketization complexity. 
To address this challenge, we formulate the design of multi-antenna placement delivery arrays (MAPDA) as a $0$–$1$ knapsack problem to maximize the achievable DoF, thereby transforming the complex combinatorial caching structure into a tractable optimization framework that yields efficient cache placement and flexible delivery strategies. 
Theoretical and numerical analyses demonstrate that: for networks with combinatorial topologies, the proposed scheme achieves a higher sum-DoF than existing schemes. Under identical cache size constraints, the subpacketization level remains comparable to existing linear subpacketization schemes. Moreover, under specific system conditions, the proposed scheme attains the theoretical maximum sum-DoF of $\min\{L+KM/N, K\}$ while achieving further reductions subpacketization. 
For particular combinatorial structures, we further derive optimized constructions that achieve even higher sum-DoF with lower subpacketization.

\end{abstract}

\section{Introduction}
With the rapid evolution of immersive, intelligent, and ubiquitous 5G and 6G services, the explosive growth of data traffic and the high-density deployment of devices are expected to impose unprecedented pressure on communication systems. 
To alleviate the resulting congestion, Maddah-Ali and Niesen (MN) proposed the concept of coded caching \cite{MN}, which effectively reduces network load by pre-caching some content during off-peak periods and transmitting coded combinations during peak periods. In the classical single-input single-output (SISO) shared-link model, a central server connects to $K$ cache-aided users via an error-free shared link. The server holds  a library  of $N$ files, and each user is equipped with a cache of size $M$ files. 
During the placement phase, each user stores a subset of file packets without prior knowledge of future demands. During the delivery phase, each user requests one file from the library, and the server multicasts messages formed by coded packets according to the users' cache contents and requests, ensuring all demands are satisfied. 
We define the coded caching gain as the average number of users served by one multicast message.
%{\blue More specifically, for a cache size $M = Kt/N$ with $t \in \{0,1,\ldots,K-1\}$, each transmitted coded packet simultaneously serves $KM/N + 1$ users, yielding a coded caching gain of $KM/N + 1$. The subpacketization level of a coded caching scheme is defined as the number of subfiles into which each file must be partitioned. The MN scheme achieves a normalized delivery load of $K(1-M/N)/(1+KM/N)$ with a subpacketization level of $\binom{K}{KM/N}$. }

However, the subpacketization level of the MN scheme grows exponentially with $K$, resulting in prohibitive system complexity and limiting its practicality in large-scale user scenarios. In order to reduce the subpacketization, several coded caching schemes have been proposed in the literature \cite{YCMTC,CJYT,CJWQY,CWZW,WCWC,LRA,CWWC,SGZG,SKTADA,YTCC,KP,CLTW}. Among these works, \cite{YCMTC} introduced the placement delivery array (PDA), a combinatorial framework in which each array entry is either an integer or the special symbol ``$*$''. The entries marked by ``$*$'' denote packets stored in users’ caches, while integer entries indicate packets jointly encoded for multicast transmission. The MN scheme can also be represented as a special case of PDA, commonly referred to as the MN-PDA.
Subsequent works have extended this concept, proposing PDA constructions based on linear block codes \cite{LRA,CWWC}, $(6,3)$-free hypergraphs \cite{SGZG}, Ruzsa–Szemerédi graphs \cite{SKTADA}, strong edge colorings of bipartite graphs \cite{YTCC}, projective spaces \cite{KP}, and other combinatorial designs \cite{CLTW}. 
%These schemes, along with the MN scheme, typically rely on uncoded cache placement and one-shot delivery. More recently, researchers have explored coded cache placement strategies \cite{CFLK,CJ,JV} to further reduce the delivery latency.

Coded caching has been extended beyond shared-link networks to wireless interference channels, where it is applied to multi–single-antenna cache-assisted transceiver systems \cite{NNMA,HNDS} with the goal of maximizing the system sum Degree of Freedom (sum-DoF). When each transmitter has access to the entire content library, the model reduces to a cache-assisted multiple-input single-output (MISO) broadcast channel (BC) with 
$L$ antennas, a setting that has been extensively investigated in \cite{SSCHB,LEEP,SMPESSEP,MSBI,SMTA,SMASKJ,PEJHC,YWCC,NKPR}.
In such systems, one-shot linear coding schemes that jointly exploits coded caching and zero-forcing (ZF) precoding can achieve a sum-DoF of $\min\{L + KM/N,\, K\}$ \cite{NNMA}, which reduces to the classical MN coded caching gain when $L=1$. 
Moreover, it has been proven in \cite{LEBAP} that, under uncoded cache placement and one-shot linear delivery, this sum-DoF expression is information-theoretically optimal.
To systematically describe the coded caching structure in multi-antenna broadcast networks, \cite{YWCC} and \cite{NKPR} introduced the multi-antenna placement delivery array (MAPDA), which generalizes the concept of PDA to accommodate one-shot linear delivery in MISO broadcast channels. Building on this framework, both studies proposed MAPDA-based schemes that achieve the optimal sum-DoF while significantly reducing the subpacketization level compared to earlier approaches.

Most existing studies assume that each user is connected to a dedicated cache node. However, in many emerging communication scenarios—such as device-to-device (D2D) networks, cell-edge environments in cellular systems, and vehicular networks—users may have access to multiple cache nodes simultaneously.
To model such situations, the authors of \cite{HKNDS} introduced the multi-access coded caching (MACC) framework under a shared-link setting, in which cache nodes are no longer dedicated to individual users but can be jointly accessed by multiple users.
The MACC model consists of a server with $N$ files, $\Lambda$ cache nodes, and $K$ cache-less users, where each user is connected to $r$ cache nodes determined by the underlying network topology.
Similar to the classical coded caching setting, each user decodes multicast messages using the content stored in their accessible cache nodes without incurring additional transmission load. 
Under the cyclic wrap-around topology where each user connects to $r$ consecutive cache nodes, substantial progress in coded caching gains was achieved in \cite{SBPEP,RKKN,SRB,MASB,RK}. Beyond this topology, \cite{PDKR} investigated combinatorial access structures in which each user is connected to a distinct subset of $r$ cache nodes, yielding $K=\binom{\Lambda}{r}$, while \cite{CWEC} further developed coded caching schemes for more general structures such as $t$-designs and $t$-group divisible designs. 
When the same concept is applied to the MISO setting, allowing users to access multiple cache nodes gives rise to the multi-access multiple-input single-output (MAMISO) network. 
Several MAMISO schemes have been proposed, including those based on cyclic wrap-around topologies \cite{PEKKS,CBWC,WCC} and others employing combinatorial access structures \cite{PERB}.

Among existing schemes, \cite{YWCC} and \cite{NKPR} achieves the maximum sum-DoF of $\min\{L + KM/N, K\}$ with lower subpacketization compared to earlier schemes \cite{NNMA,SMASKJ,PEJHC}. However, their subpacketization level remain considerably high. 
Moreover, the coded caching gain attainable under cyclic topologies remains limited \cite{PEKKS,CBWC,WCC}.  
For combinatorial topologies \cite{PERB}, the delivery phase design is only effective when $L=r+1$.
 
\subsection{Our Contributions}
This paper investigates the MAMISO coded caching problem under combinatorial topologies, aiming to design new schemes that achieve a high sum-DoF while maintaining low subpacketization. 
We propose a novel construction method for MAPDA, whose core breakthrough lies in the joint optimization of cache placement structure and DoF maximization through an optimization model inspired by the 0–1 knapsack problem, essentially formulated as an integer programming task.
Based on the solution, we further develop a flexible delivery strategy that adapts to various combinatorial network configurations. 
The proposed scheme not only achieves excellent performance, but also attains the theoretical maximum sum-DoF of $ \min\{L + KM/N, K\} $ under certain conditions while maintaining a significantly reduced subpacketization level.
The main contributions are summarized as follows:
\begin{itemize}
	\item Optimization-Driven General Framework for MAPDA Construction: 
	In the placement phase, we introduce the user-retrieve array and its subarrays to characterize the combinatorial structure of users’ accessible cache contents. By formulating the selection of these subarrays as an integer optimization problem, analogous to a $0–1$ knapsack problem, the complex combinatorial design is transformed into a tractable mathematical program aimed at maximizing the system sum-DoF.
	The resulting optimization yields an optimal MAPDA structure, which enables deep combinatorial coordination between cache resources and antenna resources. 
	Theoretically, we prove that when the system parameters $ \Lambda $ and $ L $ satisfy $\Lambda\geq 2r+\Lambda M/N$ and $L\geq{\Lambda-\Lambda M/N\choose r}-{\Lambda-\Lambda M/N-r\choose r}$, the proposed scheme attains the maximum theoretical sum-DoF of $ KM/N + L $.

	\item Exploiting Special Combinatorial Structures for Enhanced Performance: Within the proposed framework, we further identify and analyze a class of special combinatorial structures. 
	Leveraging these properties allows us to bypass complex optimization procedures and directly construct high-performance MAPDAs that offer further subpacketization reduction. Notably, this family of constructions generalizes and extends the coded caching scheme in \cite{PERB}.
\end{itemize}

\subsection{Paper Organization}
The rest of this paper is organized as follows. Section \ref{system-model} describes the system model. Section \ref{structure-of-MAPDA} proposes three constructions of MAPDA and gives the performance analysis. Section \ref{example-of-MAPDA} provides examples. Section \ref{proof-of-theorem} provides a proof of the three MAMISO coded caching schemes. Other proofs are in the Appendices.
\subsection{Notations}
Let bold capital letter, bold lowercase letter, and curlicue letter  denote array, vector, and set respectively; let $|\mathcal{A}|$ denote the cardinality of the set $\mathcal{A}$; define  $[a]=\{1,2,\ldots,a\}$ and $[a:b]$ is the set $\{a,a+1,\dots,b-1,b\}$; ${[b]\choose t} = \big\{ \mathcal{V} \subseteq [b] \, \big| \, |\mathcal{V}| = t \big\}$, i.e., ${[b]\choose t}$ is the collection of all distinct $t$-subsets of $[b]$; when $t=0$, ${[b]\choose t}=\emptyset$, ${b\choose t}=1$;  when $t>b$, ${[b]\choose t}=\emptyset$, ${b\choose t}=0$; $< \cdot >_a$ represents the modulo operation with integer quotient $a > 0$; For any $ F \times m $ array $ \mathbf{P} $, for any integers $i \in [F] $ and $j \in [m] $, $ \mathbf{P}(i,j) $ represents the element located in the $ i^{\text{th}} $ row and the $ j^{\text{th}} $ column of $ \mathbf{P}$; $ \mathbf{P}(\mathcal{V}, \mathcal{T}) $ represents the Subarray generated by the row indices in $ \mathcal{V} \subseteq [F] $ and the columns indices in $ \mathcal{T} \subseteq [m] $; The Least Common Multiple (LCM) of  integers $N_1,N_2,\cdots,N_n$, denoted as $\mathrm{LCM}(N_1,N_2,\cdots,N_n)$, is the smallest positive integer that is divisible by all $N_i$, $i \in [n]$.  

\section{Preliminaries: MAMISO Coded Caching and  Multi-Antenna Placement Delivery Array}\label{system-model}
In this section, we first review the dedicated MISO coded caching scenario in \cite{NNMA}, where each user has its own cache (i.e., the number of cache nodes and the number of users are equal, while each user accesses a different cache node). We then present the MAMISO model considered in this paper, where each user can access an arbitrary subset of cache nodes. 
\subsection{Dedicated MISO coded caching}\label{MISO-ccs}
In a $(L,K,M,N)$ dedicated MISO coded caching introduced in \cite{NNMA}, a server with $L$ antennas has access to the library containing $N$ files, denoted by $\mathcal{W} = \{\mathbf{w}_1,\mathbf{w}_2, \dots, \mathbf{w}_N\}$. The server serves $K$ users with $K\leq N$ where each user
has a cache of size $M$ files where $0 \leq M \leq N$. A $F$-division coded caching scheme contains the following two phases.
\begin{itemize}
	\item \textbf{Placement phase}: Each file is divided into $F$ equal-sized packets, i.e., $\mathbf{w}_n =(\mathbf{w}_{n,j})_{j \in [F]}$ where $n \in [N]$ and $\mathbf{w}_{n,j} \in \mathbb{F}_2^B$ satisfies the independent, identically, and uniformly distributed conditions for some integer $B$. The server places some packets into each user' cache without any prior knowledge of later demands. Let $\mathcal{Z}_{k}$ represent packet set stored by user $k$ where $k \in[K]$.
	
	\item \textbf{Delivery phase}: Each user requests an arbitrary file from the server. The file requested by user $k \in [K]$ is denoted as $\mathbf{w}_{d_k}$ where $d_k \in [N]$. Then the request vector for all users can be represented as $\mathbf{d} \triangleq (d_1, d_2, \ldots, d_K)$. The server transmits data to users via $L$ antennas based on their requests and cached content. More specifically, the server first encodes each
	packet as $\tilde{\mathbf{w}}_{n,f} \triangleq \psi(\mathbf{w}_{n,f})$ using Gaussian channel coding with rate $B / \tilde{B} = \log P + o(\log P)$ (bit per complex symbol). By assuming $P$ is large enough, it can be seen that each coded packet carries one Degree-of-Freedom (DoF). The server divides the coded packets transmission process into $S$ blocks, each of which consists of $\tilde{B}$ complex symbols (i.e., $\tilde{B}$ time slots). For each block $s \in [S]$, the server delivers a subset of requested packets to $r_s$ users. In this paper, we consider the one-shot delivery strategy, i.e., each user can decode a desired packet from at most one transmitted message with the help of its cache. Then at block $s$ the requested packets and their requesting users can be denoted by 
\begin{align*}
\mathcal{W}^{(s)} = \{\tilde{\mathbf{w}}_{d_{k_1},f_1}, \tilde{\mathbf{w}}_{d_{k_2},f_2}, \ldots, \tilde{\mathbf{w}}_{d_{k_{r_s}},f_{r_s}}\},\ \ \ 	\mathcal{K}^{(s)} = \{k_1, k_2, \ldots, k_{r_s}\} 
\end{align*}respectively. Clearly $|\mathcal{W}^{(s)}| = |\mathcal{K}^{(s)}| = r_s$. Then, at block $s$ the $l$th where $l\in[L]$ antenna sends the following linear combination 
\begin{align}
\label{sends}
\mathbf{x}^{(s)}_l =: \sum_{j \in [r_s]} v^{(s)}_{l,k_j} \tilde{\mathbf{w}}_{d_{k_j},f_j},
\end{align} to the users, where each $v_{l,k_j}(s)$ is a scalar complex coefficient from $\mathbb{C}$. Define the precoding matrix $\mathbf{V}^{(s)}=(v^{(s)}_{l,k_j})_{l\in[L],j\in[r_s]}$. The transmission by all antennas in block $s$ can be written as
\begin{align}\label{transmission}
\mathbf{X}^{(s)} = \begin{pmatrix}
	\mathbf{x}_1^{(s)} \\
	\mathbf{x}_2^{(s)} \\
	\vdots \\
	\mathbf{x}_L^{(s)}
\end{pmatrix}
=\mathbf{V}^{(s)}
\begin{pmatrix}
	\tilde{\mathbf{w}}_{d_{k_1},f_1}  \\
	\tilde{\mathbf{w}}_{d_{k_2},f_2}\\
	\vdots \\
	\tilde{\mathbf{w}}_{d_{k_{r_s}},f_{r_s}}
\end{pmatrix} =\mathbf{V}^{(s)}\tilde{\mathbf{W}}^{(s)}.
\end{align}Each user $k$ can receive the coded signal 
\begin{align}\label{received}
\mathbf{y}^{(s)}_k = \sum_{l=1}^{L} h_{k,l} \mathbf{x}^{(s)}_l + \boldsymbol{\epsilon}^{(s)}_k \in \mathbb{C}^{\tilde{B}}, 
\end{align}	where $h_{k,l}$ is the channel coefficient in an independent and identically distributed over $\mathbb{C}$, and $\boldsymbol{\epsilon}_k(s)$ represents the random noise vector at user $k$ in block $s$. Define the channel matrix $\mathbf{H}^{(s)}=(h^{(s)}_{k,l})_{k\in{K},l\in[L]}$ from the server's antennas to the users. Then from \eqref{sends} and \eqref{received}, the received signals by all users in $\mathcal{K}^{(s)}$ can be written as follows, 
\begin{align}
	\label{eq-received}
		\mathbf{Y}^{(s)} = 
		\begin{pmatrix}
			\mathbf{y}_{k_1}^{(s)} \\
			\mathbf{y}_{k_2}^{(s)} \\
			\vdots \\
			\mathbf{y}_{k_{r_s}}^{(s)}
		\end{pmatrix}
		=\mathbf{H}(\mathcal{K}^{(s)}, [L])^{(s)}\mathbf{V}^{(s)}\tilde{\mathbf{W}}^{(s)}
		+
		\begin{pmatrix}
			\boldsymbol{\epsilon}_{k_1}^{(s)} \\
			\boldsymbol{\epsilon}_{k_2}^{(s)} \\
			\vdots \\
			\boldsymbol{\epsilon}_{k_{r_s}}^{(s)}
		\end{pmatrix}.
\end{align} Since each entry of $\mathbf{H}$ is randomly independent and identically distributed  chosen in $\mathbb{C}$, any subsquare of $\mathbf{H}$ is invertible with high probability. If each user $k \in \mathcal{K}^{(s)}$ can utilize the cache content to cancel the interference from $\mathbf{y}_k^{(s)}$, it ``sees'' the output of an equivalent point-to-point Gaussian channel given by
\begin{align*}
\mathcal{L}_{k}^{(s)}\big(\mathbf{y}_k^{(s)},\tilde{\mathcal{Z}}_k\big) = \tilde{\mathbf{w}}_{d_k,f} + \boldsymbol{\epsilon}_k^{(s)},
\end{align*}where $\tilde{\mathcal{Z}}_k$ represents the coded packets cached by user $k$. Since $\tilde{\mathbf{w}}_{d_k,f}$ is encoded by a rate of $\log P + o(\log P)$, each packet can be decoded with vanishing error probability as $B$ increases.  

In the considered MISO system, each coded packet carries one DoF, and a sum-DoF achieved in block $s$ is $|\mathcal{W}^{(s)}|=r_s$. The server divides the coded packets transmission process into $S$ blocks, the sum-DoF is defined as $\frac{\sum_{s=1}^{S}r_s}{S}$. The fundamental objective is to maximize the achievable sum-DoF.

\end{itemize}

\subsection{Multi-Antenna Placement Delivery Array}
This subsection briefly reviews the concept of MAPDA and the scheme generated by MAPDA.
\begin{definition}[\cite{YWCC}]\label{def-MAPDA}\rm
	For positive integers $L$, $K$, $F$, $Z$ and $S$, an $F \times K$ array $\mathbf{P}$ composed of a specific symbol "*" and integers in $[S]$ is called a $(L,K,F,Z,S)$ multiple-antenna placement delivery array(MAPDA) if it satisfies the following conditions:
	\begin{itemize}
		\item[C$1$]The symbol ``$*$" appears $Z$ times in each column;
		\item[C$2$] Each integer occurs at least once in the array;
		\item[C$3$] Each integer $s$ appears at most once in each column;
		\item[C$4$] For any integer $s\in[S]$, define  $\mathbf{P}^{(s)} $
		to be the subarray of $\mathbf{P}$ including the rows and columns containing $s$, and let $r'_s\times r_s$ denote the dimensions of $\mathbf{P}^{(s)}$.  The number of integer entries in each row  of $\mathbf{P}^{(s)}$ is less than or equal to $L$, i.e.,
		\begin{eqnarray}\label{C4}
			\left|\{k_1\in [r_s]  |\ \mathbf{P}^{(s)}(f_1,k_1)\in[S]\}\right|\leq L, \ \forall f_1 \in [r'_s].
		\end{eqnarray}    \hfill $\square$  
	\end{itemize} 
\end{definition}

If each integer appears $g$ times in the $\mathbf{P}$, then $\mathbf{P}$ is regular, denoted by $g$-$(L, K, F, Z, S)$ MAPDA. Let us take an example to further illustrate the concept of MAPDA and the scheme generated by a MAPDA. 
\begin{example}\rm\label{exam-MAPDA}
When $(L,K,F,Z,S)=(2, 4, 4, 2, 2)$, we consider the following array
\begin{align}
\label{example-MAPDA}
\mathbf{P}= \footnotesize
\begin{blockarray}{ccccc}
\{1\} & \{2\} & \{3\} & \{4\}\\
\begin{block}{(cccc)c}
	* & * & 1 & 1 & \{1\} \\
	* & 2 & * & 2 & \{2\} \\
	2 & * & 2 & * & \{3\} \\
	1 & 1 & * & * & \{4\} \\
\end{block}
\end{blockarray},
\ \ \ \ 
\mathbf{P}^{(1)}= 
\left(\begin{array}{cccc}
	* & * & 1 & 1\\
	1 & 1 & * & *\\ 
\end{array}	\right).
\end{align}We can see that there are exactly $S=2$ different integers, the star appears exactly twice in each column, i.e., $Z=2$, and each integer occurs exactly $g=4$ times and occurs at most once in each column. Furthermore, In the sub-array $\mathbf{P}^{(1)}$, each row has exactly $L=2$ integer entries. By Definition \ref{def-MAPDA}, $\mathbf{P}$ is a $4$-$(2, 4, 4, 2, 2)$ MAPDA. \hfill $\square$ 	
\end{example}

By interpreting the rows and columns of $\mathbf{P}$ in Example \ref{exam-MAPDA} as subpacketization and users respectively, letting a star represent the user’s caching status, and letting each integer indicate both the block and the delivery strategy, we can generate a $(L,K,M,N)=(2,4,2,4)$ multiple-antenna coded caching scheme in the following way.

$\bullet$ { \textbf{Placement phase}:} Since $\mathbf{P}$ has $F=4$ rows, each file is divided into $4$ packets, i.e., $\mathbf{w}_n =(\mathbf{w}_{n,1},\mathbf{w}_{n,2},\mathbf{w}_{n,3},\mathbf{w}_{n,4})$ where $n \in [4]$. For any $j\in [F]$ and $k\in[K]$, $\mathbf{P}(j,k)=*$ means that user $k$ caches the $j^{\text{th}}$ packet of each file. Hence, the user's caches can be presented as follows. 
\begin{align}\label{MISO-MAPDA}
	\mathcal{Z}_1=\{\mathbf{w}_{n,1},\mathbf{w}_{n,2} | n \in [4]\}; \ \
	\mathcal{Z}_2=\{\mathbf{w}_{n,1},\mathbf{w}_{n,3} | n \in [4]\};\nonumber\\
	\mathcal{Z}_3=\{\mathbf{w}_{n,2},\mathbf{w}_{n,4} | n \in [4]\}; \ \
	\mathcal{Z}_4=\{\mathbf{w}_{n,3},\mathbf{w}_{n,4} | n \in [4]\}.
\end{align} Clearly each user caches exactly $NZ=4\cdot 2=8$ packets, i.e., $M=NZ/F=4\cdot 2/4$ files. That is, each user's memory ratio is $\frac{M}{N}=\frac{Z}{F}=\frac{2}{4}$.

$\bullet$ { \textbf{Delivery phase}:} Assume that the request vector $\mathbf{d} = (1, 2, 3, 4)$. If $\mathbf{P}(j,k)=s$ is an integer, it means that the $j^{\text{th}}$ packet of each file is not stored by user $k$. Then the server broadcasts a multicast message (i.e., the linear combination of all the requested packets indicated by $s$) to the users at time slot $s$ by its $L$ antennas.

For instance, when  $s = 1$, we have  $\mathbf{P}(1,4) = \mathbf{P}\mathbf(2,4) = \mathbf{P}\mathbf(3,1) = \mathbf{P}\mathbf(4,1) = 1$. The server will broadcasts the linear combination of the require packets $\mathcal{W}^{(1)} =\{\mathbf{w}_{1,4}, \mathbf{w}_{2,4},\mathbf{w}_{3,1}, \mathbf{w}_{4,1}\}$ to the users in $\mathcal{K}^{(1)}= \{1,2,3,4\}$. Using Gaussian channel coding with rate $B / \tilde{B} = \log P + o(\log P)$, the server can obtain the coded packets  $\tilde{\mathcal{W}}^{(1)} =\{\tilde{\mathbf{w}}_{1,4}, \tilde{\mathbf{w}}_{2,4},\tilde{\mathbf{w}}_{3,1}, \tilde{\mathbf{w}}_{4,1}\}$. Assume that the channel matrix
\begin{align*}
	\mathbf{H}^{(1)}=\left(\begin{array}{cc}
		1 & 1 \\
		2 & 1 \\ 
		4 & 1 \\ 
		8 & 1 
	\end{array}	\right)=\left(\begin{array}{c}
	\mathbf{h}^{(1)}_1\\
	\mathbf{h}^{(1)}_2\\ 
	\mathbf{h}^{(1)}_3\\
	\mathbf{h}^{(1)}_4\\
	\end{array}	\right).
\end{align*}In order to construct the pre-coding matrix, we will derive a column vector $\mathbf{v}^{(1)}_k$ for each user in $k\in\mathcal{K}^{(1)}$ by the following equations: 1) $\mathbf{h}^{(1)}_k\mathbf{v}^{(1)}_k=1$ and 2)  $\mathbf{h}^{(1)}_{k'}\mathbf{v}^{(1)}_k=0$ for any other $k'\in \mathcal{K}^{(1)}$ satisfying that the packet in  $\mathcal{W}^{(1)}$ which is required by user $k'$ but is not cached by user $k$. 
These two conditions ensure that the signal received by each user only contains the required packet by this user and the packets cached by this user. For instance, the packet $\tilde{\mathbf{w}}_{2,4}$ is required by user $2$ but is not cached by user $1$. Then, by solving the following equations
\begin{align*}
\left\{
\begin{array}{c}
\mathbf{h}^{(1)}_1\mathbf{v}^{(1)}_1=1\\
\mathbf{h}^{(1)}_2\mathbf{v}^{(1)}_1=0
\end{array}
\right.
\end{align*}we have $\mathbf{v}^{(1)}_1=(-1\ 2)^{\top}$. By the hypothesis that $\mathbf{H}$ is invertible with high probability and the Conditions C3 and C4 of Definition \ref{def-MAPDA}, the authors in \cite{YWCC} showed that these column vectors $\mathbf{v}^{(1)}_k$ always exist. We should point out that

Similarly we can obtain $\mathbf{v}^{(1)}_2$, $\mathbf{v}^{(1)}_3$ and $\mathbf{v}^{(1)}_4$. Then, the pre-coding matrix can be defined as  
\begin{align*}
\mathbf{V}^{(1)}=(\mathbf{v}^{(1)}_1\ \mathbf{v}^{(1)}_2\ \mathbf{v}^{(1)}_3\ \mathbf{v}^{(1)}_4)
=\left(\begin{array}{cccc}
  -1&1&-0.25&0.25  \\
   2&-1&2&-1
\end{array}	\right).
\end{align*} 

From \eqref{transmission}, the server broadcasts
\begin{align*} 
	\mathbf{X}^{(1)} = \begin{pmatrix}
		\mathbf{x}_1^{(1)} \\
		\mathbf{x}_2^{(1)}  
	\end{pmatrix}
	=\mathbf{V}^{(1)}
	\begin{pmatrix}
		\tilde{\mathbf{w}}_{1,4}\\
		\tilde{\mathbf{w}}_{2,4}\\
		\tilde{\mathbf{w}}_{3,1}\\
		\tilde{\mathbf{w}}_{4,1}
	\end{pmatrix} =\left(\begin{array}{c}
		-\tilde{\mathbf{w}}_{1,4}+\tilde{\mathbf{w}}_{2,4}-0.25\tilde{\mathbf{w}}_{3,1}+0.25\tilde{\mathbf{w}}_{4,1}  \\
		2\tilde{\mathbf{w}}_{1,4} -\tilde{\mathbf{w}}_{2,4}+2\tilde{\mathbf{w}}_{3,1}-\tilde{\mathbf{w}}_{4,1}
	\end{array}	\right)
\end{align*} by its two antennas in block $1$. From \eqref{eq-received}, we have  
\begin{align*}
	\mathbf{Y}^{(1)} = 
	\begin{pmatrix}
		\mathbf{y}_{1}^{(1)} \\
		\mathbf{y}_{2}^{(1)} \\
		\mathbf{y}_{3}^{(1)} \\
		\mathbf{y}_{4}^{(1)}
	\end{pmatrix}
	=\mathbf{H}^{(1)}\mathbf{X}^{(1)}
	+
	\begin{pmatrix}
		\boldsymbol{\epsilon}_{k_1}^{(1)} \\
		\boldsymbol{\epsilon}_{k_2}^{(1)} \\
		\boldsymbol{\epsilon}_{k_3}^{(1)} \\
		\boldsymbol{\epsilon}_{k_{4}}^{(1)}
	\end{pmatrix}
%	&=\left(\begin{array}{cc}
%		1 & 1 \\
%		2 & 1 \\ 
%		4 & 1 \\ 
%		8 & 1 
%	\end{array}	\right)\left(\begin{array}{c}
%		-\tilde{\mathbf{w}}_{1,4}+\tilde{\mathbf{w}}_{2,4}-0.25\tilde{\mathbf{w}}_{3,1}+0.25\tilde{\mathbf{w}}_{4,1}  \\
%		2\tilde{\mathbf{w}}_{1,4} -\tilde{\mathbf{w}}_{2,4}+2\tilde{\mathbf{w}}_{3,1}-\tilde{\mathbf{w}}_{4,1}
%	\end{array}	\right)+
%	\begin{pmatrix}
%		\boldsymbol{\epsilon}_{k_1}^{(1)} \\
%		\boldsymbol{\epsilon}_{k_2}^{(1)} \\
%		\boldsymbol{\epsilon}_{k_3}^{(1)} \\
%		\boldsymbol{\epsilon}_{k_{4}}^{(1)}
%	\end{pmatrix}\\
	=\left(\begin{array}{c}
		\tilde{\mathbf{w}}_{1,4}+1.75\tilde{\mathbf{w}}_{3,1}-0.75 \tilde{\mathbf{w}}_{4,1}  \\
		\tilde{\mathbf{w}}_{2,4}+1.5\tilde{\mathbf{w}}_{3,1}-0.5 \tilde{\mathbf{w}}_{4,1}  \\
		-2\tilde{\mathbf{w}}_{1,4}+3 \tilde{\mathbf{w}}_{2,4}+\tilde{\mathbf{w}}_{3,1}\\
		-6\tilde{\mathbf{w}}_{1,4}+7 \tilde{\mathbf{w}}_{2,4}+\tilde{\mathbf{w}}_{3,1}
	\end{array}\right)+
	\begin{pmatrix}
		\boldsymbol{\epsilon}_{k_1}^{(1)} \\
		\boldsymbol{\epsilon}_{k_2}^{(1)} \\
		\boldsymbol{\epsilon}_{k_3}^{(1)} \\
		\boldsymbol{\epsilon}_{k_{4}}^{(1)}
	\end{pmatrix}.
\end{align*} where user $k\in \mathcal{K}^{(1)}$ receives the signal $\mathbf{y}_{k}^{(1)}$. Let us consider the 
the signal $\mathbf{y}_{1}^{(1)}=\tilde{\mathbf{w}}_{1,4}+1.75\tilde{\mathbf{w}}_{3,1}-0.75 \tilde{\mathbf{w}}_{4,1}+\boldsymbol{\epsilon}_{k_1}^{(1)}$ received by user $1$.  Using the cached packets $\mathbf{w}_{3,1}$ and $\mathbf{w}_{4,1}$, user $1$ can obtain the required packet $\mathbf{w}_{1,4}$ by 
$$\mathcal{L}_{1}^{(1)}\big(\mathbf{y}_1^{(1)},\tilde{\mathcal{Z}}_1\big) =\mathbf{y}_{1}^{(1)}-\tilde{\mathbf{w}}_{3,1} -\tilde{\mathbf{w}}_{4,1}=
 \tilde{\mathbf{w}}_{1,4} + \boldsymbol{\epsilon}_1^{(1)}.$$ Similarly, we can check that each user can decode its required packet. From \eqref{example-MAPDA}, the server divides the coded packets transmission process into $S=2$ blocks, and in each block there are exactly $4$ required packets which are transmitted to $4$ users, i.e., $r_1=r_2=4$. Therefore, the achieved sum-DoF is $\frac{\sum_{s=1}^{2}r_s}{2}=\frac{r_1+r_2}{2}=4$.

Through the process of generating the aforementioned $(L,K,M,N)=(2,4,2,4)$ multiple-antenna coded caching scheme based on the $4$-(2, 4, 4, 2, 2) MAPDA, the following result is established.
\begin{lemma}[\cite{YWCC}]\rm\label{max-DoF}
Given an $(L, K, F, Z, S)$ MAPDA, let $r_s$ denote the occurrence number of integer $s\in[S]$. There always exists an $F$-division scheme for the $(L, K, M, N)$ multi-antenna coded caching system with the memory ratio of $M/N = Z/F$ and the sum-DoF of $\frac{\sum_{s\in[S]} r_s}{S}$.   \hfill $\square$ 
\end{lemma} 

Under uncoded placement and linear one-shot delivery strategies, the authors in \cite{LBE} derived the upper bound on som-DoF $\min\{\frac{KM}{N} + L, K\}$ of the scheme by information-theory. From the viewpoint of combinatorial theory, the authors in \cite{YWCC} derived the same upper bound of the scheme which can be generated by a MAPDA. 
\begin{lemma}[\cite{YWCC,LBE,NKPR}]\rm\label{under-DoF}
	For the $(L, K, M, N)$ multiple-antenna coded caching scheme with memory ratio $\frac{M}{N} = \frac{Z}{F}$ which can be realized by any $(L, K, F, Z, S)$ MAPDA, the sum-DoF is no more than $\min\{\frac{KM}{N} + L, K\}$. \hfill $\square$ 
\end{lemma}
There are many constructions of MAPDAs, such as in \cite{YWCC,NKPR,WCC,PERB,CTWWL}. We summarize them in Table \ref{tab-Schemes}. 
\begin{table}[http!]
	\renewcommand{\arraystretch}{2}
	\setlength\tabcolsep{1pt} 
	\centering
	\caption{The existing MAPDAs with $L$ antennas and memory ratio of $M/N = t/K$ where $K,t,C,m,r\in \mathbb{Z}^{+}$, $t \in [K]$, $t + L \leq K$, $\beta=\gcd(K,t,L)$.}
	\label{tab-Schemes}
	\begin{tabular}{|c|c|c|c|c|c|c|}\hline
		MAPDA   & sum-DoF ($g$)  & K  & F & Z & S & Constrain\\ \hline
		
%		NMA Scheme  \cite{NNMA}&$K$               & $\frac{K-t}{t+1}$& ${K\choose t}$         &$t\leq K$    \\ \hline

%		NMA Scheme  \cite{SMPESSEP}&$K$               & $\frac{K-t}{t+1}$& ${K\choose t}$         &$t\leq K$    \\ \hline
%		NMA Scheme  \cite{MSBI}&$K$               & $\frac{K-t}{t+1}$& ${K\choose t}$         &$t\leq K$    \\ \hline
          
%         LEEP MAPDA  \cite{LEEP}               & $t+L$& ${K/L \choose t/L}$         &$L\mid t$,$L\mid K$    \\ \hline 
          
         YWCC MAPDA \cite{YWCC} & $t+L$ & $K$& $\frac{t+L}{\gcd(m,L-m)}{K/m\choose t/m}$ & $\frac{t+L}{\gcd(m,L-m)}{{K/m} -1\choose {t/m} -1}$&$\frac{t+m}{\gcd(m,L-m)}{K/m\choose {t/m}+1} $& {$m\mid K$, $m\mid t$}  \\ \cline{4-7}
         & & &${K/L\choose t/L}$&${{K/L}-1\choose {t/L}-1}$&${K/L\choose {t/L}+1}$ &{$m\mid K$, $m\mid t$, $m=L$ }  \\ \hline
          
		NPR MAPDA  \cite{NKPR}               & $t+L$&K& $\frac{t+L}{\beta}{K/\beta\choose (t+L)/\beta}$ &  ${t/\beta} {K/\beta\choose {(t+L)/\beta} -1} $  & $\frac{K-t}{\beta}{K/\beta\choose (t+L)/\beta}$ &   \\ \hline
		
		& & &$2LK$&$2Lt$& $K(K-t)$& $2\nmid (K-t)$\\ \cline{4-7}
		WCC MAPDA  \cite{WCC}               & $2L$& $K$ &$LK$ & $Lt$& $K(K-t)/2$        &$2\mid(K-t)$, $L\nmid K $    \\ \cline{4-7}
		& & &$K$&$t$& $K(K-t)/{2L}$ &$2\mid(K-t)$, $L\mid K $  \\ \hline

		PR MAPDA  \cite{PERB}               & ${t+L\choose r}$& $K={C\choose r}$& ${t+L\choose t}{(C-t-r) }$  &$\left[{t+L\choose t}-{t+L-r\choose t}\right]{(C-t-r) }$  &$(t+1){(C-t-r) }$     &$L=r+1$    \\ \hline
		
%		CTWWL MAPDA  \cite{CTWWL}             & $2L\left\lfloor \frac{t+L}{K-t+L} \right\rfloor+\left \langle t+L \right \rangle_{K-t+l}$& $K$&$gK/\beta^{2}$&$gt/\beta^{2}$&$K(K-t)/\beta^{2}$         &$L\leq \left \langle t+L \right \rangle_{K-t+l} <2L$    \\ \cline{2-2} \cline{7-7}
%		 & $2L\left\lfloor \frac{t+L}{K-t+L} \right\rfloor+2L$&   &&&       &$2L\leq \left \langle t+L \right \rangle_{K-t+l} $    \\  \hline	
	\end{tabular}
\end{table}

\subsection{Multi-Access Multi-Input Single-Output Coded Caching}

\begin{figure}[http!]
	\centering
	\includegraphics[width=0.5\linewidth]{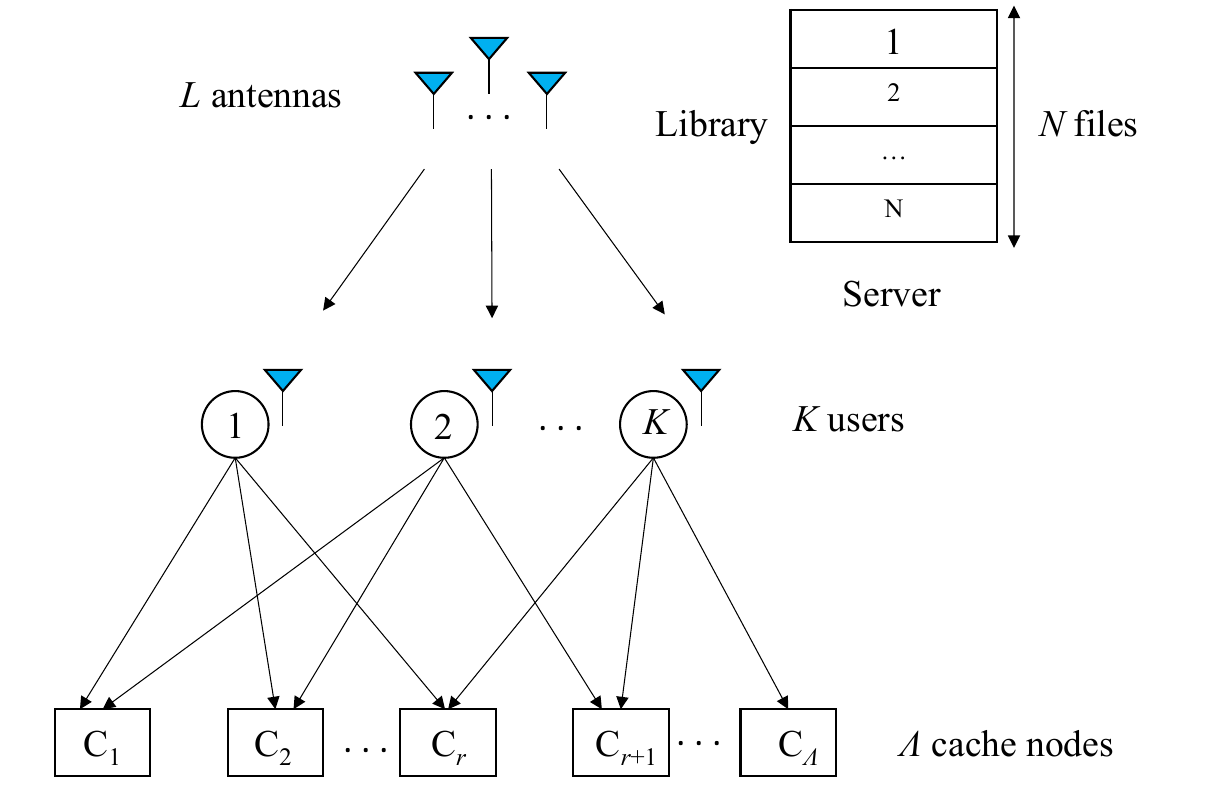}
	\caption{Multi-access multi-input single-output coded caching system}
	\label{fig:1}
\end{figure}
As listed in Fig. \ref{fig:1}, in a $(L,K,\Lambda,r,M,N)$ multi-access multi-input single-output (MAMISO) coded caching system, there is a single server with a library of $N$ equal-size files,  $\Lambda$ cache nodes (denoted by $C_1, C_2, \ldots, C_\Lambda$) each of which has a cache size of $M$ files where $0 \leq M \leq N/r$, and $K$ users without cache memory. Each user $k$ where  $k\in[K]$ can access $r$ cache nodes, denoted by $\mathcal{R}_k$.
%=\{C_{k,1},C_{k,2},\ldots,C_{k,r}\}$
 In this paper, we assume that the set of cache nodes accessible to each user is unique. When multiple users access the same set of cache nodes, no multicast opportunity can be obtained via coded caching techniques since they can only retrieve the same packets. The server transmits data to $K$ users via $L$ transmit antennas, while each user has a single receive antenna. 
An $F$-division $(L,K,\Lambda,r,M,N)$ multi-access multi-input single-output (MAMISO) coded caching scheme also consists of two phases. 

$\bullet$ { \textbf{Placement phase}:} The server also divides each file into $F$ equal-sized packets and then places some packets into each cache node without any prior knowledge of later demands. Let $\mathcal{Z}_{C_\lambda}$ denote the contents stored in cache node $C_\lambda$ where $\lambda \in[\Lambda]$, and $Z_{\mathcal{R}_k}$ denote the packets retrievable by user $k$ where $k\in[K]$. 

$\bullet$ { \textbf{Delivery phase}:} By utilizing its $L$ antennas, the server transmits linear combinations of the requested packets based on the demand vector and users' retrievable contents, thereby enabling every user to decode its desired message. Our objective is also to design an $(L,K, \Lambda, r, M, N)$ MAMISO coded caching scheme that can achieve the sum-DoF as large as possible.

In the following, we characterize the MAMISO coded caching scheme using a triple of arrays—node-placement, user-retrieve, and user-delivery in a manner analogous to the multiaccess shared-link characterization in  \cite{CWLZC}. It is worth noting that while the user-delivery array for the multiaccess scheme is required to satisfy Conditions C2-C4 of an MAPDA with only $L=1$, while the corresponding array in our MAMISO characterization must fulfill Conditions C2-C4 for an arbitrary number of antennas $L$. 
\begin{definition}[\cite{CWLZC}]\rm\label{defn:three arrays} 
%	Given integers $F$, $\lambda$ and $K$ which represent the subpacketization, the number of cache nodes and the number of users respectively, we define that
	\begin{itemize}
		\item An $F\times \Lambda$ node-placement array $\mathbf{C}$ consists of star and null, where $F$ and $\Lambda$ represent the subpacketization and the number of cache nodes, respectively. For any integers $j\in [F]$ and $\lambda\in [\Lambda]$, the entry $\mathbf{C}(j,\lambda)$ is star if and only if the cache node $C_{\lambda}$ caches the $j$th packet of each file.
		\item An $F\times K$ user-retrieve array $\mathbf{U}$ consists of star and null, where $F$ and $K$ represent the subpacketization and the number of users  respectively. For any integers $j\in [F]$ and $\lambda\in [\Lambda]$, the entry $\mathbf{U}(j,k)$ is a star if and only if the user $k$ can retrieve the $j$th packet of each file from its connected cache nodes.
		\item An $F\times K$ user-delivery array $\mathbf{Q}$ consists of $\{*\}\cup[S]$, where the stars in $\mathbf{Q}$ have the same meaning as the stars in $\mathbf{U}$. Each integer $s\in[S]$ indexes the transmitted messages at block $s$. Integer $S$ represents the total number of blocks in the delivery phase.
		\hfill $\square$
	\end{itemize}
\end{definition}
By Definition \ref{defn:three arrays}, the problem of constructing MAMISO scheme is transformed into constructing the arrays—node-placement, user-retrieve, and user-delivery, respectively.

\section{MAPDA via Knapsack Problem} \label{structure-of-MAPDA}
In this section, focusing on the full combination access topology, i.e., $K={\Lambda\choose r}$, we propose a construction framework of MAMISO schemes by using  $0$-$1$ knapsack problem. Through the study of the $0$-$1$ knapsack problem, five new schemes are obtained: the first scheme is obtained from the global optimum of the knapsack problem; the second scheme is derived from a greedily constructed solution of the knapsack problem; the third scheme achieves the maximum sum-DoF under specific system parameters; the fourth and fifth schemes improve the first scheme, under identical system parameters, in terms of sum-DoF and subpacketization, respectively.

In this paper, we use the well known MN placement strategy in \cite{MN} for the cache nodes\footnote{Our method is also applicable to other placement strategies. However, deriving an explicit expression for the sum-DoF of the MAMISO coded caching scheme may not be feasible.}. Then for any integer $t<\Lambda$ we can obtain a ${\Lambda\choose t}\times \Lambda$ node-placement $\mathbf{C}$ where each column has exactly ${\Lambda-1\choose t-1}$ stars and each row is presented by a $t$-subset of $[\Lambda]$. 
Based on the full combination access topology, the content retrieval status for each user can be determined, thereby allowing the derivation of a ${\Lambda\choose t}\times {\Lambda\choose r}$  user-retrieve array $\mathbf{U}$. Therefore, it suffices to design an filling integer strategy for the user-retrieve array to obtain the ${\Lambda\choose t}\times {\Lambda\choose r}$ user-delivery scheme, which generalizes the delivery strategy for the MAMISO coded caching scheme. 

In the following of this paper, we index the columns of $\mathbf{U}$ by the $r$-subsets of $[\Lambda]$. Let $\mathcal{K} = \binom{[\Lambda]}{r}$. The main idea of our proposed integer assignment strategy is as follows: For each integer $b\in [0: r-1]$, let us consider a subset $\mathcal{A} \in \binom{[\Lambda]}{t+r-b}$. For any integer $i \in[\max\{b-t,0\}:\min\{r,\Lambda-t-r+b\}]$, let $\mathfrak{G}_i={[\Lambda]\setminus\mathcal{A}\choose i}$. 
For any subset $\mathcal{G}_i\in \mathfrak{G}_i$, define 
\begin{align}\label{eq-B-G_i}
	\mathcal{B}_{\mathcal{G}_i}=\left\{\mathcal{D}\in \binom{[\Lambda]}{r}\ \Big|\ \mathcal{D}\setminus\mathcal{A}=\mathcal{G}_i\right\} .
\end{align}In Appendix \ref{proof-B_G_i}, we will show that all the $\mathcal{B}_{\mathcal{G}_i}$ is a partition of ${[\Lambda]\choose r}$, i.e., the partition of the column set. In addition, we assume that  ${n\choose m}=0$ if $m>n$. By studying the combinatorial structure of the sub-array $\mathbf{U}({\mathcal{A}\choose t},\mathcal{B}_{\mathcal{G}_i})$, we have the following investigations whose proofs are included in Appendix \ref{proof-U_A_B_G_i}.  
\begin{proposition}\rm
	\label{pro-1}
	For each integer $i\in[\max\{b-t,0\}:\min\{r,\Lambda-t-r+b\}]$, in the subarray $\mathbf{U}_{\mathcal{A},\mathcal{B}_{\mathcal{G}_i}}=\mathbf{U}({\mathcal{A}\choose t},\mathcal{B}_{\mathcal{G}_i})$ there are $v_{\mathcal{G}_i}={t+r-b \choose r-i}$ columns, $z_{\mathcal{G}_i}={r-b \choose r-i}$ Null entries in each row, and $u_{\mathcal{G}_i}={t+i-b \choose t}$ Null entries in each column. \hfill $\square$ 
\end{proposition}By Proposition \ref{pro-1}, we know that if $i\in[\max\{b-t,0\}:b-1]$, the entries of $\mathbf{U}_{\mathcal{A},\mathcal{B}_{\mathcal{G}_i}}$ are stars, and if $i\in[b:\min\{r,\Lambda-t-r+b\}]$ each column  of $\mathbf{U}_{\mathcal{A},\mathcal{B}_{\mathcal{G}_i}}$ has at least one Null entry. Define 
\begin{align}
\label{eq-subset}
\mathcal{I}=[b:\min\{r,\Lambda-t-r+b\}].
\end{align} It is sufficient to study the subarray $\mathbf{U}_{\mathcal{A},\mathcal{B}_{\mathcal{G}_i}}$ to construct $\mathbf{Q}$ when $i\in \mathcal{I}$. According the subset $\mathcal{B}_{\mathcal{G}_i}$ where $i\in\mathcal{I}$, we can obtain
\begin{align}
\label{eq-n}
n=\sum_{i\in\mathcal{I}}{\Lambda-t-r+b\choose i}
\end{align} sub-arrays $\mathbf{U}_{\mathcal{A},\mathcal{B}_{\mathcal{G}_i}}$ where any two have no common column. By Proposition \ref{pro-1}, we can obtain the following $(n,L,\mathbf{z},\mathbf{v})$ knapsack problem
\begin{align}
\label{eq-KP}
\max &\ \quad \phi(\mathbf{v},\mathbf{x})=\sum_{i \in \mathcal{I},\mathcal{G}_i\in \mathfrak{G}_i} v_{{\mathcal{G}_i}} x_{{\mathcal{G}_i}}\\
\text{s.t.}&\  \quad 
\begin{cases} 
	n=\sum\limits_{i\in\mathcal{I}}{\Lambda-t-r+b\choose i}\\
	\mathbf{z}=(z_{{\mathcal{G}_i}})_{i \in \mathcal{I},\mathcal{G}_i\in \mathfrak{G}_i},\ \ z_{{\mathcal{G}_i}}>0\\
	\mathbf{v}=(v_{{\mathcal{G}_i}})_{i \in \mathcal{I},\mathcal{G}_i\in \mathfrak{G}_i},\ \ v_{{\mathcal{G}_i}}>0\\
	\mathbf{x}=(x_{{\mathcal{G}_i}})_{i \in \mathcal{I},\mathcal{G}_i\in \mathfrak{G}_i}\in\{0,1\}^n\\
	\psi(\mathbf{z},\mathbf{x})=\sum\limits_{i \in \mathcal{I},\mathcal{G}_i\in \mathfrak{G}_i} z_{{\mathcal{G}_i}} x_{{\mathcal{G}_i}}\leq L.
\end{cases} \nonumber
\end{align} Given a knapsack solution $\mathbf{x}\in\{0,1\}^n$ in \eqref{eq-KP}, for each subset $\mathcal{B}_{\mathcal{G}_i}$ where $i\in \mathcal{I}$ and $\mathcal{G}_i\in \mathfrak{G}_i$ satisfying $x_{{\mathcal{G}_i}}=1$, we assign the same symbol to all null entries within each column of $\mathbf{U}_{\mathcal{A},\mathcal{B}_{\mathcal{G}_i}}$, and then append to each symbol an index indicating its order of occurrence in that column. By considering all the subset $\mathcal{A}$,  we can obtain a user-delivery array $\mathbf{Q}$, i.e., a MACMISO coded caching scheme. The authors in \cite{DBLP:books/daglib/0010031}, \cite{cormen2001introduction} pointed that the $(n,L,\mathbf{z},\mathbf{v})$ knapsack problem admits a computable globally optimal solution. When $b$ runs all the possible values, we then select a scheme with the maximum sum-DoF among the resulting schemes. That is our main result in this paper whose proof is included in Section \ref{proof-of-theorem}. 
\begin{theorem}\rm
	\label{the-olution}
	For any positive integers $\Lambda$, $r$, $t$, $L$ and $b$ such that $r+t\leq \Lambda$ and $b\in[0:r-1]$,  there exists  an $(L,\Lambda,r,M,N)$ MAMISO coded caching scheme with sum-DoF $\phi(\mathbf{v},\mathbf{x})=\sum_{i=1}^{n} v_i x_i$ where $\mathbf{x}=(x_1,x_2,\ldots,x_n)\in \{0,1\}^n$ denotes a feasible solution of  $(n,L,\mathbf{z},\mathbf{v})$ knapsack problem in Proposition \ref{pro-1}.  \hfill $\square$ 
\end{theorem}

However, for the general parameters in $\Lambda$, $r$, and $t$, the optimal solution \eqref{eq-KP} becomes intractable. Using the greedy algorithm that systematically constructs feasible solutions, we can derive a lower bound on the achievable sum-DoF of Theorem \ref{the-olution}. In order to introduce our result, the following notation is useful. For any positive integers $\Lambda$, $r$, $t$, $L$, and $b$ where $b \in [0: r-1]$, let $\delta$ be the smallest integer in $\mathcal{I}=[b: \min(r, \Lambda-t-r+b)]$ satisfying $L<\sum_{i=b}^{\delta}{\Lambda-t-r+b\choose i}\binom{r-b}{r-i}+\binom{r-b}{r-\delta-1}$, i.e., 
\begin{align}
	\label{eq-delta}
	\delta = \arg\min_{k \in \mathcal{I}} \Bigl\{ k \ \Bigl| \ L - \sum_{i=b}^{k} \binom{\Lambda-t-r+b}{i}\binom{r-b}{r-i} < \binom{r-b}{r-k-1} \Bigr\}.
\end{align}Based on the parameter $\delta$, let 
\begin{align}
	\label{eq-eta}
	\eta=L-\sum_{i=b}^{\delta}{\Lambda-t-r+b\choose i}\binom{r-b}{r-i}, \ \ \  \zeta=\min\left\{
	\left\lfloor{\Lambda-t-r+b\choose \delta }+\frac{\eta}{{r-b\choose r-\delta}}\right\rfloor,{\Lambda-t-r+b\choose \delta }\right\}.
\end{align}
Using the aforementioned notations, we can propose the following result whose proof is included in Appendix \ref{corollary 3}.
\begin{theorem}\rm\label{th-lower-DoF}
	For any positive integers $\Lambda$, $r$, $t$, $L$, and $b$ where $b \in [0: r-1]$, there exists an $(L, \Lambda, r, M, N)$ MAMISO coded caching scheme whose the sum-DoF %lower-bounded by 
	\begin{align*}
		g=\sum_{i=b}^{\delta-1}{\Lambda-t-r+b\choose i}{t+r-b\choose r-i}+\zeta {t+r-b\choose r-\delta}	
	\end{align*}
	and the subpacketization 
	\begin{align*}
		F=\Delta_1\Delta_2{\Lambda \choose t}\left[\sum_{i=b}^{\delta-1}\frac{{r\choose r-i}{\Lambda-t-r\choose i-b}}{\zeta{t+i-b\choose t}}+  \frac{{r\choose r-\delta}{\Lambda-t-r\choose \delta-b}}{{\Lambda-t-r+b\choose \delta}{t+\delta-b\choose t}}\right]
	\end{align*}where the parameters $\delta$, $\zeta$ are defined in \eqref{eq-delta} and \eqref{eq-eta} respectively. Here, $\Delta_1=\text{LCM}({\Lambda-t-r+b\choose \delta},\zeta)$, $\Delta_2=\text{LCM}{t+j-1\choose t}_{j\in[\delta-b+1]}$. \hfill $\square$  
\end{theorem}

In fact, when the parameters $\Lambda$, $r$, $t$, and $L$ satisfy some constraint, for instance when $\Lambda \geq 2r+t$ and $L \geq \binom{\Lambda-t}{r} - \binom{\Lambda-t-r}{r}$, Theorem \ref{the-olution} not only admits a closed-form expression, but also achieves the upper bound on the sum-DoF. That is the following result whose proof is included in Appendix~\ref{corollary 2}.
\begin{theorem}\rm\label{th-optimal}
	For any positive integers $\Lambda$, $r$, $t$, and $L$ satisfying $\Lambda \geq 2r+t$ and $L \geq {\Lambda-t \choose r} - {\Lambda-t-r \choose r}$, there exists an $(L,\Lambda,r,M,N)$ MAMISO coded caching scheme that achieves a sum-DoF of $KM/N + L$ and the subpacketization 
\begin{align*}
F=\begin{cases}
\Delta_3{\Lambda \choose t}\sum_{i=0}^{r-1}\frac{{r\choose r-i}{\Lambda-t-r\choose i}}{{t+i\choose t}} & \text{if}\ \ L={\Lambda-t\choose r}-{\Lambda-t-r\choose r}, \\
\Delta_4\Delta_5{\Lambda \choose t}\left(\sum_{i=0}^{r-1}\frac{{r\choose r-i}{\Lambda-t-r\choose i}}{\left(L-{\Lambda-t\choose r}+{\Lambda-t-r\choose r}\right){t+i\choose t}} +\frac{1}{{t+r\choose t}}\right)& \text{if}\ \ L>{\Lambda-t\choose r}-{\Lambda-t-r\choose r} %\\
%			{\Lambda \choose t}\left[\sum_{i=0}^{r-1}{r\choose r-i}{\Lambda-t-r\choose i} \frac{\Delta_3}{L-{\Lambda-t\choose r}+{\Lambda-t-r\choose r}}\frac{\Delta_2}{{t+i\choose t}} + {\Lambda-t-r\choose r}\frac{\Delta_3}{{\Lambda-t-r\choose r}}\frac{\Delta_2}{{t+i\choose t}}\right]& \text{if}\ \ L>{\Lambda-t\choose r}-{\Lambda-t-r\choose r}
\end{cases}
\end{align*}where $\Delta_3=\text{LCM}{t+j-1\choose t}_{j\in[r]}$, $\Delta_4=\text{LCM}{t+j-1\choose t}_{j\in[r+1]}$, and 	$\Delta_5=\text{LCM}({\Lambda-t-r\choose r},{L-{\Lambda-t\choose r}+{\Lambda-t-r\choose r}})$.  \hfill $\square$      
\end{theorem}
It is worth noting that the result in Theorem \ref{the-olution} only considers subarrays in which every column contains at least one null entry. However, for certain special parameter choices, the all-star subarrays $\mathbf{U}_{\mathcal{A},\mathcal{B}_{\mathcal{G}_i}}$ obtained with $i\in[\max\{b-t,0\}:b-1]$ can further increase the sum-DoF. That is the following result whose proof is given in Subsection \ref{double-solution-proof}.
\begin{theorem}\rm\label{double-solution}
	For any positive integers $\Lambda$, $r$, $t$, $b$, and $L$ satisfying $b<r<2b$, $t+r>2b$, $\Lambda = 2(t+r-b)$, and $L\leq {t+r-b\choose b}$, there exists an $(L,\Lambda,r,M,N)$ MAMISO coded caching scheme that achieves a sum-DoF of $2L{t+r-b\choose t}$ and the subpacketization $F={\Lambda\choose t}{r\choose b}L/\gcd({\Lambda-t-r+b\choose b},L)$. \hfill $\square$      
\end{theorem}

Finally, it is worth noting that when $|\mathcal{A}| > t+r$, we no longer partition the array into multiple subarrays, but instead construct a single subarray satisfying the required conditions to obtain some new schemes. That is the following result whose proof is included in Subsection \ref{C-L-MAPDA-proof}.
\begin{theorem}\label{C-L-MAPDA}\rm
	For any positive integers $\Lambda$, $r$, $t$, and $L$, if there exists an integer $\Lambda' \in [\Lambda]$ satisfying ${\Lambda'-t\choose r}\leq L$, then there exists an $(L, \Lambda, r, M, N)$ MAMISO coded caching scheme achieving a sum-DoF of $ {\Lambda'\choose r}$.
	
	\hfill $\square$
\end{theorem}

\subsection{Extension of the Proposed Schemes to the MISO Coded Caching Scenario}

As a by-product, we can directly extend the proposed schemes in Theorem \ref{th-lower-DoF}, Theorem \ref{th-optimal}, Theorem \ref{double-solution} and Theorem \ref{C-L-MAPDA} for the MACC model to the MISO coded caching scenario in Section \ref{MISO-ccs} with the same number of users as the MACC model; this is because we can let the retrievable content by each user in the MACC model be the cache content of one user in the MISO coded caching scenario, while the delivery phase does not change. Then we can obtain the following results.
\begin{corollary}\rm\label{corollary-1}[MAPDA via Theorem \ref{th-lower-DoF}]
For any positive integers $\Lambda$, $r$, $t$, $L$ and $b$ satisfying $r+t\leq \Lambda$, and $b\in[0:r-1]$, there exists a $(L,K={\Lambda\choose r},F=\pi{\Lambda\choose t},Z=\pi({\Lambda\choose t}-{\Lambda-r\choose t}),S=\Delta_1\Delta_2{\Lambda\choose t+r-b}/\zeta )$ MAPDA with the sum-DoF and subpacketization 
\begin{align*}
g=&\sum_{i=b}^{\delta-1}{\Lambda-t-r+b\choose i}{t+r-b\choose r-i}+\zeta {t+r-b\choose r-\delta},	\\
F=&\Delta_1\Delta_2{\Lambda \choose t}\left[\sum_{i=b}^{\delta-1}\frac{{r\choose r-i}{\Lambda-t-r\choose i-b}}{\zeta{t+i-b\choose t}}+  \frac{{r\choose r-\delta}{\Lambda-t-r\choose \delta-b}}{{\Lambda-t-r+b\choose \delta}{t+\delta-b\choose t}}\right]
\end{align*}  where the parameters $\delta$, $\zeta$ are defined in \eqref{eq-delta} and \eqref{eq-eta} respectively, and $\pi=\Delta_1\Delta_2\left[\sum_{i=b}^{\delta-1}\frac{{r\choose r-i}{\Lambda-t-r\choose i-b}}{\zeta{t+i-b\choose t}}+  \frac{{r\choose r-\delta}{\Lambda-t-r\choose \delta-b}}{{\Lambda-t-r+b\choose \delta}{t+\delta-b\choose t}}\right]$. Here, $\Delta_1=\text{LCM}({\Lambda-t-r+b\choose \delta},\zeta)$, $\Delta_2=\text{LCM}{t+j-1\choose t}_{j\in[\delta-b+1]}$. \hfill $\square$		
\end{corollary}

\begin{corollary}\rm\label{corollary-2}[MAPDA via Theorem \ref{th-optimal}]	For any positive integers $\Lambda$, $r$, $t$, $L$ and $b$ satisfying $\Lambda \geq 2r+t$ and $L \geq {\Lambda-t \choose r} - {\Lambda-t-r \choose r}$, there exists an $(L,K,F,Z,S)$ MAPDA with sum-DoF $g=KM/N+L$. If $L={\Lambda-t\choose r}-{\Lambda-t-r\choose r}$, the subpacketization $$F=\Delta_3{\Lambda \choose t}\sum_{i=0}^{r-1}\frac{{r\choose r-i}{\Lambda-t-r\choose i}}{{t+i\choose t}},$$ and the parameters are $K=\tbinom{\Lambda}{r}$, $Z=\Delta_3\sum_{i=0}^{r-1}\frac{{r\choose r-i}{\Lambda-t-r\choose i}}{{t+i\choose t}}\left(\tbinom{\Lambda}{ t}-\tbinom{\Lambda-r}{ t}\right)$, and $S=\Delta_3\tbinom{\Lambda}{t+r}$. If $L>{\Lambda-t\choose r}-{\Lambda-t-r\choose r}$, the subpacketization $$F=\Delta_4\Delta_5{\Lambda \choose t}\left(\sum_{i=0}^{r-1}\frac{{r\choose r-i}{\Lambda-t-r\choose i}}{\left(L-{\Lambda-t\choose r}+{\Lambda-t-r\choose r}\right){t+i\choose t}} +\frac{1}{{t+r\choose t}}\right),$$ and the parameters are $K=\tbinom{\Lambda}{ r}$, $Z=\pi\left(\tbinom{\Lambda}{ t}-\tbinom{\Lambda-r}{ t}\right)$, $S=\Delta_5\tbinom{\Lambda}{t+r}/(L-{\Lambda-t\choose r}+{\Lambda-t-r\choose r})$, where $$\pi=\Delta_4\Delta_5\left(\sum_{i=0}^{r-1}\frac{{r\choose r-i}{\Lambda-t-r\choose i}}{\left(L-{\Lambda-t\choose r}+{\Lambda-t-r\choose r}\right){t+i\choose t}} +\frac{1}{{t+r\choose t}}\right).$$ Here $\Delta_3=\text{LCM}{t+j-1\choose t}_{j\in[r]}$, $\Delta_4=\text{LCM}{t+j-1\choose t}_{j\in[r+1]}$ and $\Delta_5=\text{LCM}({\Lambda-t-r\choose r},{L-{\Lambda-t\choose r}+{\Lambda-t-r\choose r}})$.  \hfill $\square$	
\end{corollary}

\begin{corollary}\rm\label{corollary-3}[MAPDA via Theorem \ref{double-solution}]
	For any positive integers $\Lambda$, $r$, $t$, $b$ and $L$ satisfying $b<r<2b$,  $t+r>2b$, $\Lambda=2(t+r-b)$ and $L\leq {t+r-b\choose b}$, there exists a $(L,K={\Lambda\choose r}$, $F={r\choose b}{\Lambda\choose t}L/\beta$, $Z={r\choose b}\left({\Lambda\choose t}-{\Lambda-r\choose t}\right)L/\beta$, $S={\Lambda\choose t+r-b}{\Lambda-t-r+b\choose b}/2\beta )$ MAPDA with the sum-DoF $g=2L{t+r-b\choose t}$-and the subpacketization $F = {r\choose b}{\Lambda\choose t}L/\beta$, where $\beta=\gcd({\Lambda-t-r+b\choose b},L)$.  
	\hfill $\square$		
\end{corollary}

\begin{corollary}\rm\label{corollary-4}[MAPDA via Theorem \ref{C-L-MAPDA}]
	For any positive integers $\Lambda$, $r$, $t$ and $L$, exists a $\Lambda'\in [\Lambda]$ satisfying ${\Lambda'-t\choose r}\leq L$ and ${\Lambda'\choose r}=\phi(\mathbf{v},\mathbf{x})$, there exists a $\left(L,K={\Lambda\choose r},F={\Lambda-t-r\choose \Lambda'-t-r}{\Lambda\choose t},Z={\Lambda-t-r\choose \Lambda'-t-r}({\Lambda\choose t}-{\Lambda-r\choose t}),S={\Lambda\choose \Lambda'}{\Lambda'-r\choose t} \right)$ MAPDA with the sum-DoF $g={\Lambda'\choose r}$ and the subpacketization $F = {\Lambda-t-r\choose \Lambda'-t-r}{\Lambda\choose t}$. 
	\hfill $\square$		
\end{corollary}

%Let us consider the MAPDA in Corollary \ref{corollary-1}. Based on the $(n,L,\mathbf{z},\mathbf{v})$ Knapsack Problem, the obtained user-delivery array is a $\left(L,K={\Lambda\choose r},F=\pi{\Lambda\choose t},Z=\pi({\Lambda\choose t}-{\Lambda-r\choose t}),S=\ell\mu{\Lambda\choose t+r-b} \right)$ MAPDA.

%By Lemma \ref{max-DoF} we $(L,K,M,N)$ multiple-antenna coded caching scheme for
%multiple-input single-output broadcast channel network where\begin{align*}
%	&\frac{M}{N}=\frac{Z}{F}=\frac{{\Lambda\choose t}-{\Lambda-r\choose t}}{{\Lambda\choose t}}=1-\frac{{\Lambda-r\choose t}}{{\Lambda\choose t}},  \\
%	&\text{sum-DoF} = K(F-Z)/S.
%\end{align*}

\subsection{Theoretical comparisons}
In this subsection, we will compare the new schemes in Corollaries \ref{corollary-2}, \ref{corollary-3}, \ref{corollary-4} with existing schemes listed in Table \ref{tab-Schemes}.
\subsubsection{Comparison between Corollary \ref{corollary-2} and the YWCC scheme in \cite{YWCC}} When $K={\Lambda\choose r}$, $L={\Lambda-t\choose r}-{\Lambda-t-r\choose r}$ and $\frac{M}{N}=1-{\Lambda-r\choose t}/{\Lambda\choose t}$, the subpacketization and sum-DoF in Corollary \ref{corollary-2} are $$F_{\text{Co2}}=\text{LCM}{t+j-1\choose t}_{j\in[r]}{\Lambda \choose t}\sum_{i=0}^{r-1}\frac{{r\choose r-i}{\Lambda-t-r\choose i}}{{t+i\choose t}}$$ and sum-DoF $g_{\text{Co2}}=KM/N+L$ respectively. The YWCC scheme in \cite{YWCC} achieves subpacketization $$	F_{\text{YWCC}}=(K M/N+L){K\choose K M/N}=\left({\Lambda\choose r}-{\Lambda-t-r\choose r}\right){{\Lambda\choose r}\choose {\Lambda-t\choose r}}$$ and the same sum-DoF. Since both schemes achieve the theoretical sum-DoF upper bound, we compare their subpacketization levels.

When $r=1$, the access topology uniquely determines $L=1$, in which case both the scheme in Corollary \ref{corollary-2} and the YWCC scheme reduce to the MN scheme proposed in \cite{MN}.

When $r\ge 2$ the following subpacketization ratio can be obtained
\begin{align*}
	\frac{F_{\text{YWCC}}}{F_{\text{Co2}}} \approx \frac{2^{\Lambda H(\frac{r}{\Lambda})}2^{2^{\Lambda H(\frac{r}{\Lambda})}H({\Lambda-t\choose r}/{\Lambda\choose r})}}{2^{(\Lambda-t) H(\frac{r}{\Lambda-t})}\frac{e^{t+r}}{t+1}2^{\Lambda H(\frac{t}{\Lambda})}}=  2^{\Theta(\Lambda^{r-1}\log\Lambda)}.
\end{align*}

Compared with the YWCC scheme, both achieve the sum-DoF upper bound, while our scheme requires a lower subpacketization level.

\subsubsection{Comparison between Corollary \ref{corollary-3} and the WCC scheme in \cite{WCC}}
Let $K={\Lambda\choose r}$, $\frac{M}{N}=1-{\Lambda-r\choose t}/{\Lambda\choose t}$, $\Lambda=2(t+r-b)$ and $L\leq {t+r-b\choose b}$, the subpacketization and sum-DoF in Corollary \ref{corollary-3} are $$F_{\text{Co3}}= {r\choose b}{\Lambda\choose t}L/\gcd({\Lambda-t-r\choose b},L)$$ and $g_{\text{Co3}}=2L{t+r-b\choose t}$ respectively. We compare the WCC scheme with the proposed scheme in the following cases:
\begin{itemize}
	\item  If $2\nmid (K-KM/N)$, we have
	\begin{align*}
		F_{\text{WCC}}=2LK, \ \ \ g_{\text{WCC}}=2L.
	\end{align*}
	We have the following result.
	\begin{align*}
		&\frac{F_{\text{WCC}}}{F_{\text{Co3}}}=\frac{2LK}{{r\choose b}{\Lambda\choose t}L/\gcd({\Lambda-t-r\choose b},L)}=\frac{2{\Lambda\choose r}\gcd({t+r-b \choose b},L)}{{r\choose b}{\Lambda\choose t}}\underset{r\geq t}{\geq} \frac{2}{{r\choose b}}, \\
		&\frac{g_{\text{WCC}}}{g_{\text{Co3}}}=\frac{2L}{2L{t+r-b\choose t}}=\frac{1}{{t+r-b\choose t}}
	\end{align*}
	\item If $2\mid (K-KM/N)$ and $L\nmid K$, we can get the WCC scheme with
	\begin{align*}
		F_{\text{WCC}}=LK, \ \ \ g_{\text{WCC}}=2L.
	\end{align*}
	We have the following result.
	\begin{align*}
		\frac{F_{\text{WCC}}}{F_{\text{Co3}}}=\frac{{\Lambda\choose r}\gcd({t+r-b \choose b},L)}{{r\choose b}{\Lambda\choose t}}\underset{r\geq t}{\geq} \frac{1}{{r\choose b}}, \ \ \frac{g_{\text{WCC}}}{g_{\text{Co3}}}=\frac{2L}{2L{t+r-b\choose t}}=\frac{1}{{t+r-b\choose t}}.
	\end{align*}
	\item Otherwise, the WCWL scheme can be obtained as follow.
	\begin{align*}
		F_{\text{WCC}}=K, \ \ \ g_{\text{WCC}}=2L.
	\end{align*}
		We have the following result.
	\begin{align*}
		&\frac{F_{\text{WCC}}}{F_{\text{Co3}}}=\frac{{\Lambda\choose r}\gcd({t+r-b \choose b},L)}{L{r\choose b}{\Lambda\choose t}}\underset{r\geq t}{\geq} \frac{1}{L{r\choose b}}, \\
		&\frac{g_{\text{WCC}}}{g_{\text{Co3}}}=\frac{2L}{2L{t+r-b\choose t}}=\frac{1}{{t+r-b\choose t}}
	\end{align*}
\end{itemize}
As a result, compared to the WCC scheme in \cite{WCC}, our scheme achieves a sum-DoF gain by a factor of $\binom{t+r-b}{t}$, while increasing the subpacketization level by less than a factor of $L{r\choose b}$. when $r \geq t$. For $r<t$, although the subpacketization ratio does not admit a simple bound, the proposed scheme still achieves a strictly larger sum-DoF than the WCC scheme.

\subsubsection{Comparison between Corollary \ref{corollary-4} and the PR scheme in \cite{PERB}} When $K={\Lambda\choose r}$ and $L=r+1$. Let $\Lambda'=t+r+1$. Then $L={\Lambda'-t\choose r}={t+r+1-t\choose r}=r+1$. The subpacketization and sum-DoF in Corollary \ref{corollary-4} are $F_{\text{Co4}}=(\Lambda-t-r){\Lambda\choose t}$ and $g_\text{Co4}={\Lambda'\choose r}={t+r+1\choose r}$ respectively. The PR scheme in \cite{PERB} achieves subpacketization $F_{\text{PR}}=(\Lambda-t-r){\Lambda\choose t}$ and the sum-DoF $g_{\text{PR}}={L\choose r}={t+r+1\choose r}$. So we have
\begin{align*}
	\frac{F_{\text{PR}}}{F_{\text{Co4}}}=1, \ \ \frac{g_{\text{PR}}}{g_{\text{Co4}}}=1.
\end{align*}Therefore, the PR scheme can be viewed as a special case of Corollary \ref{corollary-4}.

\subsection{Numerical comparisons}

In this subsection, we will present numerical comparisons between our scheme in Theorem \ref{the-olution}, Corollary \ref{corollary-1}, \ref{corollary-2}, \ref{corollary-3}, \ref{corollary-4} and the existing schemes with linear subpacketization \cite{WCC}, optimal
sum-DoF \cite{YWCC}, \cite{NKPR}, and full combination access topology \cite{PERB}. These existing schemes are listed in Table \ref{tab-Schemes}.

\subsubsection{Comparison of Theorem \ref{the-olution}, Corollary \ref{corollary-1}, and the scheme in \cite{YWCC}, \cite{NKPR}, \cite{WCC}}

\begin{figure}[htbp!]
	\centering
	\begin{minipage}[t]{0.45\textwidth}
		\centering
		\includegraphics[width=\textwidth]{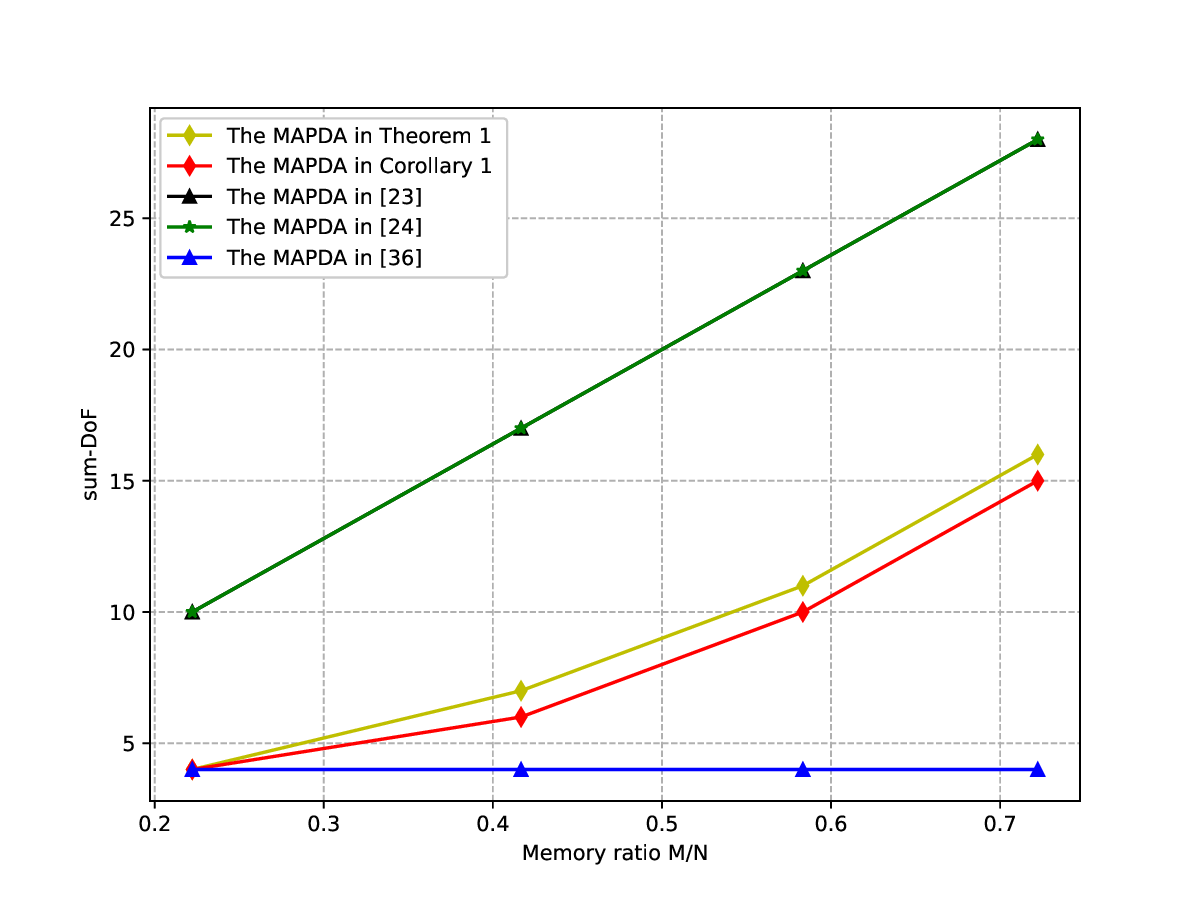}
		\caption{The subpacketization of MAPDAs in \cite{YWCC}, \cite{NKPR}, \cite{WCC}, and Theorem \ref{the-olution} and Corollary \ref{corollary-1}.} % 第一张照片的名称
		\label{fig-Th1-F}
	\end{minipage}\hspace{3ex}
	\begin{minipage}[t]{0.45\textwidth}
		\centering
		\includegraphics[width=\textwidth]{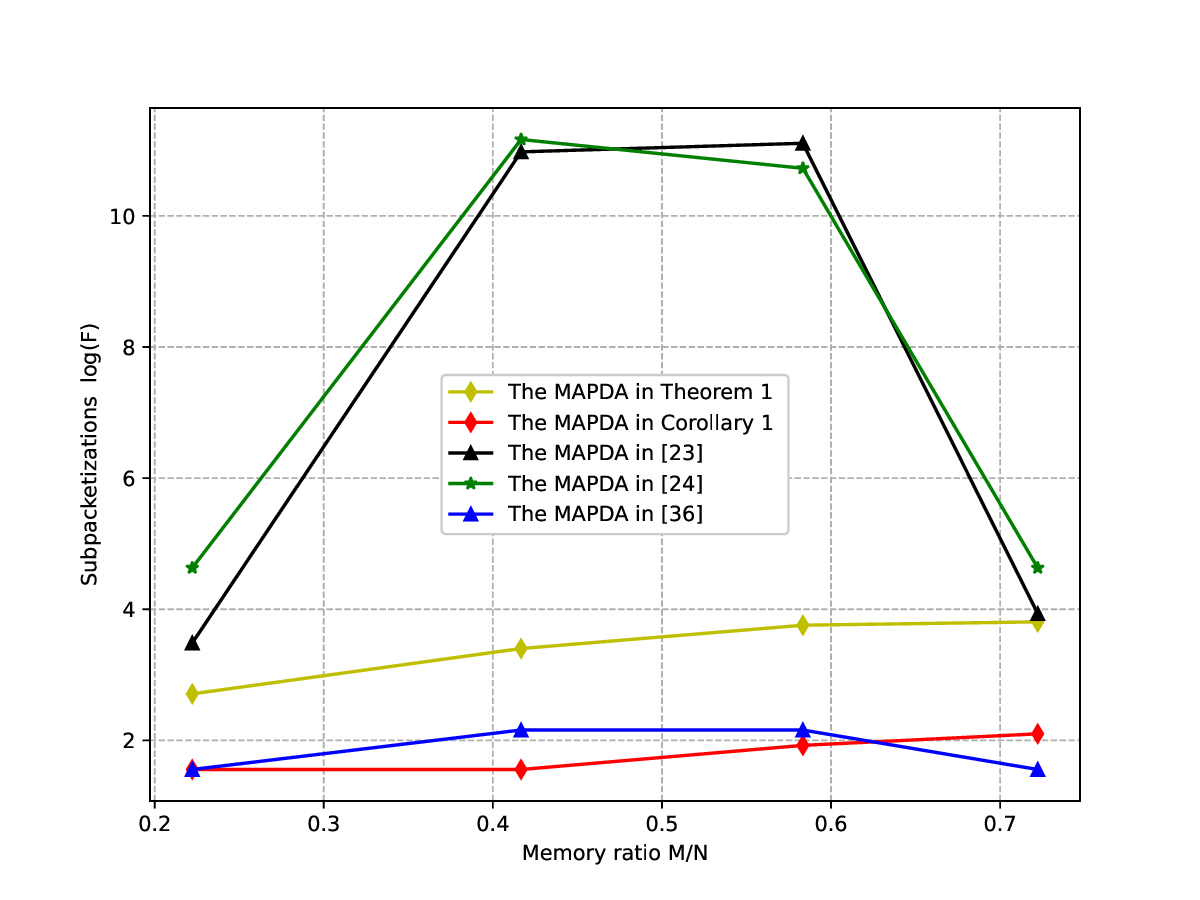}
		\caption{The sum-DoF of MAPDAs in \cite{YWCC}, \cite{NKPR}, \cite{WCC}, and Theorem \ref{the-olution} and Corollary \ref{corollary-1}.} % 第二张照片的名称
		\label{fig-Th1-g}
	\end{minipage}
\end{figure}

The sum-DoF and subpacketization comparison
of MAPDA in Theorem \ref{the-olution}, Corollary \ref{corollary-1} (when $\Lambda = 9$, $r = 2$, and $L = 2$), the MAPDA in NPR scheme\cite{NKPR} (when $K = 36$, $L = 2$), the MAPDA in YWCC scheme\cite{YWCC} (when$K = 36$, $L = 2$), the MAPDA in WCC scheme\cite{WCC} (when $K = 36$, $L = 2$) are shown in Fig. \ref{fig-Th1-F} and Fig. \ref{fig-Th1-g}. It can be seen that Theorem \ref{the-olution} and Corollary \ref{corollary-1} reduce the subpacketization level at the cost of a lower sum-DoF compared with those in \cite{YWCC} and \cite{NKPR}. In contrast to the scheme in \cite{WCC}, the scheme in Corollary \ref{corollary-1} achieves a substantially higher sum-DoF while retaining linear subpacketization. Interestingly, for some memory regimes, it also attains a higher sum-DoF together with a lower subpacketization level than the scheme in \cite{WCC}.

\subsubsection{Comparison of Corollary \ref{corollary-2} and the schemes in \cite{YWCC}, \cite{NKPR}}
Since the subpacketization expressions of these schemes involve combination numbers, powers, and products thereof, it is difficult to plot them for general $K$. We therefore provide a numerical comparison of the scheme in Corollary \ref{corollary-2} with those in \cite{YWCC} and \cite{NKPR} in Table \ref{tab-Co2-Schemes}. The table confirms that the scheme of Corollary \ref{corollary-2} attains the sum-DoF upper bound while employing a lower subpacketization level than the schemes in \cite{YWCC} and \cite{NKPR}.
\begin{table}[http!]
	\renewcommand{\arraystretch}{2}
	\setlength\tabcolsep{1pt} 
	\centering
	\caption{The numerical comparison between the scheme in Corollary \ref{corollary-2}, and the scheme in \cite{YWCC}, \cite{NKPR}}
	\label{tab-Co2-Schemes}
	\begin{tabular}{|c|c|c|c|c|c|c|}\hline
		$K$ & $M/N$ & $L$ & scheme & Parameters & $F$ & sum-DoF \\ \hline

		$15$   & $0.6$ & $5$ & YWCC scheme in \cite{YWCC}   & $m=3$  & $140$ & $14$           \\
		$15$   & $0.6$ & $5$ & Scheme in Corollary \ref{corollary-2}  & $(\Lambda,r)=(6,2)$          & $105$ & $14$                   \\ \hline
		
		$21$   & $0.5238$ & $7$ & YWCC scheme in \cite{YWCC}   & $m=1$  & $6348888$ & $18$           \\
		$21$   & $0.5238$ & $7$ & Scheme in Corollary \ref{corollary-2}  & $(\Lambda,r)=(7,2)$          & $189$ & $18$                   \\ \hline
		
		$21$   & $0.7142$ & $5$ & YWCC scheme in \cite{YWCC}   & $m=3$  & $420$ & $20$           \\
		$21$   & $0.7142$ & $5$ & Scheme in Corollary \ref{corollary-2}  & $(\Lambda,r)=(7,2)$          & $280$ & $20$                   \\ \hline
		
		$42$   & $0.6425$ & $19$ & YWCC scheme in \cite{YWCC}   & $m=4$  & $110110$ & $55$           \\
		$42$   & $0.6425$ & $19$ & Scheme in Corollary \ref{corollary-2}  & $(\Lambda,r)=(8,3)$          & $924$ & $55$                   \\ \hline
		
		$15$   & $0.6$ & $5$ & NPR scheme in \cite{NKPR}   & $\beta=1$  & $210$ & $14$           \\
		$15$   & $0.6$ & $5$ & Scheme in Corollary \ref{corollary-2}  & $(\Lambda,r)=(6,2)$          & $105$ & $14$                   \\ \hline
		
		$42$   & $0.6428$ & $19$ & NPR scheme in \cite{NKPR}   & $\beta=1$  & $3080$ & $55$           \\
		$42$   & $0.6428$ & $19$  & Scheme in Corollary \ref{corollary-2}  & $(\Lambda,r)=(8,3)$          & $924$ & $55$                   \\ \hline
		
		$36$   & $0.5833$ & $10$ & NPR scheme in \cite{NKPR}   & $\beta=1$  & $11686752$ & $31$           \\
		$36$   & $0.5833$ & $10$ & Scheme in Corollary \ref{corollary-2}  & $(\Lambda,r)=(9,2)$          & $31248$ & $31$                   \\ \hline
		$78$   & $0.6410$ & $16$ & NPR scheme in \cite{NKPR}   & $\beta=1$  & $107666559$ & $66$           \\
		$78$   & $0.6410$ & $16$ & Scheme in Corollary \ref{corollary-2}  & $(\Lambda,r)=(13,2)$          & $849420$ & $66$                   \\ \hline
	\end{tabular}
\end{table}

\subsubsection{Comparison of Corollary \ref{corollary-3} and the schemes in \cite{YWCC}, \cite{NKPR}, \cite{WCC}}
%参数特殊，要说明清楚。

\begin{figure}[htbp!]
	\centering
	\begin{minipage}[t]{0.45\textwidth}
		\centering
		\includegraphics[width=\textwidth]{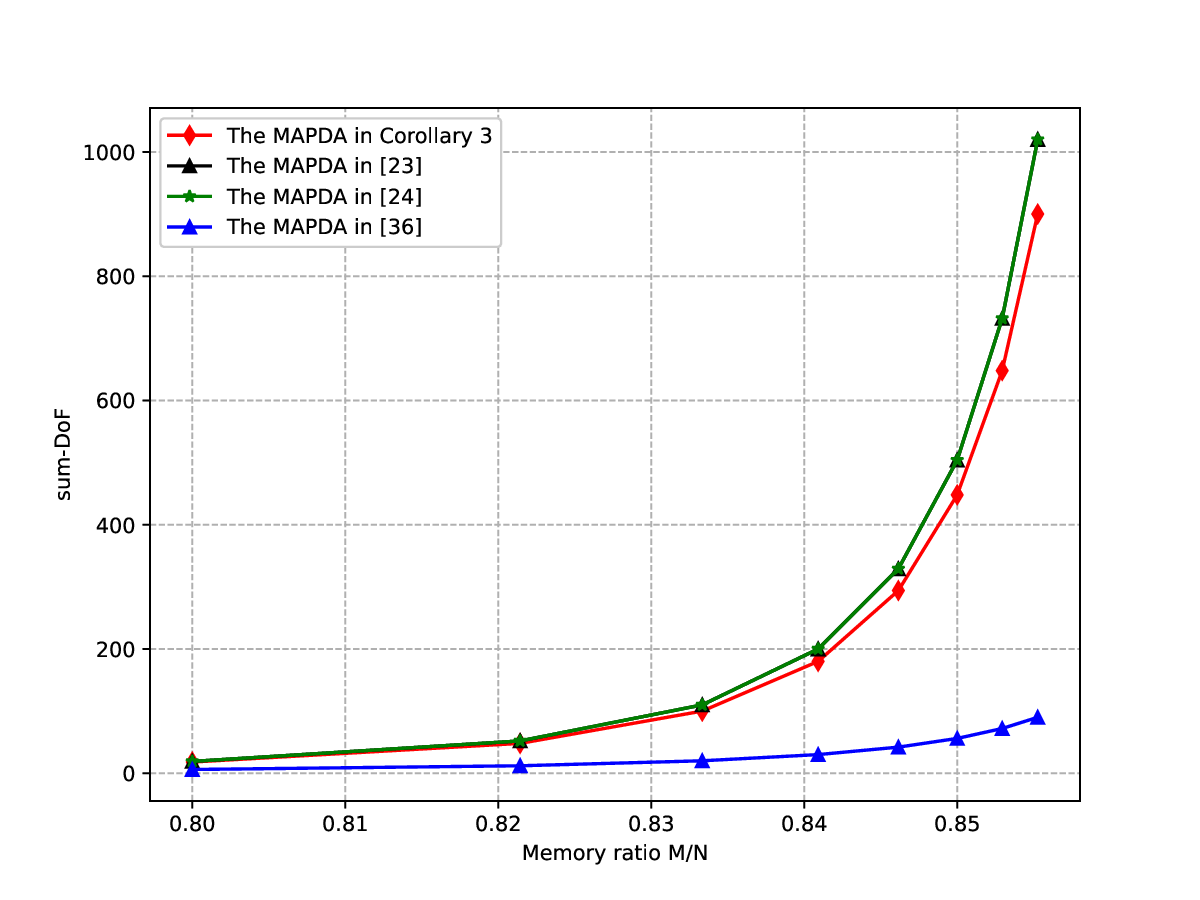}
		\caption{The subpacketization of MAPDAs in \cite{YWCC}, \cite{NKPR}, \cite{WCC}, and Corollary \ref{corollary-3}.} % 第一张照片的名称
		\label{fig-Co3-g}
	\end{minipage}\hspace{3ex}
	\begin{minipage}[t]{0.45\textwidth}
		\centering
		\includegraphics[width=\textwidth]{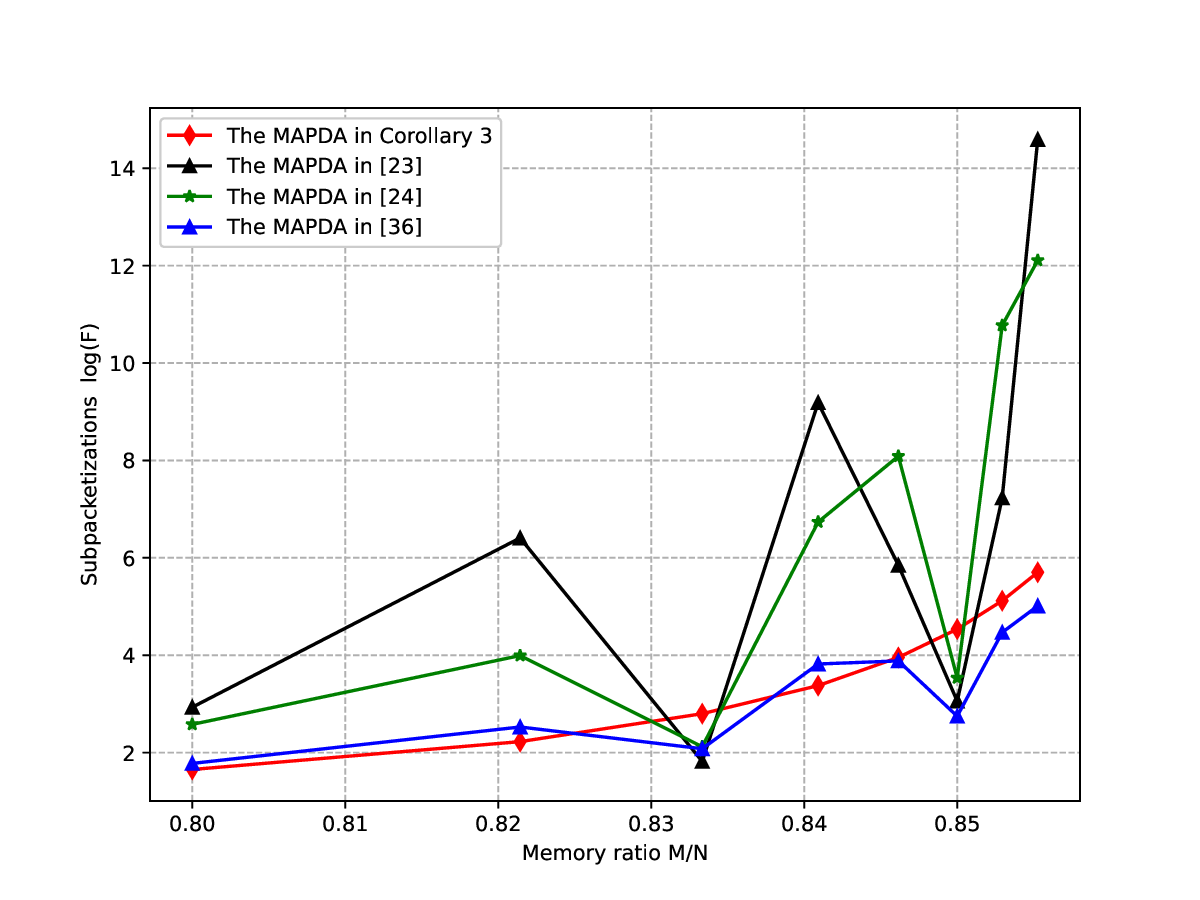}
		\caption{The sum-DoF of MAPDAs in \cite{YWCC}, \cite{NKPR}, \cite{WCC}, and Corollary \ref{corollary-3}.} % 第二张照片的名称
		\label{fig-Co3-f}
	\end{minipage}
\end{figure}

The sum-DoF and subpacketization comparison of MAPDA in Corollary \ref{corollary-3} (when $r = 3$, $b = 2$), the MAPDA in \cite{YWCC}, \cite{NKPR}, and \cite{WCC} are shown in Fig. \ref{fig-Co3-g} and Fig. \ref{fig-Co3-f}. It can be observed that the scheme in Corollary \ref{corollary-3} significantly reduces the subpacketization level while incurring only a slight loss in sum-DoF compared with the schemes in \cite{YWCC} and \cite{NKPR}. Moreover, its subpacketization is close to the linear subpacketization of the scheme in \cite{WCC}.

\subsubsection{Comparison of Theorem \ref{the-olution}, Corollary \ref{corollary-4}, and the scheme in \cite{PERB}}
\begin{figure}[htbp!]
	\centering
	\begin{minipage}[t]{0.45\textwidth}
		\centering
		\includegraphics[width=\textwidth]{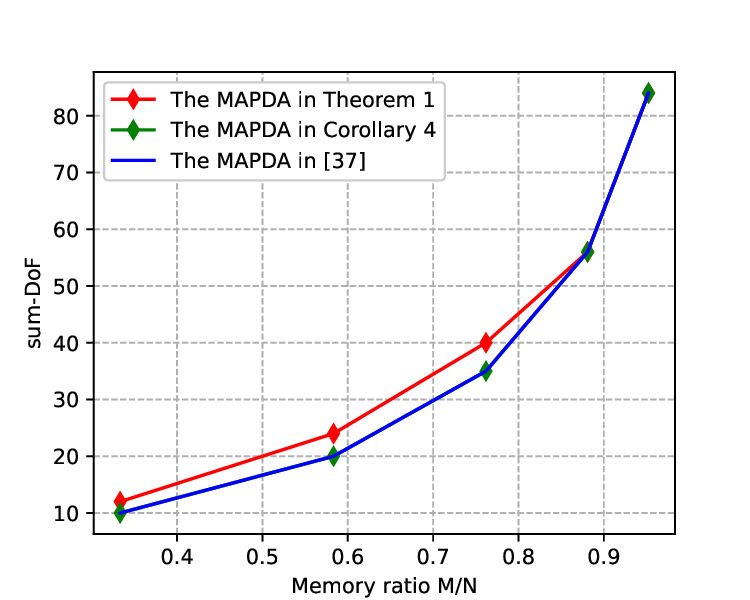}
		\caption{The subpacketization of MAPDAs in \cite{PERB}, and Theorem \ref{the-olution} and Corollary \ref{corollary-4}.} % 第一张照片的名称
		\label{fig-Co4-F}
	\end{minipage}\hspace{3ex}
	\begin{minipage}[t]{0.45\textwidth}
		\centering
		\includegraphics[width=\textwidth]{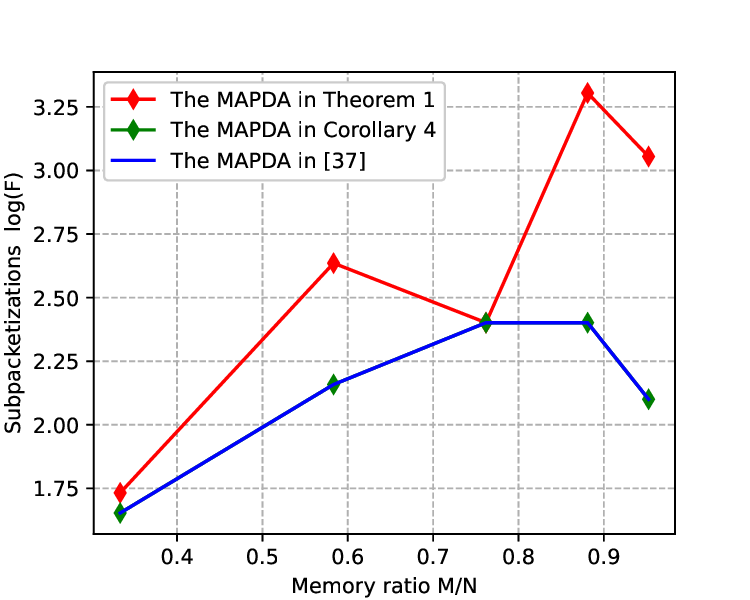}
		\caption{The sum-DoF of MAPDAs in \cite{PERB}, and Theorem \ref{the-olution} and Corollary \ref{corollary-4}.} % 第二张照片的名称
		\label{fig-Co4-g}
	\end{minipage}
\end{figure}

The sum-DoF and subpacketization comparison of MAPDA in Theorem \ref{the-olution}, Corollary \ref{corollary-4} (when $\Lambda = 9$, $r = 3$, and $L = r+1=4$), the MAPDA in PR scheme \cite{PERB} (when $K = {9\choose 3}=84$, $L = 4$) are shown in Fig. \ref{fig-Co4-F} and Fig. \ref{fig-Co4-g}. It can be seen that the scheme of Theorem \ref{the-olution} always attains a sum-DoF no lower than that of the PR scheme under the fully‑connected access topology. Moreover, the scheme in Theorem \ref{the-olution} does not increase the subpacketization level for some memory ratios. The scheme in Corollary \ref{corollary-4} maintains the same structural properties as the theoretical analysis, and thus its performance matches that of the PR scheme in \cite{PERB}.

\section{Example of Theorem Schemes}\label{example-of-MAPDA}
In this section, we propose the examples of Theorem \ref{the-olution} and Theorem \ref{double-solution} to show the main construction idea respectively. We first introduce the example of Theorem \ref{the-olution} in the following.
\subsection{Example of the Scheme in Theorem \ref{the-olution}}
\label{sub-example-th1}
Let us consider an $(L,\Lambda,r,M,N) = (2, 5, 3, 2, 10)$ MAMISO coded caching scheme for the full combination access topology. Based on the MN placement strategy in \cite{MN}, where $t = \Lambda M/N = 1$, each file is divided into $5$ packets, i.e., $\mathbf{w}_n=(\mathbf{w}_{n,j})_{j\in[5]}$ where $n\in[10]$, and the cache nodes respectively store the following packets
\begin{align}
	\mathcal{Z}_{C_1}&=\{\mathbf{w}_{n,1}\mid n\in[10]\},\ \mathcal{Z}_{C_2}=\{\mathbf{w}_{n,2}\mid n\in[10]\},\ 
	\mathcal{Z}_{C_3}=\{\mathbf{w}_{n,3}\mid n\in[10]\},\ \mathcal{Z}_{C_4}=\{\mathbf{w}_{n,4}\mid n\in[10]\},\nonumber\\
	\mathcal{Z}_{C_5}&=\{\mathbf{w}_{n,5}\mid n\in[10]\}.\label{theorem-1-C}
\end{align}To simplify the representation, in this example, a set is shortened by a string. For instance, the set $\{1,2,3\}$ is written as $123$. By Definition \ref{defn:three arrays}, we have the following $5\times 5$ node-placement $\mathbf{C}$ in Table \ref{theorem-1-C-tabel}.
\begin{table}[!ht]
	\centering
	\caption{Node - Placement Array $\mathbf{C}$}
	\label{theorem-1-C-tabel}
	\setlength{\tabcolsep}{3pt} 
	\begin{tabular}{|c|c|c|c|c|c|}
		\hline
		1& 2 & 3 & 4 & 5 & \slashbox{C}{$\mathcal{T}$}  \\ \hline
		* & ~ & ~ & ~ & ~ & 1 \\ \hline
		~ & * & ~ & ~ & ~ & 2 \\ \hline
		~ & ~ & * & ~ & ~ & 3 \\ \hline
		~ & ~ & ~ & * & ~ & 4 \\ \hline
		~ & ~ & ~ & ~ & * & 5 \\ \hline		
	\end{tabular}
\end{table}

\begin{figure}[http!]
	\centering
	\includegraphics[width=0.5\linewidth]{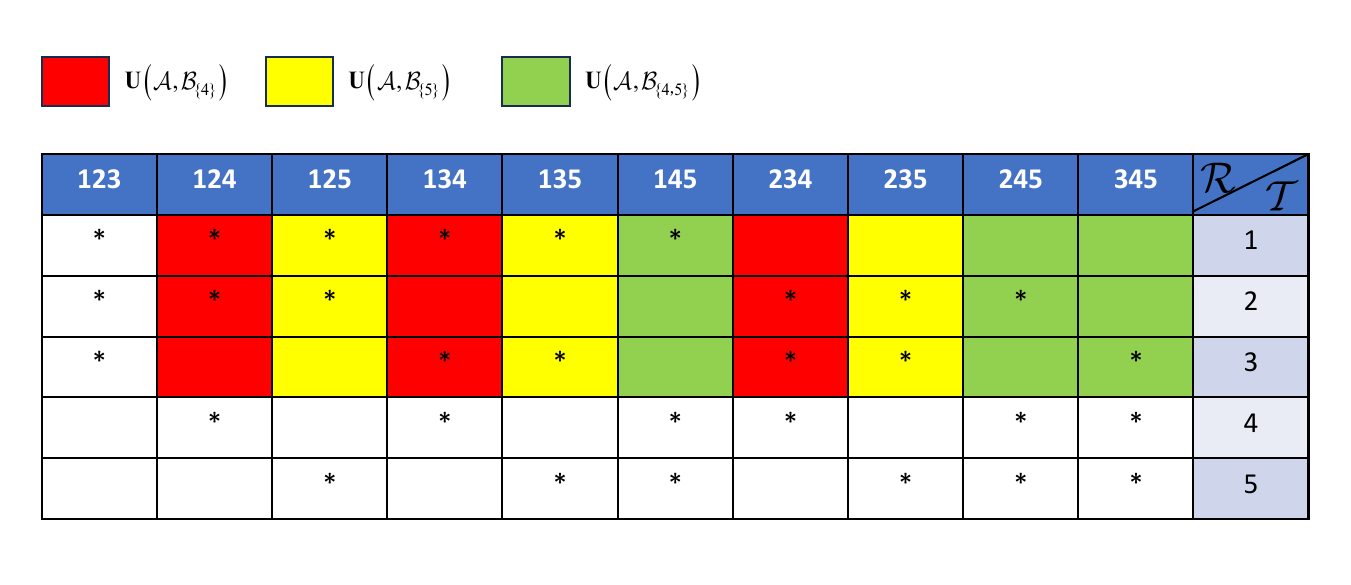}
	\caption{Sub-array partitioned when $b=1$ and $\mathcal{A} = \{1,2,3\}$.}
	\label{fig:1u}
\end{figure}

From the full combination access topology, each user $\mathcal{R}\in{[5]\choose 3}$ can retrieve the packets from $r=3$ cache nodes in $\mathcal{R}$, i.e., 
\begin{align}
	\mathcal{Z}_{123}&=\{\mathbf{w}_{n,i} \mid n\in[10], i\in \{1,2,3\}\},\ \mathcal{Z}_{124}=\{\mathbf{w}_{n,i} \mid n\in[10], i\in \{1,2,4\}\},\nonumber\\
	\mathcal{Z}_{125}&=\{\mathbf{w}_{n,i} \mid n\in[10], i\in \{1,2,5\}\},\	\mathcal{Z}_{134}=\{\mathbf{w}_{n,i} \mid n\in[10], i\in \{1,3,4\}\},\nonumber\\
	\mathcal{Z}_{135}&=\{\mathbf{w}_{n,i} \mid n\in[10], i\in \{1,3,5\}\},\ \mathcal{Z}_{145}=\{\mathbf{w}_{n,i} \mid n\in[10], i\in \{1,4,5\}\},\label{theorem-1-U}\\
	\mathcal{Z}_{234}&=\{\mathbf{w}_{n,i} \mid n\in[10], i\in \{2,3,4\}\},\ \mathcal{Z}_{235}=\{\mathbf{w}_{n,i} \mid n\in[10], i\in \{2,3,5\}\},\nonumber\\
	\mathcal{Z}_{245}&=\{\mathbf{w}_{n,i} \mid n\in[10], i\in \{2,4,5\}\},\ \mathcal{Z}_{345}=\{\mathbf{w}_{n,i} \mid n\in[10], i\in \{3,4,5\}\}.\nonumber
\end{align}By Definition \ref{defn:three arrays}, we have the following $5\times 10$  user-retrieve array $\mathbf{U}$ in Table \ref{theorem-1-U-tabel}.
\begin{table}[!ht]
	\centering
	\caption{User - Retrieve Array $\mathbf{U}$}
	\label{theorem-1-U-tabel}
	\setlength{\tabcolsep}{3pt} 
	\begin{tabular}{|c|c|c|c|c|c|c|c|c|c|c|}
		\hline
		123&123&125&134&135&145&234&235&245&345&  \slashbox{$\mathcal{R}$}{$\mathcal{T}$} \\ \hline
		* & * & * & * & * & * & ~ & ~ & ~ & ~  & 1  \\ \hline
		* & * & * & ~ & ~ & ~ & * & * & * & ~  & 2  \\ \hline
		* & ~ & ~ & * & * & ~ & * & * & ~ & *  & 3 \\ \hline
		~ & * & ~ & * & ~ & * & * & ~ & * & *  & 4 \\ \hline
		~ & ~ & * & ~ & * & * & ~ & * & * & *  & 5 \\ \hline
	\end{tabular}
\end{table}
By Definition \ref{defn:three arrays}, we only need to fill the integers into the non-star entries in $\mathbf{U}$ to obtain the user-delivery array $\mathbf{Q}$. Now, let us take the case $b=1$ as an example to introduce our filling integer strategy. We have $t+r-b=3$ and $\mathcal{I} = [1:2] $ from \eqref{eq-subset}. We first consider the $3$-subset $\mathcal{A} = \{1,2,3\} \in {[5]\choose 3}$. When $ i = 1 $, we have $\mathfrak{G}_1={[\Lambda]\setminus\mathcal{A}\choose 1}=\{\{4\},\{5\}\}$ and 
$$\mathcal{B}_4=\{123,134,234\},\ \mathcal{B}_5=\{125,135,235\}$$ from \eqref{eq-B-G_i}. 
When $ i = 2 $, we have $\mathfrak{G}_2={[\Lambda]\setminus\mathcal{A}\choose 1}=\{\{4,5\}\}$ and 
$$\mathcal{B}_{45}=\{145,245,345\}$$ 
from \eqref{eq-B-G_i}. We can see that the subsets $\mathcal{B}_4$, $\mathcal{B}_5$ and $\mathcal{B}_{45}$ partition the ${[\Lambda]\choose 3}$. Then we have 
\begin{align*}
	3=n=\sum_{i\in\mathcal{I}}{\Lambda-t-r+b\choose i}={5-1-3+1\choose 1}+{5-1-3+1\choose 2}.	
\end{align*}  Hence the array $\mathbf{U}_{\mathcal{A}} = \mathbf{U}({\mathcal{A}\choose 1}, {[\Lambda]\choose 3})$ into $n=3$ sub-arrays in Fig. \ref{fig:1u}.
In Fig. \ref{fig:1u}, the subarray $\mathbf{U}(\mathcal{A},\mathcal{B}_{\{4\}})$ consists of $v_4 = 3$ columns. In addition, each row contains $z_4 = 1$ null entry, and each column contains $u_4 = 1$ null entry. These are consistent with the statement in Proposition \ref{pro-1}. Similarly, we have $v_5 = 3$, $z_5 = 1$, $u_5 = 1$ for $\mathbf{U}(\mathcal{A},\mathcal{B}_{\{5\}})$, and $v_{45} = 3$, $z_{45} = 2$, $u_{45} = 2$ for $\mathbf{U}(\mathcal{A},\mathcal{B}_{\{4,5\}})$. Therefore, we can obtain
\begin{align}
	\label{eq-exaple-problem}
	\left\{
	\begin{array}{c}
		\mathbf{v}=(v_4=3, v_5=3, v_{45}=3),\\	\mathbf{z}=(z_{4}=1, z_{5}=1, z_{45}=2),\\ 
		\mathbf{u}=(u_{4}=1, u_{5}=1, u_{45}=2).
	\end{array}\right.
\end{align}Using dynamic programming, we obtain the optimal solution $\mathbf{x} = (x_4 = 1, x_5 = 1, x_{45} = 0)$ of the $(n,L,\mathbf{z},\mathbf{v})$ knapsack problem in \eqref{eq-KP}. Then we fill the pair $(\mathcal{A},\mathbf{x})$ into each null entry of $\mathbf{U}_{b,\mathcal{A}}^{\mathbf{x}}=\mathbf{U}(\mathcal{A},\mathcal{B}_{4}\cup \mathcal{B}_{5})$ to obtain the following subarray
\begin{align}
	\label{P_s_The1}
	{\mathbf{U}_{b,\mathcal{A}}^{\mathbf{x}}}'=\footnotesize
	\begin{blockarray}{ccccccc}
		124&125&134&135&234&235 \\  
		\begin{block}{(cccccc)c}
			* & * & * & * & (\mathcal{A},\mathbf{x}) & (\mathcal{A},\mathbf{x}) & 1 \\
			* & * & (\mathcal{A},\mathbf{x}) & (\mathcal{A},\mathbf{x}) & *  & * & 2 \\
			(\mathcal{A},\mathbf{x}) & (\mathcal{A},\mathbf{x}) & * & * & * & * & 3 \\
		\end{block}
	\end{blockarray}.
\end{align}We can see that each row of the subarray $	{\mathbf{U}_{b,\mathcal{A}}^{\mathbf{x}}}'$ contains exactly $L=2$ null entries, i.e., condition C$4$ of Definition \ref{def-MAPDA} holds.

\begin{figure}[h]
	\centering
	\includegraphics[width=0.7\linewidth]{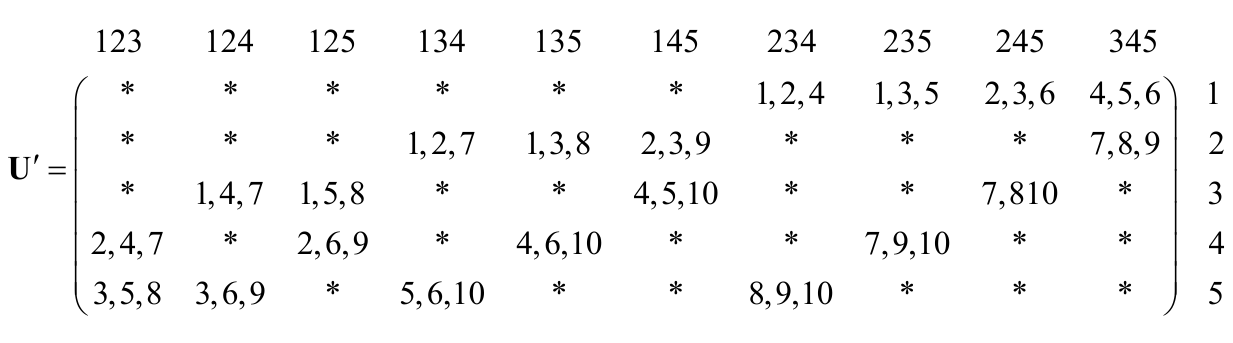}
	\caption{Filled array $\mathbf{U}'$}
	\label{fig:u}
\end{figure}

\begin{figure}[http!]
	\centering
	\includegraphics[width=0.4\linewidth]{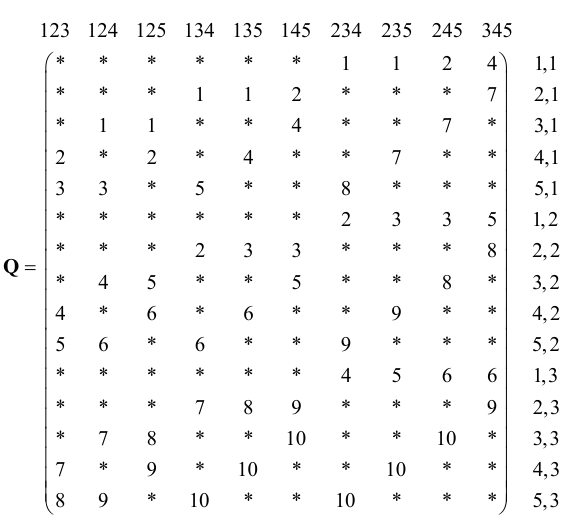}%Examp-1-Q
	\caption{User-delivery array $\mathbf{Q}$}
	\label{fig:q}
\end{figure}

Thus, after completing the filling for all $\mathcal{A}\in{[5]\choose 3}$, the filled array $\mathbf{U}'$ contains exactly $3$ vectors in each non-star position, with a total of $|{[5]\choose 3}|={5\choose 3} = 10$ vectors inserted, as illustrated in Fig. \ref{fig:u}. For the sake of clarity, in the filled array, each non-star entry is filled with an integer, instead of a vector. 
%\begin{align}
%	\label{U1_The1}
%	\mathbf{U}'=\footnotesize
%	\begin{blockarray}{ccccccccccc}
%		123&124&125&134&135&145&234&235&245&345 \\  
%		\begin{block}{(cccccccccc)c}
%			* & * & * & * & * & * &1,2,4 & 1,3,5 & 2,3,6 & 4,5,6 & 1 \\
%			* & * & * & 1,2,7 & 1,3,8 & 2,3,9 & *  & * & * & 7,8,9 & 2 \\
%			* & 1,4,7 & 1,5,8 & * & * & 4,5,10 & * & * & 7,8,10 & * & 3 \\
%			2,4,7 & * & 2,6,9 & * & 4,6,10 & * & * & 7,9,10 & * & * & 4\\
%			3,5,8  & 3,6,9 & * & 5,6,10 & * & * & 8,9,10 & * & * & * & 5\\ 
%		\end{block}
%	\end{blockarray}.
%\end{align}
Then we can obtain our desired user-delivery array $\mathbf{Q}$ by vertically replicating $\mathbf{U}'$ a total of $3$ times and filling the $3$ vectors in each non-star entry of $\mathbf{U}'$ into their corresponding replicated entries across the $3$ copies, respectively. Meanwhile, the server splits
each subfile $\mathbf{w}_{n,\mathcal{T}}$ further into $3$ equal-sized parts. Thus, $\mathbf{w}_{n,\mathcal{T}}=\{\mathbf{w}_{n,\mathcal{T}}^{\mathcal{D}}\ | \ n\in [10], \mathcal{T}\in[5], \mathcal{D}\in [3] \}$. As shown in Fig. \ref{fig:q}.

We can check that $\mathbf{Q}$ is a $(2,10,15,9,10)$ MAPDA which leads to an $(L,\Lambda,r,M,N)=(2,5, 3, 2, 10)$ MAMISO coded caching scheme  with subpacketization $15$ and $\text{DoF}=K(F-Z)/S=6$.

\begin{remark}[General case]\rm
	\label{remark-not-all}
In the above example, all the elements $\mathcal{G}_1\in \mathfrak{G}_1$, i.e., subset $\{4\},\{5\}$, satisfying that $x_{\mathcal{G}_1}=1$. We will use $\mathbf{x}$ to fill its corresponding subarray  ${\mathbf{U}_{b,\mathcal{A}}^{\mathbf{x}}}$. It should be noted that even if not every element $\mathcal{G}'_i$ satisfies $x_{\mathcal{G}'_i}=1$, the existence of an element $\mathcal{G}_i$ with $x_{\mathcal{G}_i}=1$ ensures that for any $\mathcal{G}'_i \in \mathfrak{G}_i$, we can always construct a solution $\mathbf{x}'$ such that $x'_{\mathcal{G}'_i}=1$. In this case, we will use these optimal solutions to fill each of their corresponding subarrays. The detailed proof and the detail filling strategy are included in Subsection \ref{subsect-filling}. 
\end{remark} 

\subsection{Example of the Scheme in Theorem \ref{double-solution}}
\begin{figure}
\centering
\includegraphics[width=0.8\linewidth]{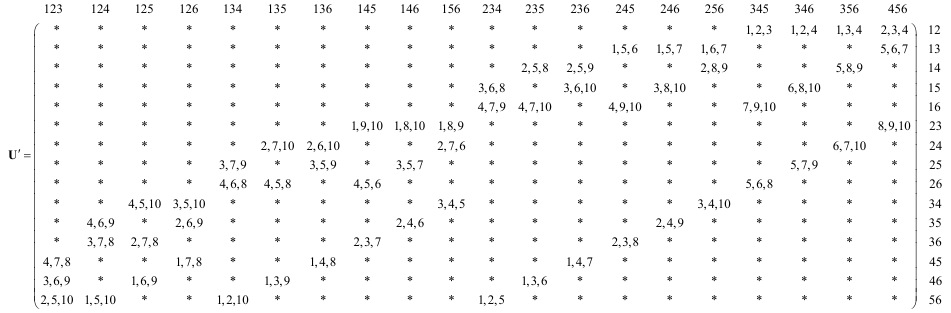}
\caption{Filled array $\mathbf{U}'$}
\label{fig:4-u}
\end{figure}

Let us consider an $(L,\Lambda,r,M,N)$ = (3, 6, 3, 20/3, 20) MAMISO coded caching scheme for the full combination access topology. We also use the MN placement strategy as in the previous example. We have $t = \Lambda M/N =2$, each file is divided into $15$ packets. Since the construction steps have already been described in detail in the previous example, here we omit the caching status of each cache node and the retrieval status of each use. We only propose the following $15\times 6$ node-placement $\mathbf{C}$ in Table \ref{theorem-3-C-tabel} and $15\times 20$  user-retrieve array $\mathbf{U}$ in Table \ref{theorem-3-U-tabel} respectively by Definition \ref{defn:three arrays}. 
\begin{table}[!ht]
	\centering
	\caption{Node - Placement Array $\mathbf{C}$}
	\label{theorem-3-C-tabel}
	\begin{tabular}{|c|c|c|c|c|c|c|c|}
		\hline
		1 & 2 & 3 & 4 & 5 & 6 &  \slashbox{C}{$\mathcal{T}$}  \\ \hline
		* & * & ~ & ~ & ~ & ~ & 12  \\ \hline
		* & ~ & * & ~ & ~ & ~ & 13  \\ \hline
		* & ~ & ~ & * & ~ & ~ & 14  \\ \hline
		* & ~ & ~ & ~ & * & ~ & 15  \\ \hline
		* & ~ & ~ & ~ & ~ & * & 16  \\ \hline
		~ & * & * & ~ & ~ & ~ & 23  \\ \hline
		~ & * & ~ & * & ~ & ~ & 24  \\ \hline
		~ & * & ~ & ~ & * & ~ & 25  \\ \hline
		~ & * & ~ & ~ & ~ & * & 26  \\ \hline
		~ & ~ & * & * & ~ & ~ & 34  \\ \hline
		~ & ~ & * & ~ & * & ~ & 35  \\ \hline
		~ & ~ & * & ~ & ~ & * & 36  \\ \hline
		~ & ~ & ~ & * & * & ~ & 45  \\ \hline
		~ & ~ & ~ & * & ~ & * & 46  \\ \hline
		~ & ~ & ~ & ~ & * & * & 56  \\ \hline
	\end{tabular}
\end{table}
\begin{table}[!ht]
	\centering
	\caption{User - Retrieve Array $\mathbf{U}$}
	\label{theorem-3-U-tabel}
	\setlength{\tabcolsep}{3pt} 
	\begin{tabular}{|c|c|c|c|c|c|c|c|c|c|c|c|c|c|c|c|c|c|c|c|c|}
		\hline
		123 & 124 & 125 & 126 & 134 & 135 & 136 & 145 & 146 & 156 & 234 & 235 & 236 & 245 & 246 & 256 & 345 & 346 & 356 & 456 &  \slashbox{$\mathcal{R}$}{$\mathcal{T}$} \\ \hline
		* & * & * & * & * & * & * & * & * & * & * & * & * & * & * & * & ~ & ~ & ~ & ~ &  12 \\ \hline
		* & * & * & * & * & * & * & * & * & * & * & * & * & ~ & ~ & ~ & * & * & * & ~ &  13\\ \hline
		* & * & * & * & * & * & * & * & * & * & * & ~ & ~ & * & * & ~ & * & * & ~ & * &  14\\ \hline
		* & * & * & * & * & * & * & * & * & * & ~ & * & ~ & * & ~ & * & * & ~ & * & * &  15\\ \hline
		* & * & * & * & * & * & * & * & * & * & ~ & ~ & * & ~ & * & * & ~ & * & * & * &  16\\ \hline
		* & * & * & * & * & * & * & ~ & ~ & ~ & * & * & * & * & * & * & * & * & * & ~ &  23\\ \hline
		* & * & * & * & * & ~ & ~ & * & * & ~ & * & * & * & * & * & * & * & * & ~ & * &  24\\ \hline
		* & * & * & * & ~ & * & ~ & * & ~ & * & * & * & * & * & * & * & * & ~ & * & * &  25\\ \hline
		* & * & * & * & ~ & ~ & * & ~ & * & * & * & * & * & * & * & * & ~ & * & * & * &  26\\ \hline
		* & * & ~ & ~ & * & * & * & * & * & ~ & * & * & * & * & * & ~ & * & * & * & * &  34\\ \hline
		* & ~ & * & ~ & * & * & * & * & ~ & * & * & * & * & * & ~ & * & * & * & * & * &  35\\ \hline
		* & ~ & ~ & * & * & * & * & ~ & * & * & * & * & * & ~ & * & * & * & * & * & * &  36\\ \hline
		~ & * & * & ~ & * & * & ~ & * & * & * & * & * & ~ & * & * & * & * & * & * & * &  45\\ \hline
		~ & * & ~ & * & * & ~ & * & * & * & * & * & ~ & * & * & * & * & * & * & * & * &  46\\ \hline
		~ & ~ & * & * & ~ & * & * & * & * & * & ~ & * & * & * & * & * & * & * & * & * &  56\\ \hline
	\end{tabular}
\end{table} 
Since $r = 3$ and $b \in [0:r-1]$, the condition $b < r < 2b$ implies that $b = 2$ is the only feasible value. In this case, we have $ t + r - b = 3 $, and $\mathcal{I} = [b : \min\{r,\Gamma - t - r + b\}] = [2:3]$. In addition, the condition $ t + r > 2b $ and $\Lambda=2(t+r-b)$ hold. We first consider the $3$-subset $\mathcal{A} = \{1,2,3\} \in {[6]\choose 3}$. When $i=2$, we have $\mathfrak{G}_2={[\Lambda]\setminus\mathcal{A}\choose 2}=\{45,46,56\}$ and 	
\begin{align*}
	\mathcal{B}_{45}=\{145,245,345\}, \ \mathcal{B}_{46}=\{146,246,346\}, \ \mathcal{B}_{56}=\{156,256,356\}
\end{align*}from \eqref{eq-B-G_i}. When $i=3$, we have $\mathfrak{G}_3={[\Lambda]\setminus\mathcal{A}\choose 3}=\{456\}$ and $\mathcal{B}_{456}=\{456\}$ from \eqref{eq-B-G_i}. Similar to the above subsection, we can obtain 
\begin{align*}
\mathbf{v}=(v_{45}=3, v_{46}=3, v_{56}=3, v_{456}=1),\\
\mathbf{z}=(z_{45}=3, z_{46}=3, z_{56}=3, z_{456}=1),\\
\mathbf{u}=(u_{45}=1, u_{46}=1, u_{56}=1, u_{456}=3).
\end{align*} which is consistent with Proposition \ref{pro-1}. By dynamic programming, we can obtain the optimal solution $\mathbf{x}=(x_{45}=1,x_{46}=1, x_{56}=1, x_{456}=0)$ of the knapsack problem in \eqref{eq-KP}. Based on this solution, we can obtain the subarray $\mathbf{U}_{b,\mathcal{A}}^{\mathbf{x}}=\mathbf{U}({\mathcal{A}\choose 2},\mathcal{B}_{45}\cup \mathcal{B}_{46}\cup\mathcal{B}_{56})$ and
\begin{align}\label{Exa-Th4-b=2}
\mathbf{U}\left({\mathcal{A}\choose 2},\mathcal{B}_{45}\cup \mathcal{B}_{46}\cup\mathcal{B}_{56}\right)=\footnotesize
\begin{blockarray}{cccccccccc}
	145 & 146 & 156 & 245 & 246 & 256 & 345 & 346 & 356 \\
	\begin{block}{(ccccccccc)c}
		* & * & * & * & * & * & \square & \square & \square &   12\\ 
		* & * & * & \square & \square & \square & * & * & * &   13\\ 
		\square & \square & \square & * & * & * & * & * & * &   23\\ 
	\end{block}
\end{blockarray}.
\end{align}It is worth noting that when $i=1$, using \eqref{eq-B-G_i} we can obtain the subarray
\begin{align}\label{Exa-Th4-b=1}
\mathbf{U}\left({\mathcal{A}\choose 2},\mathcal{B}_{4}\cup \mathcal{B}_{5}\cup\mathcal{B}_{6}\right)=\footnotesize
\begin{blockarray}{cccccccccc}
	124 & 125 & 126 & 134 & 135 & 136 & 234 & 235 & 236 \\
	\begin{block}{(ccccccccc)c}
		* & * & * & * & * & * & * & * & * &   12\\ 
		* & * & * & * & * & * & * & * & * &   13\\ 
		* & * & * & * & * & * & * & * & * &   14\\ 
	\end{block}
\end{blockarray}.
\end{align}
Let us consider the other $3$-subset $\mathcal{A}' =[\Lambda]\setminus \mathcal{A}=\{4,5,6\}$. Similar to the above subsection, we can also obtain 
\begin{align*}
	\mathbf{v}'&=(v_{12}=3, v_{13}=3, v_{23}=3, v_{123}=1),\\
	\mathbf{z}'&=(z_{12}=3, z_{13}=3, z_{23}=3, z_{123}=1),\\
	\mathbf{u}'&=(u_{12}=1, u_{13}=1, u_{23}=1, u_{123}=3).
\end{align*}By replacing the integers $1$, $2$ and $3$ by $4$, $5$ and $6$ respectively, we can also obtain the optimal solution $\mathbf{x}' = (x_{12}=1, x_{13}=1, x_{23}=1, x_{123}=0)$ of the $(n,L,\mathbf{z}', \mathbf{v}')$ knapsack problem in \eqref{eq-KP}. Then we can obtain the subarray $\mathbf{U}_{b,\mathcal{A}'}^{\mathbf{x}'}=\mathbf{U}({\mathcal{A}'\choose 2},\mathcal{B}_{12}'\cup \mathcal{B}_{13}'\cup\mathcal{B}_{23}')$ and
\begin{align}\label{Exa-Th4-b=2-2}
	\mathbf{U}\left({\mathcal{A}\choose 2},\mathcal{B}_{12}\cup \mathcal{B}_{13}\cup\mathcal{B}_{23}\right)=\footnotesize
	\begin{blockarray}{cccccccccc}
		124 & 125 & 126 & 134 & 135 & 136 & 234 & 235 & 236 \\
		\begin{block}{(ccccccccc)c}
			* & * & * & * & * & * & \square & \square & \square &   45\\ 
			* & * & * & \square & \square & \square & * & * & * &   46\\ 
			\square & \square & \square & * & * & * & * & * & * &   56\\ 
		\end{block}
	\end{blockarray}.
\end{align} 
When $i=1$, using \eqref{eq-B-G_i} we also can obtain the subarray\begin{align}\label{Exa-Th4-b=1-2}
	\mathbf{U}\left({\mathcal{A}'\choose 2},\mathcal{B}_{1}'\cup \mathcal{B}_{2}'\cup\mathcal{B}_{3}'\right)=\footnotesize
	\begin{blockarray}{cccccccccc}
		145 & 146 & 156 & 245 & 246 & 256 & 345 & 346 & 356 \\
		\begin{block}{(ccccccccc)c}
			* & * & * & * & * & * & * & * & * &   45\\ 
			* & * & * & * & * & * & * & * & * &   46\\ 
			* & * & * & * & * & * & * & * & * &   56\\ 
		\end{block}
	\end{blockarray}.
\end{align}
Since the subarrays in \eqref{Exa-Th4-b=1} and \eqref{Exa-Th4-b=1-2} consist entirely of star entries, and they respectively have the same column labels with the subarrays in \eqref{Exa-Th4-b=2-2} and \eqref{Exa-Th4-b=2}, we can merge them to obtain a new subarray
\begin{align*}
\mathbf{U}\left({\mathcal{A}\choose 2}\cup {\mathcal{A}'\choose 2},\mathcal{B}_{45}\cup \mathcal{B}_{46}\cup\mathcal{B}_{56}\cup \mathcal{B}_{12}'\cup \mathcal{B}_{13}'\cup\mathcal{B}_{23}' \right).	
\end{align*}It can be checked that every row of this new subarray contains exactly $L=3$ null entries, i.e., condition C$4$ of Definition \ref{def-MAPDA} holds. The difference from the previous example is that we perform the filling on the newly obtained subarray instead of on $\mathbf{U}_{b,\mathcal{A}}^{\mathbf{x}}$. Consequently, the number of distinct vector types (equivalently, the parameter $S$) is reduced by half. When $\mathcal{A}$ runs all the possible element in ${[6]\choose 3}$, each non-star entry of the obtained array $\mathbf{U}'$ has exactly $3$ different vectors, as illustrated in Fig \ref{fig:4-u}. We can check that there are exactly $|{[6]\choose 3}|/2={6\choose 3}/2= 10$ different vectors in $\mathbf{U}'$.
Finally, we can obtain our desired user-delivery array $\mathbf{Q}$ by vertically replicating $\mathbf{U}'$ a total of $3$ times and filling the $3$ vectors in each non-star entry of $\mathbf{U}'$ into their corresponding replicated entries across the $3$ copies, respectively. Therefore, $\mathbf{Q}$ is a $(3,20, 45, 36, 10)$ MAPDA which leads to a $(3,6,3,20/3, 20)$  MAMISO coded caching scheme  with subpacketization $45$ and $\text{DoF}=K(F-Z)/S=18$.

\section{Proof of Theorems}\label{proof-of-theorem}

We use the MN placement strategy in \cite{MN}. That is, each file $\mathbf{w}_n \in \mathcal{W}$ is divided into ${\Lambda\choose t}$ non-overlapping equal-sized subfiles, i.e., $\mathbf{w}_n =(\mathbf{w}_{n,\mathcal{T}})_{\mathcal{T}\in{[\Lambda]\choose t}}$ where $t=\Lambda M/N \in [\Lambda]$.  The cache node $C_\lambda$ where $\lambda \in [\Lambda]$ stores the following packets
$$\mathcal{Z}_{C_\lambda} = \left\{\mathbf{w}_{n,\mathcal{T}}\ \Big|\ n\in [N], \lambda\in\mathcal{T}, \mathcal{T}\in{[\Lambda]\choose t}\right\}.$$
Recall that each user $\mathcal{R}\in{[\Lambda]\choose r}$ can retrieve the packets from $r$ cache nodes in $\mathcal{R}$, i.e., $\mathcal{Z}_{\mathcal{R}}=\cup_{\lambda \in \mathcal{R}} \mathcal{Z}_{C_\lambda}$. Then we have the node-placement array $\mathbf{C}=(\mathbf{C}(\mathcal{T}, \lambda))_{\mathcal{T}\in{[\Lambda]\choose t},\lambda\in[\Lambda]}$, user-retrieve array $\mathbf{U}=(\mathbf{U}(\mathcal{T}, \mathcal{R}))_{\mathcal{T}\in{[\Lambda]\choose t},\mathcal{R}\in{[\Lambda]\choose r}}$ where for each $\mathcal{T}\in{[\Lambda]\choose t}$, $\lambda\in[\Lambda]$ and $\mathcal{R}\in{[\Lambda]\choose r}$, the entries are respectively defined as follows.
\begin{align}
	\label{eq-caching-retrieval-arrays}
	\mathbf{C}(\mathcal{T}, \lambda) = 
	\begin{cases} 
		\ast &  \text{if}\ \lambda \in \mathcal{T} \\
		Null &  \text{Otherwise} 
	\end{cases},\ \ \ \text{and}\ \ \ 
	\mathbf{U}(\mathcal{T}, \mathcal{R}) = 
	\begin{cases} 
		\ast &  \text{if}\ \mathcal{T}\cap \mathcal{R} \neq \emptyset\\
		Null &  \text{Otherwise} 
	\end{cases}.
\end{align} 
In order to realize the delivery phase, the null positions of the user-retrieve array $\mathbf{U}$ will be filled with integers such that the user-delivery array $\mathbf{Q}$ is an MAPDA. In the following, we will focus on constructing $\mathbf{Q}$ by filling some integers into the Null entries of $\mathbf{U}$.

By Definition \ref{def-MAPDA}, given a $\mathbf{Q}$ let us consider the subarray $\mathbf{Q}^{(s)}$ where $s\in[S]$. Denote its row label set by $\mathcal{F}_s$ and column label set by $\mathcal{K}_s$. Using the row label set $\mathcal{F}_s$ and column label set $\mathcal{K}_s$, we have its related subarray $\mathbf{U}(\mathcal{F}_s,\mathcal{K}_s)$ of $\mathbf{U}$. By condition C$3$ of Definition \ref{def-MAPDA}, the subarray $\mathbf{U}(\mathcal{F}_s,\mathcal{K}_s)$ has the following the property. 
\begin{align}
\label{P-C4}
\left|\{k \in \mathcal{K}_s \mid \mathbf{U}(f,k) = \text{Null} \}\right| \leq L, \quad \forall f \in \mathcal{F}_s.
\end{align}
It is difficult to directly extract from the array $\mathbf{U}$ a subarray $\mathbf{U}(\mathcal{F}_s,\mathcal{K}_s)$ that satisfies \eqref{P-C4}. To address this, we adopt the following approach. Fix a subset $\mathcal{A}\in\binom{[\Lambda]}{t+r-b}$ with $b\in[0:r-1]$. We partition $\binom{\mathcal{A}}{r}$ according to \eqref{eq-B-G_i}, obtaining $\mathcal{B}_{\mathcal{G}_i}$ where $i\in[b:\min\{\Lambda-t-r+b,r\}]$, $\mathcal{G}_i\in \mathfrak{G}_i$. Each $\mathcal{B}_{\mathcal{G}_i}$ then gives rise to a subarray $\mathbf{U}({\mathcal{A}\choose t},\mathcal{B}_{\mathcal{G}_i})$. Finally, from these partitioned subarrays we select some subarrays whose horizontal concatenation obtains subarray $\mathbf{U}(\mathcal{F}_s,\mathcal{K}_s)$ satisfying \eqref{P-C4}. 

From Lemma \ref{max-DoF}, we know that the sum-DoF depends on $|\mathcal{K}_s|$, whose value corresponds to the number of columns in $\mathbf{Q}^{(s)}$. To maximize the achievable sum-DoF, we prefer to obtain a selection of subarrays such that the horizontally concatenated subarray $\mathbf{U}(\mathcal{F}_s,\mathcal{K}_s)$ satisfies \eqref{P-C4} and contains as many columns as possible. To this end, we formulate the selection process as a knapsack problem based on the structure of the subarrays, thereby transforming it into an optimization problem. Based on the solution of the knapsack problem, we can obtain a subarray $\mathbf{U}(\mathcal{F}_s,\mathcal{K}_s)$ that satisfies \eqref{P-C4}. Owing to the strong symmetry of the MN placement strategy, for any $t+r-b$-subset we can obtain our desired array $\mathbf{Q}$. The proof of Theorem \ref{the-olution} is divided into the following two parts. 
\subsection{Selecting subarrays}
\label{subsect-subarray-selction}
For any positive integers $\Lambda$, $r$, $t$ and $L$, let us consider each integer $b\in[0:r-1]$ and any subset $\mathcal{A} \in {[\Lambda]\choose t+r-b}$ for the user-retrieve array $\mathbf{U}$. Recall that the parameters  $n=\sum_{i\in\mathcal{I}}{\Lambda-t-r+b\choose i}$, $\mathbf{z}=(z_{\mathcal{G}_i})_{i\in\mathcal{I}, \mathcal{G}_i\in \mathfrak{G}_i}$ and $\mathbf{v}=(v_{\mathcal{G}_i})_{i\in\mathcal{I}, \mathcal{G}_i\in \mathfrak{G}_i}$ only depend on the integer $b$ and the subset $\mathcal{A}$. Consequently, the solution $\mathbf{x}$ of a $(n,L,\mathbf{z},\mathbf{v})$ knapsack problem can be determined by the parameters $b$ and $\mathcal{A}$.  
%In addition, given the integer $b$ and the subset $\mathcal{A}$, all the solutions satisfy the following useful properties.
Based on the solution $\mathbf{x}$, we can obtain a subarray $\mathbf{U}_{b,\mathcal{A}}^{\mathbf{x}}$ of $\mathbf{U}$ where the row labels are the subsets in ${\mathcal{A}\choose t}$ and the columns labels are the subsets each of which is the element in $\mathcal{B}_{\mathcal{G}_i}$ with  $x_{\mathcal{G}_i} = 1$ for some integer $i\in\mathcal{I}$. Thus, the set of column labels of subarray $\mathbf{U}_{b,\mathcal{A}}^{\mathbf{x}}$  can be expressed as $\cup_{x_{\mathcal{G}_i}= 1} \mathcal{B}_{\mathcal{G}_i}$.
Now, let us count the number of Null entries in each row of $\mathbf{U}_{b,\mathcal{A}}^{\mathbf{x}}$. For each integer $i \in \mathcal{I}$ and any subset $\mathcal{G}_i \in \mathfrak{G}_i$, if $x_{\mathcal{G}_i} = 1$, we have  $z_{\mathcal{G}_i}={r-b \choose r-i}$ Null entries in each row of $\mathbf{U}_{\mathcal{A},\mathcal{B}_{\mathcal{G}_i}}$. So there are exactly
\begin{align}
\label{eq-L}
\sum_{i\in\mathcal{I}}\sum_{\mathcal{G}_i\in \mathfrak{G}_i}z_{\mathcal{G}_i}x_{\mathcal{G}_i}=
\sum_{i\in\mathcal{I}}\sum_{\mathcal{G}_i\in \mathfrak{G}_i}{r-b \choose r-i}x_{\mathcal{G}_i}
\leq L
\end{align}Null entries in each row. 

For instance, let us consider the parameters, i.e.,  $L=2$, $\Lambda=5$, $r=3$, $t=1$, $b=1$ and $\mathcal{A} = \{1,2,3\}$, in Subsection \ref{sub-example-th1}.  The partition of sub-arrays and their detailed structures are shown in Fig. \ref{fig:1u}. Form \eqref{eq-exaple-problem}, we can obtain a solution $\mathbf{x} =(x_4=1,x_5 = 1,x_{45}=0)$ of the $(n,L,\mathbf{z},\mathbf{v})$ knapsack problem. Since $\mathcal{B}_4=\{123,134,234\}$ and $ \mathcal{B}_5=\{125,135,235\}$, we have
\begin{align*}
	\mathbf{U}_{b,\mathcal{A}}^{\mathbf{x}}=\mathbf{U}\left({\mathcal{A}\choose 1}, \mathcal{B}_4\cup \mathcal{B}_5\right)
=\footnotesize
	\begin{blockarray}{ccccccc}
		124&125&134&135&234&235 \\  
		\begin{block}{(cccccc)c}
			* & * & * & * & \square & \square & 1 \\
			* & * & \square & \square & *  & * & 2 \\
			\square & \square & * & * & * & * & 3 \\
		\end{block}
	\end{blockarray}.
\end{align*} 

%\subsection{Constructing a set system}
%\label{Constructing-set-system}
%Recall that for any $i \in \mathcal{I}$ and all subsets $\mathcal{G}_i$ in $\mathfrak{G}_i$, the corresponding values of $v_{\mathcal{G}_i}$, $z_{\mathcal{G}_i}$, and $u_{\mathcal{G}_i}$ are all the same, so we can obtain the other solutions from one solution. We construct $\mathbf{Q}$ via the solution $\mathbf{x}$.

By Proposition \ref{pro-1}, we know that $z_{\mathcal{G}_i}=z_{\mathcal{G}_i'}$ and $v_{\mathcal{G}_i}=v_{\mathcal{G}_i'}$ for any two different subsets ${\mathcal{G}_i},{\mathcal{G}_i'} \in \mathfrak{G}_i$ where $i\in \mathcal{I}$. Then the following result can be directly obtained. 
\begin{proposition}\rm \label{swap}
Given a  solution $\mathbf{x}=(x_{\mathcal{G}_i})_{i\in\mathcal{I} ,\mathcal{G}_i\in\mathfrak{G}_i}$ of the $(n,L,\mathbf{z},\mathbf{v})$ knapsack problem, the vector $\mathbf{x}'$, which is generated by exchanging the values of any two coordinates labeled by ${\mathcal{G}_i},{\mathcal{G}_i'} \in \mathfrak{G}_i$ for some integer $i\in\mathcal{I}$, is also a solution and $\phi(\mathbf{v},\mathbf{x})=\phi(\mathbf{v},\mathbf{x}')$. \hfill $\square$  
\end{proposition}

Given a solution $\mathbf{x}=(x_{\mathcal{G}_i})_{i\in\mathcal{I} ,\mathcal{G}_i\in\mathfrak{G}_i}$ of the $(n,L,\mathbf{z},\mathbf{v})$ knapsack problem, assume that there are exactly $m$ integers  $i_1$, $i_2$, $\ldots$, $i_m\in\mathcal{I}$ such that there exits at least one non-zero entry labeled by the subset in ${[\Lambda]\setminus \mathcal{A} \choose i_j}$ of $\mathbf{x}$. In addition, for each $j\in[m]$ we assume that there are $q_{j}$ non-zero entries labeled by the subsets in ${[\Lambda]\setminus \mathcal{A} \choose i_j}$, and let $p_j=|{[\Lambda]\setminus\mathcal{A}\choose i_j}|={\Lambda-t-r+b\choose i_j}$ and $\ell_j =\text{LCM}(q_j,p_j)$. By Proposition \ref{swap}, we can obtain another solution of the $(n,L,\mathbf{z},\mathbf{v})$ knapsack problem by setting $x_{\mathcal{G}_{i_j}}=1$ for selected $q_j$ $\mathcal{G}_{i_j}$ among the $p_j$ elements of $\mathfrak{G}_{i_j}$ where $j\in[m]$ and  $x_{\mathcal{G}_i}=0$ for the left elements $\mathcal{G}_i$. In the following, we employ a regular design to derive some different solutions from $\mathbf{x}$, ensuring that each element of $\mathfrak{G}_{i_j}$ is selected an equal number of times for every $j \in [m]$. A pair $(\mathcal{V}, \mathfrak{R})$ is called a design where $\mathcal{V}$ is a set of elements called points, and  $\mathfrak{R}$ is a collection of non-empty subsets of $\mathcal{V}$ called blocks. A design is called $r$-regular if each point occurs in exactly $r$ blocks. A $r$-regular design containing $v$ points and $k$ blocks each of which has zies $z$ is denoted by $r$-regular $(v,z,k)$ design. For any parameters $v$ and $k$, the following result is useful in this paper where the detailed proof is included in Appendix\ref{proof-r-regular}.
\begin{lemma}\rm\label{LCM-r-regular}
	For any positive integers $v$ and $k$, there exists a $r$-regular $(v,k)$ design where $r=\text{LCM}(v,k)  /v$.   \hfill $\square$ 
\end{lemma}

For each integer $j\in[m]$, by Lemma \ref{LCM-r-regular} we can obtain a $(\ell_j  /p_j)$-regular $(p_j,q_j)$ design $(\mathcal{V}_j={[\Lambda]\setminus\mathcal{A}\choose i_j}, \mathcal{R}_j)$ where $\mathcal{R}_j=\{\mathcal{R}_{j,1},\mathcal{R}_{j,2},\ldots,\mathcal{R}_{j,\ell_j  /q_j}\}$. Let $\ell=\text{LCM}(\ell_j/q_j)_{j\in[m]}$. We can construct a set system 
$(\mathcal{V}=\cup_{j\in[m]}\mathcal{V}_j,\mathcal{R}=\{\mathcal{R}'_h| h\in[\ell]\})$ where for each $h\in[\ell]$ the following equation holds.
\begin{align}
	\label{eq-support-set}
	\mathcal{R}_h'=\mathcal{R}_{i_1,<h>_{\ell_1/q_1}}\bigcup \mathcal{R}_{i_2,<h>_{\ell_2/q_2}}\bigcup \cdots\bigcup \mathcal{R}_{i_m,<h>_{\ell_m/q_m}}.
\end{align}Based on the set system $(\mathcal{V},\mathcal{R})$, Proposition \ref{swap} yields $\ell$ distinct solutions to the $(n,L,\mathbf{z},\mathbf{v})$ knapsack problem. Denote these solutions by $\mathcal{X} = \{\mathbf{x}_1 = \mathbf{x}, \mathbf{x}_2, \dots, \mathbf{x}_\ell\}$. The selection frequency of each element in $\mathfrak{G}_{i_j}$ across $(\mathcal{V},\mathcal{R})$ can then be computed. That is the following result whose proof is included in Appendix \ref{appendix-lemma-a}.  
\begin{lemma}\rm\label{element-set-numbers}
In the set system $(\mathcal{V},\mathcal{R})$, each subset $\mathcal{G}_{i_j} \in {[\Lambda]\setminus \mathcal{A} \choose i_j}$ occurs exactly $\ell q_{j}/p_j$ blocks where $j\in[m]$. 
\end{lemma}

For instance, let us also consider the parameters in Subsection \ref{sub-example-th1}. Under $\mathcal{A}=\{1,2,3\}$ we have a solution $\mathbf{x} = (x_4=1, x_5=1,x_{45}=0)$. Only subsets $\mathcal{G}_1 \in \mathfrak{G}_1 = \binom{[\Lambda]\setminus\mathcal{A}}{1}$ satisfy $x_{\mathcal{G}_1}=1$; hence $m=1$ and $i_1=1$. Moreover, $p_1 = q_1 = 2$, so \(\ell_1 = \text{LCM}(p_1,q_1)=2\). Consequently, we have $\ell = \text{LCM}(\ell_j/q_j)_{j\in[m]} = \text{LCM}(\ell_1/q_1) = 1$.
From \eqref{eq-support-set}, we obtain the set system $(\mathcal{V}, \mathcal{R})$ with $\mathcal{V} = \{4,5\}$ and $\mathcal{R} = \{\{4,5\}\}$. Therefore $\mathcal{X}$ contains exactly $\ell = 1$ solution, i.e., $\mathcal{X} = \{\mathbf{x}\}$.

\subsection{Construction of $\mathbf{Q}$ via Knapsack Problem}
\label{subsect-filling}

In the following, we will use these solutions in $\mathcal{X}$ to complete the filling strategy for the user-retrieval array $\mathbf{U}$. Conequently, our  desired user-delivery array $\mathbf{Q}$ can be obtained. 

Recall that given a solution $\mathbf{x}\in \mathcal{X}$, we obtain the subarray $\mathbf{U}_{b,\mathcal{A}}^{\mathbf{x}}$ whose column index set is $\cup_{i\in\mathcal{I} ,\mathcal{G}_i\in\mathfrak{G}_i,x_{\mathcal{G}_i} = 1} \mathcal{B}_{\mathcal{G}_i}$, and each column of $\mathbf{U}_{b,\mathcal{A}}^{\mathbf{x}}$ has at least one Null entry. In particular, each column labeled by the subset in $\mathcal{B}_{\mathcal{G}_{i_j}}$ has exactly $u_{\mathcal{G}_{i_j}}$ Null entries. Let $\mu=\text{LCM}(u_{\mathcal{G}_{i_j}})_{j\in[m]}$. For each $\mathcal{T}\in {\mathcal{A}\choose t}$ and $\mathcal{D}\in \mathcal{B}_{\mathcal{G}_{i_j}}$ where $j\in[m]$, if $\mathcal{D}\cap \mathcal{T} \neq \emptyset$ we fill the following $\mu/u_{\mathcal{G}_{i_j}}$ vectors 
\begin{align}
	\label{fill-ruler}
	\left((n-1)\frac{\mu}{u_{\mathcal{G}_{i_j}}}+1,\mathcal{A},\mathbf{x}\right), \left((n-1)\frac{\mu}{u_{\mathcal{G}_{i_j}}}+2,\mathcal{A},\mathbf{x}\right), \ldots,\left(n\frac{\mu}{u_{\mathcal{G}_{i_j}}},\mathcal{A},\mathbf{x}\right) 
\end{align}into the entry $\mathbf{U}_{b,\mathcal{A}}^{\mathbf{x}}(\mathcal{T},\mathcal{D})$ where $n$ is the order of the Null entries in the column labeled by $\mathcal{D}$. From \eqref{eq-L}, each row of  $\mathbf{U}_{b,\mathcal{A}}^{\mathbf{x}}$ has 
$\phi(\mathbf{v},\mathbf{x})\leq L$ Null entries, i.e., the Condition $4$ of Definition \ref{def-MAPDA} holds. In addition, there are exactly $\cup_{i\in\mathcal{I} ,\mathcal{G}_i\in\mathfrak{G}_i,x_{\mathcal{G}_i} = 1} \mathcal{B}_{\mathcal{G}_i}=\phi(\mathbf{v},\mathbf{x})$ columns in $\mathbf{U}_{b,\mathcal{A}}^{\mathbf{x}}$. Then we have the sum-DoF $g=\phi(\mathbf{v},\mathbf{x})$.

For instance, let us consider the parameters in Subsection \ref{sub-example-th1} and the solution $\mathbf{x} = (x_4 = 1,x_5 = 1,x_{45}=0)\in\mathcal{X}$ under $\mathcal{A}=\{1,2,3\}$. Because $m=1$, we have $\mu/u_{\mathcal{G}{i_1}}=1$. Hence, we fill the single vector $(\mathcal{A},\mathbf{x})$ into the first non‑star entry of each column of $\mathbf{U}_{b,\mathcal{A}}^{\mathbf{x}}$, yielding
\begin{align*}
\footnotesize
	\begin{blockarray}{ccccccc}
		124&125&134&135&234&235 \\  
		\begin{block}{(cccccc)c}
			* & * & * & * & (\mathcal{A},\mathbf{x}) & (\mathcal{A},\mathbf{x}) & 1 \\
			* & * & (\mathcal{A},\mathbf{x}) & (\mathcal{A},\mathbf{x}) & *  & * & 2 \\
			(\mathcal{A},\mathbf{x}) & (\mathcal{A},\mathbf{x}) & * & * & * & * & 3 \\
		\end{block}
	\end{blockarray}.
\end{align*}

By the filling strategy in \eqref{fill-ruler}, we can obtain a new array, denoted by $\mathbf{U}'$, where each vector satisfies the Conditions $3$-$4$ of Definition \ref{def-MAPDA}. In addition, the following statements can be obtained, where the proof is included in Appendix \ref{appendex-lemma 5}.
\begin{lemma}\label{element-number}
Given the integer $b\in[0:r-1]$, each non-star entry contains exactly 
\begin{align*}
\pi=\sum_{j \in [m]} {r\choose r-i_j}{\Lambda-t-r\choose i_j-b}\cdot \frac{ q_{j}}{p_j}\cdot  \frac{\mu}{u_{\mathcal{G}_{i_j}}}
\end{align*}vectors, and there are exactly $S=\ell\mu{\Lambda\choose t+r-b}$ different vectors in $\mathbf{U}'$.
\end{lemma} 

For instance, when $L=2$, $\Lambda=5$, $r=3$, $t=1$ and $b=1$, from \eqref{eq-choose-A} and \eqref{eq-A-number} we obtain 
$$\pi={r\choose r-i_1}{\Lambda-t-r\choose i_1-b}\cdot \frac{ q_{j}}{p_j}\cdot  \frac{\mu}{u_{\mathcal{G}_{i_j}}}={3\choose 3-1}{5-1-3\choose 1-1}=3,$$
 and from lemma \ref{element-number} we have $\ell\mu{\Lambda\choose t+r-b}=1\cdot 1\cdot{5\choose 1+3-1}=10$. After completing the filling process for each $\mathcal{A}\in\binom{[5]}{3}$, we obtain the array $\mathbf{U}'$ displayed in Fig. \ref{fig:u}. We can see that each non-star position of $\mathbf{U}'$ contains $\pi=3$ vectors, and the total number of distinct vectors filled in $\mathbf{U}'$ is $10$. We can obtain the user-delivery array $\mathbf{Q}$ in Fig. \ref{fig:q} by replicating the array $\mathbf{U}'$ $\pi=3$ times.

By Lemma \ref{element-number}, each non-star entry of $\mathbf{U}'$ has the same number $\pi$ of vectors. Then we can obtain our desired user-delivery array $\mathbf{Q}$ by vertically replicating $\mathbf{U}'$ a total of $\pi$ times and filling the $\pi$ vectors of each non-star entry of $\mathbf{U}'$ into their corresponding replicated entries across the $\pi$ copies, respectively. In addition, $\mathbf{Q}$ has $F=\pi\cdot {\Lambda\choose t}$ rows since $\mathbf{U}'$ has exactly ${\Lambda\choose t}$ rows. Each column of $\mathbf{Q}$ has $Z=\pi\cdot ({\Lambda\choose t}-{\Lambda-r\choose t})$. So, $\mathbf{Q}$ is a $(L,K,F,Z,S)$ MAPDA. In conclusion, the proof of Theorem \ref{the-olution} is completed.
 
\subsection{Proof of Theorem \ref{double-solution}}\label{double-solution-proof} 
%For any positive integers $\Lambda$, $r$, $t$, $b$ and $L$ with $b\in[0:r-1]$. Based on Theorem \ref{the-olution}, we can obtain a $$\left(L,K={\Lambda\choose r},F=\pi{\Lambda\choose t},Z=\pi\left({\Lambda\choose t}-{\Lambda-r\choose t}\right),S=\ell\mu{\Lambda\choose t+r-b} \right)$$ MAPDA.

Let $b<r<2b$, $t+r>2b$, $\Lambda=2(t+r-b)$ and $L\leq{t+r-b\choose b}$. For any $\mathcal{A}\in {[\Lambda]\choose t+r-b}$, according to Proposition \ref{pro-1}, we have $v_{\mathcal{G}_i}={t+r-b\choose r-i}$ and $z_{\mathcal{G}_i}={r-b\choose r-i}$ where  $i\in\mathcal{I}=[b:\min(r,t+r-b)]$, $\mathcal{G}_i\in \mathfrak{G}_i$, and $\mathfrak{G}_i={[\Lambda]\setminus\mathcal{A}\choose i}$.

Suppose $L={t+r-b\choose b}$. When $i=b$, we obtain $\mathfrak{G}_b$. For every $\mathcal{G}_b\in \mathfrak{G}_b$, the subarray $\mathbf{U}_{\mathcal{A},\mathcal{B}_{\mathcal{G}_b}}$ has $v_{\mathcal{G}_b}= {t+r-b\choose r-b}$ columns, and each of its rows contains exactly $z_{\mathcal{G}_b}= {r-b\choose r-b}=1$ Null entry. There are $|\mathfrak{G}_b|={t+r-b\choose b}$ such subarrays, i.e., contains ${t+r-b\choose r-b}$ columns, and each of its rows contains exactly $z_{\mathcal{G}_b}= {r-b\choose r-b}=1$ Null entry. When $i=r-b<b$, we have $\mathfrak{G}_{r-b}$. For any $\mathcal{G}_{r-b}\in\mathfrak{G}_{r-b}$, the subarray $\mathbf{U}_{\mathcal{A},\mathcal{B}_{\mathcal{G}_{r-b}}}$ is an all-star subarray; it contains ${t+r-b\choose r-i}={t+r-b\choose b}$ columns. The total number of such all‑star subarrays is $|\mathfrak{G}_{r-b}|={t+r-b\choose r-b}$. Consequently, both the subarray $\mathbf{U}({\mathcal{A}\choose t},\cup_{\mathcal{G}_b\in \mathfrak{G}_b}\mathcal{B}_{\mathcal{G}_b})$ and the subarray $\mathbf{U}({\mathcal{A}\choose t},\cup_{\mathcal{G}_{r-b}\in \mathfrak{G}_{r-b}}\mathcal{B}_{\mathcal{G}_{r-b}})$ contain ${t+r-b\choose b}{t+r-b\choose r-b}$ columns.

Let $\mathcal{A}'=[\Lambda]\setminus\mathcal{A}$. From \eqref{eq-B-G_i}, every column label $\mathcal{D}\in\cup_{\mathcal{G}_{r-b}\in \mathfrak{G}_{r-b}}\mathcal{B}_{\mathcal{G}_{r-b}}$ satisfies $|\mathcal{D}\cap\mathcal{A}'|=b$; therefore the column-label set of the all-star subarray, $\cup_{\mathcal{G}_{r-b}\in\mathfrak{G}_{r-b}}\mathcal{B}_{\mathcal{G}_{r-b}}$, coincides with the column-label set $\cup_{\mathcal{G}_b'\in\mathfrak{G}_b'}\mathcal{B}_{\mathcal{G}_b'}$ under $\mathcal{A}'$, i.e., $\cup_{\mathcal{G}_{r-b}\in\mathfrak{G}_{r-b}}\mathcal{B}_{\mathcal{G}_{r-b}}=\cup_{\mathcal{G}_b'\in\mathfrak{G}_b'}\mathcal{B}_{\mathcal{G}_b'}$. Similarly, we can also have $\cup_{\mathcal{G}_{r-b}'\in\mathfrak{G}_{r-b}'}\mathcal{B}_{\mathcal{G}_{r-b}'}= \cup_{\mathcal{G}_b\in\mathfrak{G}_b}\mathcal{B}_{\mathcal{G}_b}$. Because subarray $\mathbf{U}({\mathcal{A}\choose t},\cup_{\mathcal{G}_{r-b}\in \mathfrak{G}_{r-b}}\mathcal{B}_{\mathcal{G}_{r-b}})$ and subarray $\mathbf{U}({\mathcal{A}'\choose t},\cup_{\mathcal{G}_{r-b}'\in \mathfrak{G}_{r-b}'}\mathcal{B}_{\mathcal{G}_{r-b}'})$ are both all-star subarrays, $\mathbf{U}(\binom{\mathcal{A}}{t},\cup_{\mathcal{G}_b\in\mathfrak{G}_b}\mathcal{B}_{\mathcal{G}_b})$ and $\mathbf{U}(\binom{\mathcal{A}'}{t},\cup_{\mathcal{G}_b'\in\mathfrak{G}_b'}\mathcal{B}_{\mathcal{G}_b'})$ can be combined to form a new subarray$$\mathbf{U}\left({\mathcal{A}\choose t}\cup {\mathcal{A}'\choose t},\left(\cup_{\mathcal{G}_b\in\mathfrak{G}_b}\mathcal{B}_{\mathcal{G}_b}\right)\cup \left(\cup_{\mathcal{G}_b'\in\mathfrak{G}_b'}\mathcal{B}_{\mathcal{G}_b'}\right)\right),$$
which still satisfies \eqref{P-C4}. If $L<{t+r-b\choose b}$, we have $|\mathcal{X}|>1$. Then for a solution $\mathbf{x}\in\mathcal{X}$ under $\mathcal{A}$, there exists a corresponding solution $\mathbf{x}'\in\mathcal{X}'$ under $\mathcal{A}'$. In this case the subarrays$$\mathbf{U}\left({\mathcal{A}\choose t}\cup {\mathcal{A}'\choose t},\left(\cup_{x_{\mathcal{G}_i} = 1} \mathcal{B}_{\mathcal{G}_i}\right)\cup \left(\cup_{x_{\mathcal{G}_i'}' = 1} \mathcal{B}_{\mathcal{G}_i}\right)\right)$$ also satisfy \eqref{P-C4}, since $\mathbf{U}_{\mathcal{A},\mathbf{x}}^b$ and $\mathbf{U}_{\mathcal{A}',\mathbf{x}'}^b$ are respectively contained in $\mathbf{U}({\mathcal{A}\choose t},\cup_{\mathcal{G}_b\in \mathfrak{G}_b}\mathcal{B}_{\mathcal{G}_b})$ and $\mathbf{U}({\mathcal{A}’\choose t},\cup_{\mathcal{G}_b'\in \mathfrak{G}_b'}\mathcal{B}_{\mathcal{G}_b'})$.

For any $\mathcal{A}\in {[\Lambda]\choose t+r-b}$, according to Proposition \ref{pro-1}, we can obtain a solution $\mathbf{x}$ for the $(n,L,\mathbf{z},\mathbf{v})$ knapsack problem. Subsequently, we will describe the structure of the solution $\mathbf{x}$ under the conditions that $b<r<2b$, $t+r>2b$, $\Lambda=2(t+r-b)$ and $L\leq{t+r-b\choose b}$. For $i=b$, we select $L$ elements with $x_{\mathcal{G}_b}=1$ in $\mathfrak{G}_b$, and $x_{\mathcal{G}_b}=0$ for the remaining elements; For all $i\in[b+1:\min(r,t+r-b)]$ and all elements $\mathcal{G}_i\in \mathfrak{G}_i$, let $x_{\mathcal{G}_i}$=0. This completes the construction of the solution vector $\mathbf{x}$. Since
\begin{align*}
	\psi(\mathbf{z},\mathbf{x})=\sum\limits_{i \in \mathcal{I},\mathcal{G}_i\in \mathfrak{G}_i} z_{{\mathcal{G}_i}} x_{{\mathcal{G}_i}}&=\sum\limits_{i =b,\mathcal{G}_i\in \mathfrak{G}_i} z_{{\mathcal{G}_i}} x_{{\mathcal{G}_i}}+\sum\limits_{i\in [b+1:\min(r,t+r-b)],\mathcal{G}_i\in \mathfrak{G}_i} z_{{\mathcal{G}_i}} x_{{\mathcal{G}_i}} \\
	&=\sum\limits_{\mathcal{G}_b\in \mathfrak{G}_b} z_{{\mathcal{G}_b}} x_{{\mathcal{G}_b}} =L{r-b\choose r-b} =L,
\end{align*}$\mathbf{x}$ is a solution to the $(n, L, \mathbf{z}, \mathbf{v})$ knapsack problem. Then we can obtain subarray $\mathbf{U}_{b,\mathcal{A}}^{\mathbf{x}}$. According to Theorem \ref{the-olution}, we can obtain an $(L,\Lambda,r,M,N)$ MAMISO coded caching scheme with sum-DoF \begin{align*}
	\phi(\mathbf{v},\mathbf{x})=\sum_{i \in \mathcal{I},\mathcal{G}_i\in \mathfrak{G}_i} v_{{\mathcal{G}_i}} x_{{\mathcal{G}_i}}=\sum\limits_{i =1,\mathcal{G}_i\in \mathfrak{G}_i} v_{{\mathcal{G}_i}} x_{{\mathcal{G}_i}}++\sum\limits_{i\in [b+1:\min(r,t+r-b)],\mathcal{G}_i\in \mathfrak{G}_i} v_{{\mathcal{G}_i}} x_{{\mathcal{G}_i}} =L{t+r-b\choose r-b}.
\end{align*}
Since there is only one index $i=b$ that corresponds to the non-zero entries in $\mathbf{x}$, we have $p_1={\Lambda-t-r+b\choose b}$ and $q_1=L$. Hence,  $\ell_1=\text{LCM}(p_1,q_1)=\text{LCM}({\Lambda-t-r+b\choose b},L)$,  
\begin{align}\label{eq-th4-ell}
	\ell=\text{LCM}(\ell_1/q_1)=\frac{\text{LCM}({\Lambda-t-r+b\choose b},L)}{L}=\frac{{\Lambda-t-r+b\choose b}}{\gcd({\Lambda-t-r+b\choose b},L)}, 	
\end{align}
and
\begin{align}\label{eq-th4-mu}
	\mu=\text{LCM}(u_{\mathcal{G}_b})=1.
\end{align} can be derived. 
By Lemma \ref{element-number}, we have
\begin{align}\label{eq-th4-pi}
	\pi={r\choose r-b}\frac{\ell q_1}{p_1}\frac{\mu}{u_{\mathcal{G}_b}}={r\choose b}\frac{\text{LCM}({\Lambda-t-r+b\choose b},L)}{{\Lambda-t-r+b\choose b}}=\frac{{r\choose b}L}{\gcd({\Lambda-t-r+b\choose b},L)}.
\end{align}

When $\Lambda=6$, $r=3$, $t=2$, $b=2$ and $L=3$, we take $\mathcal{A}=\{1,2,3\}$ and $\mathcal{A}' = [\Lambda]\setminus\mathcal{A}=\{4,5,6\}$. The corresponding solutions are $\mathbf{x}=(x_{\{4,5\}}=1,x_{\{4,6\}}=1,x_{\{5,6\}}=1,x_{\{4,5,6\}}=0)$ and $\mathbf{x}'=(x_{\{1,2\}}=1,x_{\{1,3\}}=1,x_{\{2,3\}}=1,x_{\{1,2,3\}}=0)$. The relative positions of the subarrays $\mathbf{U}_{b,\mathcal{A}}^{\mathbf{x}}$ and $\mathbf{U}_{b,\mathcal{A}'}^{\mathbf{x}'}$ are illustrated in \eqref{Exa-Th4-b=2}, \eqref{Exa-Th4-b=1}, \eqref{Exa-Th4-b=2-2} and \eqref{Exa-Th4-b=1-2}. The two subarrays can be merged to form a new, larger multicast transmission. 

According to the filling strategy in \ref{subsect-filling}, we will change the vector $(1,\mathcal{A}',\mathbf{x}')$ to $(1,\mathcal{A},\mathbf{x})$. Due to $[\Lambda] = \mathcal{A}' \cup \mathcal{A}$, the number of vectors is reduced by half. So, the number of vectors in $\mathbf{U}'$ is
\begin{align*}
	S=\frac{\ell\mu{\Lambda\choose t+r-b}}{2}=\frac{{\Lambda\choose t+r-b}{\Lambda-t-r+b\choose b}}{2\gcd\left({\Lambda-t-r+b\choose b},L\right)}.
\end{align*}
Then we can obtain our desired user-delivery array $\mathbf{Q}$ by vertically replicating $\mathbf{U}'$ ${r\choose b}L/\gcd({\Lambda-t-r+b\choose b},L)$ times and filling the ${r\choose b}L/\gcd({\Lambda-t-r+b\choose b},L)$ vectors in each non-star entry of $\mathbf{U}'$ into their corresponding replicated entries across the ${r\choose b}L/\gcd({\Lambda-t-r+b\choose b},L)$ copies, respectively. In addition, $\mathbf{Q}$ has $F={\Lambda\choose t}{r\choose b}L/\gcd({\Lambda-t-r+b\choose b},L)$ rows since $\mathbf{U}'$ has exactly ${\Lambda\choose t}$ rows. Each column of $\mathbf{Q}$ has $Z= ({\Lambda\choose t}-{\Lambda-r\choose t}){r\choose b}L/\gcd({\Lambda-t-r+b\choose b},L)$. Therefore, $\mathbf{Q}$ is a $(L,K={\Lambda\choose r},F=\pi {\Lambda\choose t},Z=\pi ({\Lambda\choose t}-{\Lambda-r\choose t}),S=\ell\mu{\Lambda\choose t+r-b}/2 )$ MAPDA, where the parameters $\ell$, $\mu$ and $\pi$ are defined in \eqref{eq-th4-ell}, \eqref{eq-th4-mu} and \eqref{eq-th4-pi} respectively.
\subsection{Proof of Theorem \ref{C-L-MAPDA}}\label{C-L-MAPDA-proof}
Let $\mathbf{x}$ be a solution of the knapsack problem, and suppose there exists $\Lambda'\in[\Lambda]$ such that $\binom{\Lambda'-t}{r}\leq L$. Following the filling strategy described in Section \ref{subsect-filling}, we replace each subarray $\mathbf{U}_{b,\mathcal{A}}^{\mathbf{x}}$ by the subarray $\mathbf{U}({\mathcal{H}\choose t}, {\mathcal{H}\choose r})$, where $\mathbf{U}({\mathcal{H}\choose t}, {\mathcal{H}\choose r})$ is uniquely determined when $\mathcal{H}\in {[\Lambda]\choose \Lambda'}$.

For each $\mathcal{H}\in {[\Lambda]\choose \Lambda'}$, the subarray $\mathbf{U}({\mathcal{H}\choose t}, {\mathcal{H}\choose r})$ has exactly ${\Lambda'-t\choose r}$ Null entries in each row, and exactly ${\Lambda'-r\choose t}$ Null entries in each column. According to the filling strategy in \ref{subsect-filling}, $\mu=\text{LCM}({\Lambda'-r\choose t},{\Lambda'-r\choose t},\ldots, {\Lambda'-r\choose t})={\Lambda'-r\choose t}$. For each $\mathcal{T}\in {\mathcal{H}\choose t}$ and $\mathcal{D}\in {\mathcal{H}\choose r}$ with $\mathcal{D}\cap \mathcal{T} \neq \emptyset$, we fill the single vector (since $(\mu/{\Lambda'-r\choose t})=1$)
\begin{align}\label{fill-theorem1}
	(n,\mathcal{H})
\end{align} into the entry $\mathbf{U}_{\mathcal{T},\mathcal{D}}$ where $n$ is the order of the Null entries in the column labeled by $\mathcal{D}$. Each row of  $\mathbf{U}({\Lambda'\choose t}, {\Lambda'\choose r})$ has 
${\Lambda'-t\choose r}\leq L$ Null entries. So each vector $(n,\mathcal{H})$ satisfies Condition $4$ of Definition \ref{def-MAPDA}. 

Recall that for any $\mathcal{D}\in {\Lambda\choose r}$ and any $\mathcal{T}\in {\Lambda\choose t}$, the entry  $\mathbf{U}(\mathcal{T},\mathcal{D})$ is Null if and only if $ \mathcal{D}\cap \mathcal{T}=\emptyset$. We count all $\Lambda'$-subset $\mathcal{H}$ satisfying $\mathcal{T} \subset \mathcal{H}$, $\mathcal{D} \subset \mathcal{H}$ and $ \mathcal{D}\cap \mathcal{T}=\emptyset$. A straightforward combination yields
\begin{align}\label{fill-theorem5}
	{\Lambda-t-r \choose \Lambda'-t-r}
\end{align}such subsets $\mathcal{H}$. From \eqref{fill-theorem1} and \eqref{fill-theorem5}, we put exactly $\pi={\Lambda-t-r \choose \Lambda'-t-r}$ vectors in the entry $\mathbf{U}(\mathcal{T},\mathcal{D})$. 

Finally, we count the number of different vectors in $\mathbf{U}'$. According to \eqref{fill-ruler} $\mu={\Lambda'-r\choose t}$ vectors are filled into $\mathbf{U}({\mathcal{H}\choose t}, {\mathcal{H}\choose r})$. So, the number of vectors in $\mathbf{U}'$ is
\begin{align}
	\label{eq-vectors-number1}
	S={\Lambda'-r\choose t} {\Lambda\choose \Lambda'}.
\end{align}
Then we can obtain our desired user-delivery array $\mathbf{Q}$ by vertically replicating $\mathbf{U}'$ a total of ${\Lambda-t-r \choose \Lambda'-t-r}$ times and filling the ${\Lambda-t-r \choose \Lambda'-t-r}$ vectors in each non-star entry of $\mathbf{U}'$ into their corresponding replicated entries across the ${\Lambda-t-r \choose \Lambda'-t-r}$ copies, respectively. In addition, $\mathbf{Q}$ has $F={\Lambda-t-r \choose \Lambda'-t-r}\cdot {\Lambda\choose t}$ rows since $\mathbf{U}'$ has exactly ${\Lambda\choose t}$ rows. Each column of $\mathbf{Q}$ has $Z={\Lambda-t-r \choose \Lambda'-t-r}\cdot ({\Lambda\choose t}-{\Lambda-r\choose t})$. Therefore, $\mathbf{Q}$ is an $(L,K,F,Z,S)$ MAPDA with parameters$$K={\Lambda\choose r},F={\Lambda-t-r \choose \Lambda'-t-r}{\Lambda\choose t},Z={\Lambda-t-r \choose \Lambda'-t-r}\left({\Lambda\choose t}-{\Lambda-r\choose t}\right),S={\Lambda'-r\choose t} {\Lambda\choose \Lambda'}.$$

\section{Conclusion}\label{Conclusion} 
In this paper, we investigated the coded caching problem in the MAMISO network with a combinatorial topology. A novel MAPDA construction framework was proposed based on the knapsack problem. Under the full combination access topology, the proposed scheme achieves a higher sum-DoF than existing schemes, and attains the maximum sum-DoF when the parameters satisfy $L \geq {\Lambda-t \choose r} - {\Lambda-t-r \choose r}$. Furthermore, under the same full combination access topology, existing schemes can also be derived as special cases within this construction framework.

\appendices
\section{Proof of the partition on $\mathcal{B}_{\mathcal{G}_i}$ in \eqref{eq-B-G_i}}
\label{proof-B_G_i}

For any $\mathcal{A} \in \binom{[\Lambda]}{t+r-b}$ with $b \in [0:r-1]$ and any $\mathcal{D} \in \binom{[\Lambda]}{r}$, we have $|\mathcal{A} \cup \mathcal{D}| \leq \Lambda$ which gives $|\mathcal{D} \setminus \mathcal{A}| \leq \Lambda - t - r + b$. Together with $|\mathcal{D}| = r$, we obtain the upper bound $|\mathcal{D} \setminus \mathcal{A}| \leq \min\{r, \Lambda - t - r + b\}$. On the other hand, $|\mathcal{D} \setminus \mathcal{A}| \ge r - (t+r-b) = b-t$ and is nonnegative, yielding the lower bound $|\mathcal{D} \setminus \mathcal{A}| \ge \max\{b-t,0\}$. Consequently, we can obtain $|\mathcal{D} \setminus \mathcal{A}| \in [\max\{b-t,0\},\; \min\{r,\; \Lambda - t - r + b\}]$. 
 
For each integer $i \in[\max\{b-t,0\}:\min\{r,\Lambda-t-r+b\}]$, and for any $i$-subset $\mathcal{G}_i \in \binom{[\Lambda] \setminus \mathcal{A}}{i}$, we have define
\begin{align*}
	\mathcal{B}_{\mathcal{G}_i}=\left\{\mathcal{D}\in \binom{[\Lambda]}{r}\ \Big|\ \mathcal{D}\setminus\mathcal{A}=\mathcal{G}_i\right\}.
\end{align*}
Consider any $\mathcal{D}_1 \in \mathcal{B}_{\mathcal{G}_{i_1}}$ and $\mathcal{D}_2 \in \mathcal{B}_{\mathcal{G}_{i_2}}$, where $\mathcal{G}_{i_1}, \mathcal{G}_{i_2} \in \binom{[\Lambda] \setminus \mathcal{A}}{i}$ and $\mathcal{G}_{i_1} \ne \mathcal{G}_{i_2}$. By the definition of $\mathcal{B}_{\mathcal{G}_i}$, we have $\mathcal{D}_1 \setminus \mathcal{A} = \mathcal{G}_{i_1}$ and $\mathcal{D}_2 \setminus \mathcal{A} = \mathcal{G}_{i_2}$. Since $\mathcal{G}_{i_1} \ne \mathcal{G}_{i_2}$, it follows that $\mathcal{D}_1 \ne \mathcal{D}_2$ which implies $\mathcal{B}_{\mathcal{G}_{i_1}} \cap \mathcal{B}_{\mathcal{G}_{i_2}} = \emptyset$. For any $\mathcal{D} \in \binom{[\Lambda]}{r}$, we have $|\mathcal{D}\setminus\mathcal{A}| \in [\max\{b-t,0\}:\min\{r,\Lambda-t-r+b\}]$. Thus, there exists an integer $i \in [\max\{b-t,0\}:\min\{r,\Lambda-t-r+b\}]$ such that $\mathcal{D}\setminus\mathcal{A} \in \binom{[\Lambda]\setminus\mathcal{A}}{i}$, which means $\mathcal{D} \in \mathcal{B}_{\mathcal{D}\setminus\mathcal{A}}$. Consequently, the collection of all $\mathcal{B}_{\mathcal{G}_i}$ partitions $\binom{[\Lambda]}{r}$ where $i \in[\max\{b-t,0\}:\min\{r,\Lambda-t-r+b\}]$ and $\mathcal{G}_i \in \binom{[\Lambda] \setminus \mathcal{A}}{i}$. The proof of the Partition is complete.

%Moreover, for any $\mathcal{B}_{\mathcal{G}_{i_1}} \in \mathcal{B}_{i_1}$ and $\mathcal{B}_{\mathcal{G}_{i_2}} \in \mathcal{B}_{i_2}$ with $i_1 \ne i_2$, it follows from the definition that for any $\mathcal{D}_1 \in \mathcal{B}_{\mathcal{G}_{i_1}}$, we have $|\mathcal{D}_1 \setminus \mathcal{A}| = i_1$, and similarly for $\mathcal{D}_2 \in \mathcal{B}_{\mathcal{G}_{i_2}}$, we have $|\mathcal{D}_2 \setminus \mathcal{A}| = i_2$. Since $i_1 \ne i_2$, we conclude that
%\begin{align*}
%\mathcal{B}_{i_1} \cap \mathcal{B}_{i_2} = \emptyset.
%\end{align*}

\section{Proof of Proposition \ref{pro-1}}
\label{proof-U_A_B_G_i}
Let $\mathcal{A}\in {[\Lambda]\choose t+r-b}$ with $b\in [0:r-1]$. For each $i\in[ \max\{b-t,0\}:\min\{r,\Lambda-t-r+b\}]$, let $\mathfrak{G}_i={[\Lambda]\setminus\mathcal{A}\choose i}$. For any $\mathcal{G}_i\in \mathfrak{G}_i$, by \eqref{eq-B-G_i} have 
\begin{align*}
	\mathcal{B}_{\mathcal{G}_i}=\left\{\mathcal{D}\in \binom{[\Lambda]}{r}\ \Big|\ \mathcal{D}\setminus\mathcal{A}=\mathcal{G}_i\right\}.
\end{align*}Every $\mathcal{D} \in \mathcal{B}_{\mathcal{G}_i}$ can be written uniquely as $\mathcal{D} = \mathcal{D}' \cup \mathcal{G}_i$, where $\mathcal{D}'\subset \mathcal{A}$ and $\mathcal{D}'\cap \mathcal{G}_i=\emptyset$. Since $|\mathcal{G}_i|=i$ and $|\mathcal{D}|=r$, we have $|\mathcal{D}'|=r-i$; hence $\mathcal{D}'\in {\mathcal{A}\choose r-i}$. Because $|\mathcal{A}|=t+r-b$, the number of possible subsets $\mathcal{D}'$ is ${t+r-b\choose r-i}$. Consequently, $|\mathcal{B}_{\mathcal{G}_i}|={t+r-b\choose r-i}$. 
Thus the subarray $\mathbf{U}({\mathcal{A}\choose t},\mathcal{B}_{\mathcal{G}_i})$ contains exactly $v_{\mathcal{G}_i}={t+r-b\choose r-i}$ columns. Fix $\mathcal{T} \in \binom{\mathcal{A}}{t}$ and consider $\mathcal{B}_{\mathcal{G}_i}$. Every $\mathcal{D} \in \mathcal{B}_{\mathcal{G}_i}$ can be written uniquely as $\mathcal{D} = \mathcal{D}' \cup \mathcal{G}_i$, where $\mathcal{D}'\in {\mathcal{A}\choose r-i}$. The condition $\mathcal{D} \cap \mathcal{T} = \emptyset$ is equivalent to $\mathcal{D}' \cap \mathcal{T} = \emptyset$, i.e., $\mathcal{D}' \in \binom{\mathcal{A} \setminus \mathcal{T}}{r-i}$. Since $|\mathcal{A} \setminus \mathcal{T}| = r - b$, the number of such $\mathcal{D}'$ is $
{r-b\choose r-i}$. Hence, exactly ${r-b\choose r-i}$ elements of $\mathcal{B}_{\mathcal{G}_i}$ are disjoint from $\mathcal{T}$. By \eqref{eq-caching-retrieval-arrays}, each row indexed by $\mathcal{T}$ in the subarray $\mathbf{U}({\mathcal{A}, \mathcal{B}_{\mathcal{G}_i}})$ contains $z_{\mathcal{G}_i}=\binom{r - b}{r - i}$ Null entries. Take $\mathcal{D} \in \mathcal{B}_{\mathcal{G}_i}$. Since $|\mathcal{A} \cap \mathcal{D}| = r - i$, we have
\begin{align}\label{set-size-pro-th4}
	|\mathcal{A} \setminus \mathcal{D}|=t+r-b-(r-i)=t+i-b.
\end{align}
For any $\mathcal{T} \in \binom{\mathcal{A}}{t}$ satisfies $\mathcal{D} \cap \mathcal{T} = \emptyset$ if and only if $\mathcal{T} \in {\mathcal{A} \setminus \mathcal{D} \choose t}$. By \eqref{set-size-pro-th4}, the number of such $\mathcal{T}$ is $\binom{t+i-b}{t}$. Therefore, each column of $\mathbf{U}_{\mathcal{A}, \mathcal{B}_{\mathcal{G}_i}}$ contains $u_{\mathcal{G}_i}={t+i-b\choose t}$ Null entries. Then the proof of the Proposition \ref{pro-1} is completed.

\section{Proof of Theorem \ref{th-lower-DoF} }\label{corollary 3}
Consider any subset $\mathcal{A}\in {[\Lambda]\choose t+r-b}$ with $b\in [0:r-1]$. For each integer $i\in\mathcal{I}=[b:\min\{r,\Lambda-t-r+b\}]$ let $\mathfrak{G}_i={[\Lambda]\setminus\mathcal{A}\choose i}$. For any $\mathcal{G}_i\in \mathfrak{G}_i$, we have \begin{align}\label{decrease}
	v_{\mathcal{G}_i}/z_{\mathcal{G}_i}={t+r-b\choose r-i}/{r-b\choose r-i}&=\frac{(t+r-b)!}{(r-i)!(t-b+i)!}/\frac{(r-b)!}{(r-i)!(i-b)!}\nonumber  \\
	&=\frac{(t+r-b)!}{(t-b+i)!}/\frac{(r-b)!}{(i-b)!}\nonumber \\
	&=\frac{(t+r-b)!}{(r-b)!}\frac{(i-b)!}{(t-b+i)!}\nonumber \\
	&=\frac{(t+r-b)!}{(r-b)!}\frac{1}{(i-b+1)\cdot(i-b+2)\cdot \dots \cdot (i-b+t) }.
\end{align} According to \eqref{decrease}, $v_{\mathcal{G}_i}/z_{\mathcal{G}_i}$ decreases as $i$ increases, so $\mathbf{z}$ and $\mathbf{v}$ are already sorted in descending order by ratio. 

For any subset $\mathcal{A}\in {[\Lambda]\choose t+r}$, according to Proposition \ref{pro-1}, we obtain $\mathbf{z}=(z_{\mathcal{B}_{\mathcal{G}_i}})_{i \in \mathcal{I},\mathcal{G}_i\in \mathfrak{G}_i}$ and $
\mathbf{v}=(v_{\mathcal{B}_{\mathcal{G}_i}})_{i \in \mathcal{I},\mathcal{G}_i\in \mathfrak{G}_i}$. For each $\mathcal{G}_i\in\mathfrak{G}_i$, the associated parameters are given by $v_{\mathcal{G}_i}={t+r-b\choose r-i}$, $z_{\mathcal{G}_i}={r-b\choose r-i}$ and $u_{\mathcal{G}_i}={t+i-b\choose t}$. From \eqref{eq-delta}, $\delta$ is the smallest integer in $\mathcal{I}=[b: \min(r, \Lambda-t-r+b)]$ satisfying $L<\sum_{i=b}^{\delta}{\Lambda-t-r+b\choose i}\binom{r-b}{r-i}+\binom{r-b}{r-\delta-1}$. For all $i \in [b:\delta-1]$ and each subset $\mathcal{G}_i \in \mathfrak{G}_i$, let $x_{\mathcal{G}_i} = 1$; In $\mathfrak{G}_\delta$, we select 
\begin{align*}
\zeta=\min\left\{
\left\lfloor{\Lambda-t-r+b\choose \delta }+\frac{L-\sum_{i=b}^{\delta}{\Lambda-t-r+b\choose i}\binom{r-b}{r-i}}{{r-b\choose r-\delta}}\right\rfloor,{\Lambda-t-r+b\choose \delta }\right\}
\end{align*}elements with $x_{\mathcal{G}_g}=1$, and $x_{\mathcal{G}_\delta}=0$ for the remaining elements; For all $i \in [\delta+1:\min\{r,\Lambda-t-r+b\}]$ and all $\mathcal{G}_i \in \mathfrak{G}_i$, we set $x_{\mathcal{G}_i} = 0$. This completes the construction of the solution vector $\mathbf{x}$. If $L-\sum_{i=b}^{\delta}{\Lambda-t-r+b\choose i}\binom{r-b}{r-i}\geq 0$, we have 
\begin{align*}
	\psi(\mathbf{z},\mathbf{x})=\sum\limits_{i \in \mathcal{I},\mathcal{G}_i\in \mathfrak{G}_i} z_{{\mathcal{G}_i}} x_{{\mathcal{G}_i}}=\sum\limits_{i \in [b:\delta],\mathcal{G}_i\in \mathfrak{G}_i} z_{{\mathcal{G}_i}} x_{{\mathcal{G}_i}}
	=\sum_{i=b}^{\delta}{\Lambda-t-r+b\choose i}\binom{r-b}{r-i} 
	\leq L.
\end{align*}
If $L-\sum_{i=b}^{\delta}{\Lambda-t-r+b\choose i}\binom{r-b}{r-i}< 0$, we have
\begin{align*}
	\psi(\mathbf{z},\mathbf{x})=\sum\limits_{i \in \mathcal{I},\mathcal{G}_i\in \mathfrak{G}_i} z_{{\mathcal{G}_i}} x_{{\mathcal{G}_i}}&=\sum\limits_{i \in [b:\delta-1],\mathcal{G}_i\in \mathfrak{G}_i} z_{{\mathcal{G}_i}} x_{{\mathcal{G}_i}}+\sum\limits_{\mathcal{G}_r\in \mathfrak{G}_\delta} z_{{\mathcal{G}_\delta}} x_{{\mathcal{G}_\delta}} \\
	&=\sum_{i=b}^{\delta-1}{\Lambda-t-r+b\choose i}\binom{r-b}{r-i}+\left\lfloor{\Lambda-t-r+b\choose \delta }+\frac{L-\sum_{i=b}^{\delta}{\Lambda-t-r+b\choose i}\binom{r-b}{r-i}}{{r-b\choose r-\delta}}\right\rfloor{r-b\choose r-\delta} \\
	&\leq \sum_{i=b}^{\delta-1}{\Lambda-t-r+b\choose i}\binom{r-b}{r-i}+\left({\Lambda-t-r+b\choose \delta }+\frac{L-\sum_{i=b}^{\delta}{\Lambda-t-r+b\choose i}\binom{r-b}{r-i}}{{r-b\choose r-\delta}}\right){r-b\choose r-\delta} \\
	&=\sum_{i=b}^{\delta-1}{\Lambda-t-r+b\choose i}\binom{r-b}{r-i}+{\Lambda-t-r+b\choose \delta }{r-b\choose r-\delta}+L-\sum_{i=b}^{\delta}{\Lambda-t-r+b\choose i} \\
	&=L.
\end{align*}Thus $\psi(\mathbf{z},\mathbf{x})\le L$ in both cases, confirming that $\mathbf{x}$ is a feasible solution. By Theorem \ref{the-olution}, we can obtain an $(L,\Lambda,r,M,N)$ MAMISO coded caching scheme with sum-DoF 
\begin{align*}
	\phi(\mathbf{v},\mathbf{x})=\sum_{i \in \mathcal{I},\mathcal{G}_i\in \mathfrak{G}_i} v_{{\mathcal{G}_i}} x_{{\mathcal{G}_i}}&=\sum\limits_{i \in [b:\delta-1],\mathcal{G}_i\in \mathfrak{G}_i} v_{{\mathcal{G}_i}} x_{{\mathcal{G}_i}}+\sum\limits_{\mathcal{G}_\delta\in \mathfrak{G}_\delta} v_{{\mathcal{G}_\delta}} x_{{\mathcal{G}_\delta}}  \\
	&=\sum_{i=b}^{\delta-1}{\Lambda-t-r+b\choose i}{t+r-b\choose r-i}+\zeta {t+r-b\choose r-\delta}.
\end{align*}From the structure of $\mathbf{x}$, the number of indices $i$ with non-zero entries is $m = \delta-b+1$. These indices are $i_j = b-1+j$ for $j\in[m]$. 
For $j\in[m-1]$, we have $p_j={\Lambda-t-r+b\choose i_j}={\Lambda-t-r+b\choose b-1+j}$ and $q_j={\Lambda-t-r+b\choose i_j}={\Lambda-t-r+b\choose b-1+j}$. Since $p_j = q_j$, 
\begin{align*}
	\ell_j=\text{LCM}(p_j,q_j)=q_j={\Lambda-t-r+b\choose i_j}={\Lambda-t-r+b\choose b-1+j}
\end{align*} and $q_j/p_j=1$. For $j=m$, we have $i_m=\delta$, $p_m={\Lambda-t-r+b\choose \delta}$, $q_m=\zeta$. Then $\ell_m=\text{LCM}(p_m,q_m)=\text{LCM}({\Lambda-t-r+b\choose \delta},\zeta)$ and $q_m/p_m = \zeta/\binom{\Lambda-t-r+b}{\delta}$. Therefore, we have 
\begin{align}
	\label{Th2-prof-ell}
	\ell=\text{LCM}(\ell_j/q_j)_{j\in[m]}=\ell_m/q_m=\text{LCM}({\Lambda-t-r+b\choose \delta},\zeta)/\zeta.
\end{align} 
The parameter $\mu$ is given by 
\begin{align}
	\label{Th2-prof-mu}
	\mu=\text{LCM}(u_{\mathcal{G}_{i_j}})_{j\in[m]}=\text{LCM}{t+i_j-b\choose t}_{j\in[m]}=\text{LCM}{t+j-1\choose t}_{j\in[\delta-b+1]}.
\end{align}
By Lemma \ref{element-number}, we have
\begin{align}
	\label{Th2-prof-pi}
	\pi=\sum_{i=b}^{\delta-1}{r\choose r-i}{\Lambda-t-r\choose i-b} \frac{\text{LCM}\left({\Lambda-t-r+b\choose \delta},\zeta\right)}{\zeta }\frac{\text{LCM}{t+j-1\choose t}_{j\in[\delta-b+1]}}{{t+i-b\choose t}}\nonumber \\
	+ {r\choose r-\delta}{\Lambda-t-r\choose \delta-b} \frac{\text{LCM}\left({\Lambda-t-r+b\choose \delta},\zeta\right)}{{\Lambda-t-r+b\choose \delta}}\frac{\text{LCM}{t+j-1\choose t}_{j\in[\delta-b+1]}}{{t+\delta-b\choose t}}.
\end{align}
By Theorem \ref{the-olution}, we obtain an $(L,K,F,Z,S)$ MAPDA $\mathbf{Q}$, with parameters $$K={\Lambda\choose r},F=\pi{\Lambda\choose t},Z=\pi({\Lambda\choose t}-{\Lambda-r\choose t}),S=\mu\ell{\Lambda\choose t+r-b},$$where $\ell$, $\mu$, $\pi$ are defined in \eqref{Th2-prof-ell}, \eqref{Th2-prof-mu} and \eqref{Th2-prof-pi} respectively. The proof of the Theorem \ref{th-lower-DoF} is complete.

\section{Proof of Theorem \ref{th-optimal} }\label{corollary 2}
Set $b=0$ and assume $\Lambda\geq 2r+t$. Then the index set is  $$\mathcal{I}=[b:\min\{r,\Lambda-t-r+b\}]=[0:r].$$
For any subset $\mathcal{A}\in {[\Lambda]\choose t+r}$ and $\mathfrak{G}_i=\binom{[\Lambda]\setminus\mathcal{A}}{i}$ for $i\in\mathcal{I}$. By Proposition \ref{pro-1}, we have
\begin{align*}
	v_{\mathcal{G}_i}={t+r\choose r-i},\ \ z_{\mathcal{G}_i}={r\choose r-i},\ \ u_{\mathcal{G}_i}={t+i-b\choose t}
\end{align*}for each $\mathcal{G}_i\in\mathfrak{G}_i$. Note that the collection $\mathcal{B}_{\mathcal{G}_i}$ ($i\in \mathcal{I}$ and $\mathcal{G}_i \in \mathfrak{G}_i$) partitions ${[\Lambda]\choose r}$. Consequently, we have
\begin{align*}
	\sum_{i=0}^{r}v_{\mathcal{G}_i}|\mathfrak{G}_i|&=\sum_{i=0}^{r}{t+r\choose r-i}{\Lambda-t-r\choose i}={\Lambda \choose r},   \\
	\sum_{i=0}^{r}z_{\mathcal{G}_i}|\mathfrak{G}_i|&=\sum_{i=0}^{r}{r\choose r-i}{\Lambda-t-r\choose i}={\Lambda-t\choose r}.
\end{align*}For $L\geq {\Lambda-t\choose r} - {\Lambda-t-r\choose r}$, construct $\mathbf{x}$ as follows: \begin{itemize}
	\item For all $i \in [0:r-1]$ and all $\mathcal{G}_i \in \mathfrak{G}_i$, set $x_{\mathcal{G}_i} = 1$.
	\item In $\mathfrak{G}_r$ , select exactly $L-{\Lambda-t\choose r}+{\Lambda-t-r\choose r}$ elements and set $x_{\mathcal{G}_r}=1$  for them; set $x_{\mathcal{G}_r}=0$ for the remaining elements in $\mathfrak{G}_r$. 
\end{itemize}
Then, we can obtain
\begin{align*}
	\psi(\mathbf{z},\mathbf{x})=\sum\limits_{i \in \mathcal{I},\mathcal{G}_i\in \mathfrak{G}_i} z_{{\mathcal{G}_i}} x_{{\mathcal{G}_i}}&=\sum\limits_{i \in [0:r-1],\mathcal{G}_i\in \mathfrak{G}_i} z_{{\mathcal{G}_i}} x_{{\mathcal{G}_i}}+\sum\limits_{\mathcal{G}_r\in \mathfrak{G}_r} z_{{\mathcal{G}_r}} x_{{\mathcal{G}_r}} \\
	&=\sum\limits_{i \in \mathcal{I},\mathcal{G}_i\in \mathfrak{G}_i} z_{{\mathcal{G}_i}} -|\mathfrak{G}_r|z_{{\mathcal{G}_r}}+\left(L-{\Lambda-t\choose r}+{\Lambda-t-r\choose r}\right)z_{{\mathcal{G}_r}} \\
	&={\Lambda-t\choose r}-{\Lambda-t-r\choose r}\cdot 1+\left(L-{\Lambda-t\choose r}+{\Lambda-t-r\choose r}\right)\cdot 1 \\
	&=L
\end{align*} so $\mathbf{x}$ is a solution to the $(n, L, \mathbf{z}, \mathbf{v})$ knapsack problem. By Theorem \ref{the-olution}, we can obtain an $(L,\Lambda,r,M,N)$ MAMISO coded caching scheme with sum-DoF 
\begin{align*}
	\phi(\mathbf{v},\mathbf{x})=\sum_{i \in \mathcal{I},\mathcal{G}_i\in \mathfrak{G}_i} v_{{\mathcal{G}_i}} x_{{\mathcal{G}_i}}&=\sum\limits_{i \in [0:r-1],\mathcal{G}_i\in \mathfrak{G}_i} v_{{\mathcal{G}_i}} x_{{\mathcal{G}_i}}+\sum\limits_{\mathcal{G}_r\in \mathfrak{G}_r} v_{{\mathcal{G}_r}} x_{{\mathcal{G}_r}} \\
	&=\sum\limits_{i \in \mathcal{I},\mathcal{G}_i\in \mathfrak{G}_i} v_{{\mathcal{G}_i}} -|\mathfrak{G}_r|v_{{\mathcal{G}_r}}+\left(L-{\Lambda-t\choose r}+{\Lambda-t-r\choose r}\right)v_{{\mathcal{G}_r}} \\
	&={\Lambda \choose r}-{\Lambda-t-r\choose r}\cdot 1+\left(L-{\Lambda-t\choose r}+{\Lambda-t-r\choose r}\right)\cdot 1 \\
	&={\Lambda \choose r}+L-{\Lambda-t\choose r} \\
	&={\Lambda\choose r}\frac{{\Lambda\choose t}-{\Lambda-r\choose t}}{{\Lambda\choose t}}+L\\
	&=KZ/F+L=KM/N+L.
\end{align*}
When $L={\Lambda-t\choose r}-{\Lambda-t-r\choose r}$. Since $\mathbf{x}$ has non-zero entries only for $i=0,\dots,r-1$, thus $m=r$. For $j\in[m]$, setting  $i_j=j-1$, we have 
\begin{align*}
	p_j=q_j={\Lambda-t-r\choose i_j}, \ell_j=q_j, \ell_j/q_j=1,
\end{align*}thus $\ell=\text{LCM}(\ell_j/q_j)_{j\in[m]}=1$ and $q_j/p_j=1$.
In addition, we have $\mu=\text{LCM}(u_{\mathcal{G}_{i_j}})_{j\in[m]}=\text{LCM}{t+j-1\choose t}_{j\in[r]}$. By Lemma \ref{element-number}, we have 
\begin{align*}
\pi=\sum_{i=0}^{r-1}{r\choose r-i}{\Lambda-t-r\choose i}\frac{\text{LCM}{t+j-1\choose t}_{j\in[r]}}{{t+i\choose t}}.
\end{align*}
By Theorem \ref{the-olution}, we obtain an $(L,K,F,Z,S)$ MAPDA $\mathbf{Q}$, with parameters $$K={\Lambda\choose r},F=\pi{\Lambda\choose t},Z=\pi\left({\Lambda\choose t}-{\Lambda-r\choose t}\right),S=\mu\ell{\Lambda\choose t+r-b}.$$
When $L>{\Lambda-t\choose r}-{\Lambda-t-r\choose r}$. Since all indices $i=0,\dots,r$ appear with non-zero entries, so $m=r+1$. For $j\in[m-1]$, let $i_j=j-1$, then $p_j=q_j={\Lambda-t-r\choose i_j}$ and $\ell_j/q_j=\text{LCM}(p_j,q_j)/q_j=1$. For $j=m$, we have $i_m=r$, $p_m={\Lambda-t-r\choose i_m}$, $q_m=L-{\Lambda-t\choose r}+{\Lambda-t-r\choose r}$ and $\ell_m=\text{LCM}(p_m,q_m)=\text{LCM}\left({\Lambda-t-r\choose i_m},L-{\Lambda-t\choose r}+{\Lambda-t-r\choose r}\right)$, which implies 
\begin{align*}
	 \ell&=\text{LCM}(\ell_j/q_j)_{j\in[m]}=\ell_m/q_m=\frac{\text{LCM}({\Lambda-t-r\choose r},{L-{\Lambda-t\choose r}+{\Lambda-t-r\choose r}})}{L-{\Lambda-t\choose r}+{\Lambda-t-r\choose r}}, \\
\mu&=\text{LCM}(u_{\mathcal{G}_{i_j}})_{j\in[m]}=\text{LCM}{t+j-1\choose t}_{j\in[r+1]}.	 
\end{align*}By Lemma \ref{element-number}, we have 
\begin{align*}
	\pi=\sum_{i=0}^{r-1}{r\choose r-i}{\Lambda-t-r\choose i} \frac{\text{LCM}({\Lambda-t-r\choose r},{L-{\Lambda-t\choose r}+{\Lambda-t-r\choose r}})}{L-{\Lambda-t\choose r}+{\Lambda-t-r\choose r} }\frac{\text{LCM}{t+j-1\choose t}_{j\in[r+1]}}{{t+i\choose t}}+  \\ 
	{r\choose r-r}{\Lambda-t-r\choose r} \frac{\text{LCM}({\Lambda-t-r\choose r},{L-{\Lambda-t\choose r}+{\Lambda-t-r\choose r}})}{{\Lambda-t-r\choose r}}\frac{\text{LCM}{t+j-1\choose t}_{j\in[r+1]}}{{t+i\choose t}}.
\end{align*}
In both cases, Theorem \ref{the-olution} provides an $(L,K,F,Z,S)$ MAPDA $\mathbf{Q}$, with parameters $$K={\Lambda\choose r},F=\pi{\Lambda\choose t},Z=\pi\left({\Lambda\choose t}-{\Lambda-r\choose t}\right),S=\mu\ell{\Lambda\choose t+r-b}.$$
The proof of Theorem \ref{th-optimal} is complete.
%$\mathcal{L},\mathfrak{L},\mathscr{C},\ell$

\section{Proof of Proposition \ref{swap} }
\label{proof-obatian-structure}
Assume that vector $\mathbf{x}$ is a solution of the $(n,L,\mathbf{z},\mathbf{v})$ knapsack problem and satisfies that there exists two different subsets ${\mathcal{G}_i},{\mathcal{G}_i'} \in {[\Lambda]\setminus \mathcal{A} \choose i}$ with $x_{{\mathcal{G}_i}}=1$ and $x_{{\mathcal{G}_i'}}=0$. We can obtain a new vector $\mathbf{x}'=(x'_{\mathcal{G}_i})_{\mathcal{G}_i\in{[\Lambda]\setminus \mathcal{A} \choose i},i\in\mathcal{I} }$ by letting $x'_{{\mathcal{G}_i}}=x_{{\mathcal{G}_i'}}=0$, $x'_{{\mathcal{G}_i}}=x_{\mathcal{G}_i}=1$, and $x'_{{\mathcal{G}_i}}=x_{{\mathcal{G}_i''}}=0$ for each ${\mathcal{G}_i''\in {[\Lambda]\setminus \mathcal{A} \choose i}\setminus\{\mathcal{G}_i},{\mathcal{G}_i'} \}$ where $i\in\mathcal{I}$. That is, $\mathbf{x}'$ is obtained by exchanging the values in coordinates labeled by ${\mathcal{G}_i}$ and ${\mathcal{G}_i'}$, and the values of the other coordinates are unchanged. By Proposition\ref{pro-1} ${\mathcal{G}_i}$ and ${\mathcal{G}_i}'$ have $v_{\mathcal{G}_i}=v_{{\mathcal{G}_i}'}$ and $z_{\mathcal{G}_i}=z_{{\mathcal{G}_i}'}$, we can check that $\phi(\mathbf{v},\mathbf{x})=\phi(\mathbf{v},\mathbf{x}')$ and $\psi(\mathbf{z},\mathbf{x})=\psi(\mathbf{z},\mathbf{x}') \leq L$. Thus $\mathbf{x}'$ is also a solution.

\section{Proof of Lemma \ref{LCM-r-regular}}
\label{proof-r-regular}
For any positive integers $v$ and $k$ with $v\geq k$. Let $G =\{g^0, g^1, \ldots, g^{v-1}\}$ be a multiplicative cyclic group of order $v$. Set the point set $\mathcal{V} = G$ and define $z = \text{LCM}(v,k)$. Construct a collection of blocks $\mathfrak{R} = {\mathcal{R}_1, \mathcal{R}_2, \ldots, \mathcal{R}_{z/k}}$ as follows: for each $i \in [z/k]$
\begin{align*}
	\mathcal{R}_i=\{g^{(i-1)k}, g^{(i-1)k+1}, \ldots, g^{ik-1}\}.
\end{align*} We now verify that $(\mathcal{V}, \mathfrak{R})$ is an $r$-regular $(v,k)$ design with $r = z/v$. Consider an arbitrary point $g^n \in \mathcal{V}$ where $n \in [0:v-1]$. In the cyclic group $G$, $g^m = g^n$ if and only if $m \equiv n \pmod{v}$. Within the range $m \in [0:z-1]$, there are exactly $z/v$ integers $m$ satisfying $m \equiv n \pmod{v}$. Since $v \geq k$, the elements of each block $\mathcal{R}_i$ are distinct consecutive powers of $g$; hence $g^n$ can appear at most once in any block $\mathcal{R}_i$. Therefore, the point $g^n$ occurs in precisely $z/v$ distinct blocks of $\mathfrak{R}$.

Since every point $g^j \in \mathcal{V}$ (where $j\in [0:v-1]$) occurs in exactly $r = z/v$ blocks, and each block has size $k$, $(\mathcal{V}, \mathfrak{R})$ constitutes an $r$-regular $(v,k)$ design. Here $r = \text{LCM}(v,k)/v$.

\section{Proof of Lemma \ref{element-set-numbers}}
\label{appendix-lemma-a}   
Since the design $(\mathcal{V}_j, \mathcal{R}_j)$ is a $(\ell_j  /p_j)$-regular $(p_j,q_j)$ design, every point in $\mathcal{V}_j$ appears in exactly $\ell_j / p_j$ blocks. That is, any subset $\mathcal{G}_{i_j}$ in ${[\Lambda]\setminus \mathcal{A} \choose i_j}$ occurs $\ell_j / p_j$ times. By the cyclic selection defined in \eqref{eq-support-set}, each block in $\mathcal{R}_j$ is selected $\frac{\ell}{\ell_j/q_j}$ times. Therefore, there exist $\ell q_{j}/p_j$ blocks containing subsets $\mathcal{G}_{i_j}$.

\section{Proof of Lemma \ref{element-number}}
\label{appendex-lemma 5}  
Recall that for any $\mathcal{D}\in {\Lambda\choose r}$ and any $\mathcal{T}\in {\Lambda\choose t}$, the entry  $\mathbf{U}(\mathcal{T},\mathcal{D})$ is Null if and only if $ \mathcal{D}\cap \mathcal{T}=\emptyset$. From our filling strategy in \eqref{fill-ruler}, given the integer $b\in[0:r-1]$ we consider all the possible $(r+t-b)$-subset $\mathcal{A}$ satisfying $\mathcal{T} \subset \mathcal{A}$. Then according to the solution of $\mathbf{x}=(x_{\mathcal{G}_i})_{i\in\mathcal{I} ,\mathcal{G}_i\in\mathfrak{G}_i}$ where $x_{\mathcal{G}_{i_j}}=1$ for each integer $j\in[m]$, we have $|\mathcal{D}\setminus \mathcal{A}|\in \{i_1, i_2, \ldots, i_m\}$. In the following, we will first count the number of all possible sets $\mathcal{A}$, and then we will count the total number of vectors fill in $\mathbf{U}(\mathcal{T},\mathcal{D})$ for each given $\mathcal{A}$. It is not difficult to obtain that there are 
\begin{align}
\label{eq-choose-A}
\pi_1=\sum_{j \in [m]} {r\choose r-i_j}\cdot {\Lambda-r-t\choose i_j-b}
\end{align} possible subsets $\mathcal{A}$ since the intersection of $\mathcal{A}$ and $\mathcal{D}$ has exactly $r-i_j$ elements for each integer $j\in[m]$.   
	
Now, given an appropriate $\mathcal{A}$, without loss of generality we assume that $|\mathcal{D}\setminus \mathcal{A}|=i_1$. Let us count the total number of vectors fill in $\mathbf{U}(\mathcal{T},\mathcal{D})$. From \eqref{fill-ruler} we fill $\frac{\mu}{u_{\mathcal{G}_{i_j}}}$ vectors in $\mathbf{U}(\mathcal{T},\mathcal{D})$. In addition, given the subset $\mathcal{A}$ we consider exactly $\ell$ solution of the $(n,L,\mathbf{z},\mathbf{v})$ knapsack problem based on the set system $(\mathcal{V},\mathcal{R})$. By Lemma \ref{element-set-numbers}, the subset $\mathcal{D}$ in $\mathcal{G}_{i_1}$ is selected exactly $\ell q_{j}/p_j$ times since $\mathcal{G}_{i_1}$ occurs exactly $\ell q_{j}/p_j$ blocks. So, given a subset $\mathcal{A}$ the total number of vectors fill in $\mathbf{U}(\mathcal{T},\mathcal{D})$ is 
\begin{align}
\label{eq-A-number}	
\pi_2=\frac{\ell q_{j}}{p_j}\cdot\frac{\mu}{u_{\mathcal{G}_{i_j}}}.
\end{align}Clearly, it is sufficient to consider the number of all the possible $\mathcal{A}$. From \eqref{eq-choose-A} and \eqref{eq-A-number}, there are exactly $\pi=\pi_1\cdot \pi_2$ vectors fill in $\mathbf{U}(\mathcal{T},\mathcal{D})$. 
	
Finally, let us count the number of different vectors in $\mathbf{U}'$. From \eqref{fill-ruler} we fill $\mu$ vectors in $\mathbf{U}_{b,\mathcal{A}}^{\mathbf{x}}$. In addition, for each $\mathcal{A}\in{\Lambda\choose t+r-b}$ we consider exactly $\ell$ solution of the $(n,L,\mathbf{z},\mathbf{v})$ knapsack problem based on the set system $(\mathcal{V},\mathcal{R})$. So, the number of vectors in $\mathbf{U}'$ is
\begin{align}
\label{eq-vectors-number}
S=\ell\mu{\Lambda\choose t+r-b}.
\end{align}

\bibliographystyle{IEEEtran}
\bibliography{references}
\end{document}